\documentclass[12pt]{article}
\usepackage{graphicx}
\usepackage{amsmath,amssymb}
\usepackage{mathrsfs}
\usepackage[nosort]{cite}
\usepackage{color}

\allowdisplaybreaks

\begin{document}

\baselineskip=17pt

\begin{titlepage}
\rightline{\tt arXiv:1307.7110}
\rightline{\tt RIKEN-MP-73}

\begin{center}
\vskip 3cm
{\Large \bf {Primordial spectra from sudden turning trajectory }}\\[5mm]
\vskip 1cm
{\large {Toshifumi Noumi$^1$ and Masahide Yamaguchi$^2$}}
\vskip 1.0cm
$^1${\it {Mathematical Physics Laboratory, RIKEN Nishina Center, Saitama 351-0198, JAPAN}}\\[0.5cm]
$^2${\it Department of Physics, Tokyo Institute of Technology, Tokyo
152-8551, Japan}\\[0.5cm]
toshifumi.noumi``at"riken.jp, gucci``at"phys.titech.ac.jp

\vskip 2.0cm

{\bf Abstract}
\end{center}

\noindent

Effects of heavy fields on primordial spectra of curvature perturbations
are discussed in inflationary models with a sudden turning trajectory.
When heavy fields are excited after the sudden turn and oscillate around
the bottom of the potential, the following two effects are generically
induced: deformation of the inflationary background spacetime and
conversion interactions between adiabatic and isocurvature
perturbations, both of which can affect the primordial density
perturbations. In this paper, we calculate primordial spectra in
inflationary models with sudden turning potentials taking into account
both of the two effects appropriately. We find that there are some
non-trivial correlations between the two effects in the power spectrum
and, as a consequence, the primordial scalar power spectrum has a peak
around the scale exiting the horizon at the turn. Though both effects
can induce parametric resonance amplifications, they are shown to be
canceled out for the case with the canonical kinetic terms. The peak
feature and the scale dependence of bispectra are also discussed.

\end{titlepage}

\newpage

\section{Introduction}
\setcounter{equation}{0}

Inflation is strongly supported by recent observations of cosmic
microwave background (CMB) anisotropies
\cite{Bennett:2012zja,Ade:2013ktc}. In particular, single field
slow-roll inflation predicts almost adiabatic, Gaussian, and scale
invariant primordial curvature perturbations, and these predictions well
fit the observational results.
On the other hand,
high energy theories such as
supergravity and superstring theory
generically predict additional scalar fields
other than inflaton.
To reconcile
such a generic prediction of theories
with recent observations,
it may be suggested that only one light field plays a
role of inflaton while the others are heavy
and their effects are negligible.
In fact,
propagations of such heavy fields
are generically suppressed by
their mass
and completely negligible
in the heavy mass limit:
heavy fields can affect
inflationary dynamics
only via their background energy
and
their effects
can be absorbed into
the inflaton potential~\cite{Yamaguchi:2005qm}.

\medskip
However,
heavy field propagations
can lead to non-negligible effects
if their interactions with inflaton
are comparable to their mass.
In particular,
it was pointed out
in~\cite{Tolley:2009fg}
that
heavy fields can reduce
the effective sound speed of
adiabatic perturbations significantly.
Then, a lot of works
are devoted to the investigation on the effects of heavy fields
\cite{Achucarro:2010jv,Jackson:2010cw,Cremonini:2010ua,Chen:2011zf,Chen:2011tu,Shiu:2011qw,Cespedes:2012hu,Achucarro:2012sm,Avgoustidis:2012yc,Chen:2012ge,Pi:2012gf,Achucarro:2012yr,Gao:2012uq,Saito:2012pd,Burgess:2012dz,Gwyn:2012mw,Kobayashi:2012kc,Noumi:2012vr,Achucarro:2012fd,Battefeld:2013xka,Saito:2013aqa,Cespedes:2013rda,Assassi:2013gxa,Gong:2013sma,Gao:2013ota}.
As discussed in~\cite{Achucarro:2012sm,Avgoustidis:2012yc,Gwyn:2012mw},
even if heavy field propagations have non-negligible effects,
they can be integrated out simply
and
inflationary dynamics can be described
by effective single-field models~\cite{Cheung:2007st}
as long as time-dependence of interactions
are negligible compared to the mass of heavy fields.\footnote{See also
Appendix A in Ref. \cite{Noumi:2012vr} for the condition of integrating
out heavy fields.}
On the other hand,
when time-dependence of interactions are non-negligible,
it is necessary to follow the full dynamics of heavy fields
in general.

\medskip
One typical example for the latter
is the case when
heavy fields are excited
for example by the sudden turn of the potential
or the phase transition
and oscillates with high frequency.
Such oscillations of heavy fields
can affect inflationary dynamics
and can be a probe of high energy physics~\cite{Chen:2011zf,Chen:2011tu,Shiu:2011qw,Cespedes:2012hu,Gao:2012uq,Saito:2012pd,Kobayashi:2012kc,Battefeld:2013xka,Saito:2013aqa,Gao:2013ota,Ashoorioon:2008qr,Gao:2013hn}.
In general,
oscillations of heavy fields
generate the following two significant effects:
the modification of the Hubble parameter
and the conversion effect,
that is, the mixing between
adiabatic and isocurvature (heavy field) modes.
Theses effects can leave an imprint on the primordial curvature perturbations.
In particular, the parametric amplification of the
curvature perturbations~\cite{Chen:2011zf,Chen:2011tu,Saito:2012pd,Battefeld:2013xka,Saito:2013aqa}
and the peak at the scale exiting the
horizon during the heavy field oscillations may be observable~\cite{Shiu:2011qw,Cespedes:2012hu,Gao:2012uq,Kobayashi:2012kc,Gao:2013ota}.

\medskip
In this paper,
we investigate these effects in detail and evaluate the
power spectra and bispectra of the primordial curvature perturbations
in inflationary models with sudden turning potentials.
For simplicity,
we concentrate on the case with the canonical kinetic terms
in this paper.\footnote{
Although some results
depend on this setting,
it is straightforward to
extend our discussions
to more general setups
such as those with non-canonical kinetic terms
or derivative interactions.
The results will be given
elsewhere~\cite{NY}.}
In our calculations,
we take the kinetic basis of primordial perturbations,
where
scalar perturbations
are described in terms of
adiabatic perturbations (which are along the background trajectory)
and massive isocurvature perturbations
(which are orthogonal to the trajectory).
As explained before, the deviation from the
single-field slow-roll inflation caused by the presence of heavy field oscillations
is characterized by the following two effects: the deformation effect of
the Hubble parameter and the conversion effect between the adiabatic and
the isocurvature modes.
We first evaluate
the primordial power spectrum
and show that
the parametric
resonance amplification
occurs from both of the two effects
and
the peak at the turning scale
arises from the conversion effect.
It is, however, explicitly shown that
resonance effects from the two effects
accidentally cancel each other out
for the case with the canonical kinetic terms.\footnote{
Very recently in~\cite{Gao:2013ota},
the effects of sudden turning trajectory on the primordial power spectrum
were discussed in the potential basis
taking into account both of the deformation of the Hubble parameter and the conversion effect.
It was shown there
that the parametric resonance effects
are significantly suppressed for the models with the canonical kinetic terms.
Although the potential basis
is useful to discuss the absence of resonance effects,
we would like to point out
that the potential basis calculation
suffers from spurious singularities associated
with the suddenness of the turning potential.
In this paper
we take the kinetic basis
to avoid this kind of spurious singularities
and explicitly show that
the resonance cancellation appears
for the canonical kinetic term case.
See also footnote~\ref{footnote:kinetic_basis}
in Sec~\ref{subsubsec:action_kin}
and footnote~\ref{footnote:comments_on_GLM} in Sec~\ref{subsec:totalPS} for related comments.}
As a consequence,
the peak at the turning scale
becomes clear and
this feature characterizes this class of models
with heavy field oscillations.

\medskip
We also evaluate primordial bispectra induced by
the sudden turning potentials.
The main source of the bispectra
comes again from
the deformation of the Hubble parameter
and the conversion effect.
The shape and the scale dependence of the bispectrum for each effect is investigated.
We find
resonance and peak features
in the bispectra as in the case of power spectra.\footnote{
Resonance features from the Hubble deformation effects
were discussed in~\cite{Saito:2013aqa}
using the potential basis.}
Although the size of bispectra
is not necessarily large,
our results may be useful for probing these effects
observationally.

\medskip
The organization of this paper is as follows:
In the next section
we
give our setup to characterize the sudden turning trajectory, in which there are a slow-roll direction and a heavy field direction.
We then investigate the background trajectory and
discuss the deformation effect of the Hubble parameter caused by the
oscillation of the heavy field.
By employing the kinetic basis and
introducing the Goldstone boson
based on the effective field theory approach to inflation
\cite{Cheung:2007st,Senatore:2010wk,Noumi:2012vr},
we write down the action for the
adiabatic and the isocurvature (heavy field) modes.
In Sec.~\ref{sec:PS},
the power spectrum of the primordial curvature
perturbations is evaluated using the in-in formalism.
The effects of the deformation of
the Hubble parameter and the conversion interaction
between the adiabatic and isocurvature perturbations are discussed in detail.
In particular,
it is explicitly shown that
resonance effects from the two effects
cancel each other out
in the case with canonical kinetic terms
and
an analytic expression
for the power spectrum
in the heavy mass approximation
is also given.
In Sec.~\ref{sec:bispectrum},
the bispectrum of the primordial curvature perturbations
is evaluated
and
its shape and scale dependence is investigated.
Final section is devoted to the summary and discussion.
Some technical details are summarized in Appendices.

\section{Setup}
\label{sec:setup}
\setcounter{equation}{0}

\subsection{Modeling the potential}
\begin{figure}[t]
\begin{center}
\includegraphics[width=100mm, bb=0 0 580 285]{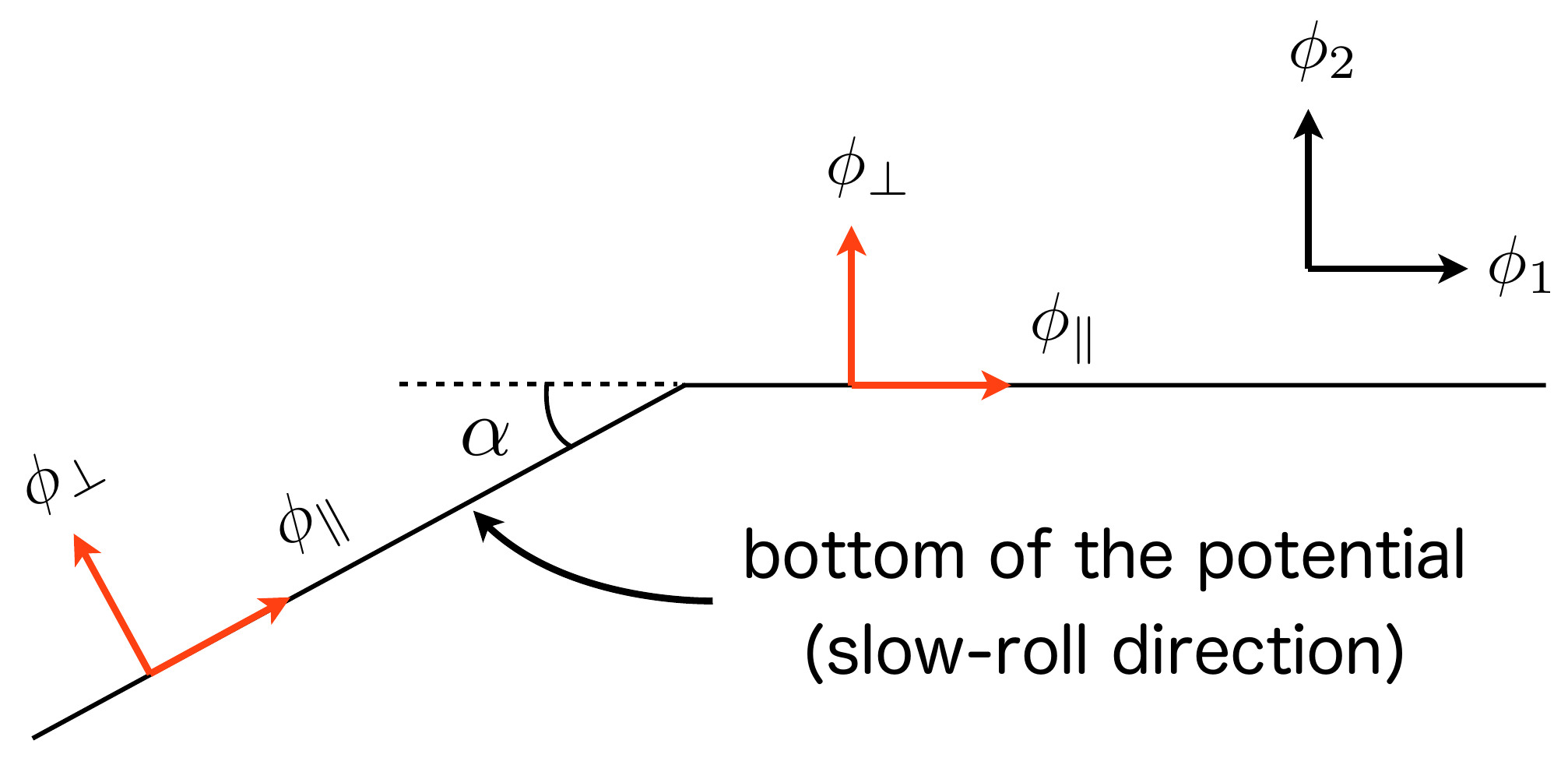}
\end{center}
\vspace{-5mm}
\caption{Sudden turning potential with turning angle $\alpha$.}
\label{fig:pot}
\end{figure}

In this paper
we consider a class of two-field models
with
the Einstein-Hilbert term,
the canonical kinetic terms of scalar fields,
and a sudden turning potential:
\begin{align}
\label{total_action}
S=\int d^4x\sqrt{-g}\left[\frac{1}{2}M_{\rm Pl}^2R-\sum_{i=1,2}\frac{1}{2}\partial_\mu\phi_i\partial^\mu\phi_i-V(\phi_i)\right]\,.
\end{align}
As depicted in~Fig.~\ref{fig:pot},
the potential $V(\phi_i)$ is decomposed into
the slow-roll direction $\phi_\parallel$
and the massive direction $\phi_\bot$ as
\begin{align}
V(\phi_i)=V_{\rm sr}(\phi_\parallel)+V_\bot(\phi_\bot)\,.
\end{align}
Here $V_{\rm sr}(\phi_\parallel)$ is a slow-roll potential
and the massive potential $V_\bot$
takes its minimum value at $\phi_\bot=0$.
The slow-roll direction turns at $(\phi_\parallel,\phi_\bot)=(0,0)$
and we identify this point with $(\phi_1,\phi_2)=(0,0)$ in the $\phi_i$ basis.
Before and after the turn,
the slow-roll direction is expressed in the $\phi_i$ basis as
$(\cos \alpha,\sin \alpha)$ and $(1,0)$, respectively.
The relations between the potential basis labeled by $(\phi_\parallel,\phi_\bot)$
and the $\phi_i$ basis are given by
\begin{align}
\label{pot_to_i1}
(\phi_\parallel,\phi_\bot)&=(\cos \alpha\,\phi_1+\sin\alpha\,\phi_2,
-\sin\alpha\,\phi_1+\cos\alpha\,\phi_2)
\quad\text{before the turn,}\\
\label{pot_to_i2}
(\phi_\parallel,\phi_\bot)&=(\phi_1,
\phi_2)
\quad\text{after the turn.}
\end{align}
Then, the potential $V(\phi_i)$ takes the following form
in the $\phi_i$ basis:\footnote{
The coordinate system $(\phi_\parallel,\phi_\bot)$
is not continuously connected
before and after the turn
so that the expression~(\ref{potential_i})
may not seem to be well-defined around the turning point.
However, as discussed in Appendix~\ref{app:sudden},
the potential~(\ref{potential_i}) can be regularized
and defined appropriately
by introducing a one-parameter family of turning potentials
and taking its sudden turning limit.}
\begin{align}
\nonumber
V(\phi_i)&=
\theta(-\phi_1)\Big[V_{\rm sr}(\cos\alpha\,\phi_1+\sin\alpha\,\phi_2)
+V_\bot(-\sin\alpha\,\phi_1+\cos\alpha\,\phi_2)\Big]\\
\label{potential_i}
&\quad+\theta(\phi_1)\Big[V_{\rm sr}(\phi_1)+V_\bot(\phi_2)\Big]\,.
\end{align}
In the following,
we assume the following form of the massive potential $V_\bot$:
\begin{align}
\label{massive_potential}
V_\bot(\phi_\bot)=\frac{m^2}{2}\phi_\bot^2+\frac{\lambda}{4!}\phi_\bot^4\,,
\end{align}
which preserves a $Z_2$ symmetry $\phi_\bot\to-\phi_\bot$
and seems to be a natural choice.

\subsection{Background evolutions}
Let us then summarize the classical dynamics of this class of models.
Suppose that
the background trajectory is along the slow-roll direction
before arriving at the turning point:
\begin{align}
\bar{\phi}_1(t)=\cos \alpha\, \bar{\phi}_{\rm sr}(t)\,,
\quad
\bar{\phi}_2(t)=\sin\alpha\,\bar{\phi}_{\rm sr}(t)
\quad
{\rm for}
\quad
t<t_\ast\,.
\end{align}
Here
$\bar{\phi}_{\rm sr}$ is the background evolution
in single field slow-roll inflation
and
$t_\ast$ is the time
when the background trajectory passes the turning point:
$\bar{\phi}_1(t_\ast)=\bar{\phi}_2(t_\ast)=\bar{\phi}_{\rm sr}(t_\ast)=0$.
We also introduce the Hubble parameter $H_{\rm sr}$
corresponding to $\bar{\phi}_{\rm sr}$ as
\begin{align}
\label{eom_slow-roll}
\ddot{\bar{\phi}}_{\rm sr}+3H_{\rm sr}\,\dot{\bar{\phi}}_{\rm sr}
+V^\prime(\bar{\phi}_{\rm sr})=0\,,
\quad
\frac{1}{2}\dot{\bar{\phi}}_{\rm sr}^2
+V_{\rm sr}(\bar{\phi}_{\rm sr})=3M_{\rm Pl}^2H_{\rm sr}^2\,,
\quad
\dot{\bar{\phi}}_{\rm sr}^2=-2M_{\rm Pl}^2\dot{H}_{\rm sr}\,,
\end{align}
and assume that
$H_{\rm sr}$ satisfies the slow-roll conditions:
$\displaystyle\epsilon_{\rm sr}=-\frac{\dot{H}_{\rm sr}}{H^2_{\rm sr}}\ll1$
and $\displaystyle\eta_{\rm sr}=\frac{\dot{\epsilon}_{\rm sr}}{H_{\rm sr}\epsilon_{\rm sr}}\ll1$.
After passing the turning point,
the background trajectory
first deviates from the slow-roll direction
and then
would be getting to approach that direction
because of the cosmic expansion:
\begin{align}
\bar{\phi}_1(t)\sim\bar{\phi}_{\rm sr}(t)\,,
\quad
\bar{\phi}_2(t)\sim0
\quad
{\rm for}
\quad
t\gg t_\ast\,.
\end{align}
We therefore decompose the background trajectory as
\begin{align}
\bar{\phi}_1(t)&=\theta(t_\ast-t)\cos \alpha\, \bar{\phi}_{\rm sr}(t)
+\theta(t-t_\ast)\big(\bar{\phi}_{\rm sr}(t)+\varphi_1(t)\big)\,,\\
\bar{\phi}_2(t)&=\theta(t_\ast-t)\sin \alpha\, \bar{\phi}_{\rm sr}(t)
+\theta(t-t_\ast)\varphi_2(t)\,.
\end{align}
Here $\varphi_i$'s describe the deviation of the background trajectory
from the slow-roll direction of the potential
and characterize the turning background trajectory.
In the following discussions,
we assume that $\alpha$ and
$\varphi_i$'s can be treated as perturbations.

\paragraph{Background evolution of $\phi_i$'s}
We then solve the background equations of motion
after passing the turning point $(t>t_\ast)$.
In terms of $\bar{\phi}_i$'s,
they are given by
\begin{align}
\label{background_eom_i1}
&\ddot{\bar{\phi}}_1+3H\dot{\bar{\phi}}_1+V^\prime_{\rm sr}(\bar{\phi}_1)=0\,,
\quad
\ddot{\bar{\phi}}_2+3H\dot{\bar{\phi}}_2+V^\prime_\bot(\bar{\phi}_2)=0\,,\\
\label{background_eom_i2}
&\frac{1}{2}\big(\dot{\bar{\phi}}_1^2+\dot{\bar{\phi}}_2^2\big)
+V_{\rm sr}(\bar{\phi}_1)+V_\bot(\bar{\phi}_2)
=3M_{\rm Pl}^2H^2\,,
\quad
\dot{\bar{\phi}}_1^2+\dot{\bar{\phi}}_2^2=-2M_{\rm Pl}^2\dot{H}\,.
\end{align}
Using the equations of motion~(\ref{eom_slow-roll}) in single field inflation,
we can rewrite~(\ref{background_eom_i1}) and (\ref{background_eom_i2})
in terms of $\varphi_i$'s as
\begin{align}
\label{background_eom_var1}
\ddot{\varphi}_1+3H_{\rm sr}\dot{\varphi}_1=0\,,
\quad
\ddot{\varphi}_2+3H_{\rm sr}\dot{\varphi}_2+m^2\varphi_2=0\,,\\*
\label{background_eom_H}
\dot{\bar{\phi}}_{\rm sr}\dot{\varphi}_1
+V^\prime_{\rm sr}(\bar{\phi}_{\rm sr})\varphi_1
+\frac{1}{2}\dot{\varphi}_2^2
+\frac{1}{2}V_\bot^{\prime\prime}(0)\varphi_2^2
=6M_{\rm Pl}^2H_{\rm sr}\delta H\,,\\*
\label{background_eom_Hd}
2\dot{\bar{\phi}}_{\rm sr}\dot{\varphi}_1
+\dot{\varphi}_2^2
=-2M_{\rm Pl}^2\delta\dot{H}
\,,
\end{align}
where $\delta H$
is defined by $\delta H=H-H_{\rm sr}$.
The initial conditions on $\varphi_i$'s at $t=t_\ast$ are given by
\begin{align}
\varphi_1(t_\ast)=\varphi_2(t_\ast)=0\,,
\quad
\dot{\varphi}_1(t_\ast)=-\frac{\alpha^2}{2}\dot{\bar{\phi}}_{\rm sr}(t_\ast)\,,
\quad
\dot{\varphi}_2(t_\ast)=\alpha\,\dot{\bar{\phi}}_{\rm sr}(t_\ast)\,.
\end{align}
Here and in what follows,
we drop higher order terms in $\epsilon_{\rm sr}$, $\eta_{\rm sr}$, $\varphi_i$'s, and $\alpha$.
The equations of motion (\ref{background_eom_var1}) can be then solved as follows:
\begin{align}
\varphi_1(t)
&=\frac{\alpha^2\dot{\bar{\phi}}_{\rm sr}}{6H_{\rm sr}}\big(e^{-3H_{\rm sr}(t-t_\ast)}-1\big)\,,\\[2mm]
\label{varphi2_phi}
\varphi_2(t)
&=\left\{\begin{array}{c}
\displaystyle\frac{\alpha\dot{\bar{\phi}}_{\rm sr}}{\mu H_{\rm sr}} e^{-\frac{3}{2}H_{\rm sr}(t-t_\ast)}\sin [\mu H_{\rm sr}(t-t_\ast)]
\quad
{\rm for}
\quad
m>\frac{3}{2}H_{\rm sr}\,,
\\[5mm]
\displaystyle\frac{\alpha\dot{\bar{\phi}}_{\rm sr}}{\nu H_{\rm sr}} e^{-\frac{3}{2}H_{\rm sr}(t-t_\ast)}\sinh [\nu H_{\rm sr}(t-t_\ast)]
\quad
{\rm for}
\quad
m<\frac{3}{2}H_{\rm sr}\,,
\end{array}\right.
\end{align}
where $\displaystyle\mu=\sqrt{\frac{m^2}{H_{\rm sr}^2}-\frac{9}{4}}$
for $ m>\frac{3}{2}H_{\rm sr}$
and
$\displaystyle\nu=\sqrt{\frac{9}{4}-\frac{m^2}{H_{\rm sr}^2}}$
for $ m<\frac{3}{2}H_{\rm sr}$.
As is clear from~(\ref{varphi2_phi}),
the expression for $m<\frac{3}{2}H_{\rm sr}$
can be obtained by the replacement $\mu\to-i\nu$ in that for $m>\frac{3}{2}H_{\rm sr}$.
Therefore, we do not write the expression for $m<\frac{3}{2}H_{\rm sr}$ explicitly
in the following discussions.
Using $\dot{\bar{\phi}}_{\rm sr}=\sqrt{2}\epsilon_{\rm sr}^{1/2}M_{\rm Pl}H_{\rm sr}$,
$\varphi_i$'s can be rewritten as
\begin{align}
\varphi_1(t)
&=\frac{\sqrt{2}\alpha^2}{6}\epsilon_{\rm sr}^{1/2}M_{\rm Pl}\big(e^{-3H_{\rm sr}(t-t_\ast)}-1\big)\,,\\[2mm]
\label{vaphi2_in_epsilon}
\varphi_2(t)
&=
\displaystyle\frac{\sqrt{2}\alpha}{\mu} \epsilon_{\rm sr}^{1/2}M_{\rm Pl}\, e^{-\frac{3}{2}H_{\rm sr}(t-t_\ast)}\sin [\mu H_{\rm sr}(t-t_\ast)]\,.
\end{align}
Notice that $\varphi_1$ and $\varphi_2$
are second and first order in $\alpha$, respectively.
Time-derivatives of $\varphi_i$'s are given by
\begin{align}
\nonumber
\dot{\varphi}_1(t)&=-\frac{1}{2}\alpha^2\dot{\bar{\phi}}_{\rm sr}\,
e^{-3H_{\rm sr}(t-t_\ast)}\\
&=-\frac{\sqrt{2}}{2}\alpha^2\epsilon_{\rm sr}^{1/2}M_{\rm Pl}H_{\rm sr}e^{-3H_{\rm sr}(t-t_\ast)}\,,\\
\nonumber
\dot{\varphi}_2(t)
&=\alpha \dot{\bar{\phi}}_{\rm sr}\,
e^{-\frac{3}{2}H_{\rm sr}(t-t_\ast)}\Big(-\frac{3}{2\mu}\sin [\mu H_{\rm sr}(t-t_\ast)]+\cos [\mu H_{\rm sr}(t-t_\ast)]\Big)\\
&=\sqrt{2}\alpha \epsilon_{\rm sr}^{1/2}M_{\rm Pl}H_{\rm sr}\,
e^{-\frac{3}{2}H_{\rm sr}(t-t_\ast)}\Big(-\frac{3}{2\mu}\sin [\mu H_{\rm sr}(t-t_\ast)]+\cos [\mu H_{\rm sr}(t-t_\ast)]\Big)\,.
\end{align}

\paragraph{Background spacetime}
We next discuss the background spacetime
and the deviation $\delta H(t)$ of the Hubble parameter
from single field slow-roll inflation.
First,
it follows
from the equation of motion (\ref{background_eom_Hd})
that $\delta\dot{H}$ for $t\geq t_\ast$ is given by
\begin{align}
\nonumber
\delta \dot{H}&=-\frac{1}{2M_{\rm Pl}^2}(2\dot{\bar{\phi}}_{\rm sr}\dot{\varphi}_1
+\dot{\varphi}_2^2)\\*
\nonumber
&=\alpha^2\frac{\dot{\bar{\phi}}_{\rm sr}^2}{M_{\rm Pl}^2}
e^{-3H_{\rm sr}(t-t_\ast)}
\left[\Big(\frac{1}{4}-\frac{9}{16\mu^2}\Big)
\big(1-\cos[2\mu H_{\rm sr}(t-t_\ast)]\big)
+\frac{3}{4\mu}\sin[2\mu H_{\rm sr}(t-t_\ast)]
\right]\\*
\label{before_kappa}
&=\alpha^2\epsilon_{\rm sr}H_{\rm sr}^2
e^{-3H_{\rm sr}(t-t_\ast)}
\left[\Big(\frac{1}{2}-\frac{9}{8\mu^2}\Big)
\big(1-\cos[2\mu H_{\rm sr}(t-t_\ast)]\big)
+\frac{3}{2\mu}\sin[2\mu H_{\rm sr}(t-t_\ast)]
\right]\,,
\end{align}
which is second order in $\alpha$.
It is convenient to introduce a function $\kappa(t)$ defined by
\begin{align}
\nonumber
\kappa(t)
&=2\frac{\dot{\varphi}_1}{\dot{\bar{\phi}}_{\rm sr}}+\left(\frac{\dot{\varphi}_2}{\dot{\bar{\phi}}_{\rm sr}}\right)^2\\*
&=-\alpha^2
e^{-3H_{\rm sr}(t-t_\ast)}
\left[\Big(\frac{1}{2}-\frac{9}{8\mu^2}\Big)
\big(1-\cos[2\mu H_{\rm sr}(t-t_\ast)]\big)
+\frac{3}{2\mu}\sin[2\mu H_{\rm sr}(t-t_\ast)]
\right]\,.
\end{align}
Then,
$\delta\dot{H}$ can be expressed in terms of $\kappa$ as
\begin{align}
\delta \dot{H}=\theta(t-t_\ast)\kappa\dot{H}_{\rm sr}\,.
\end{align}
It is also straightforward
to evaluate the deviation $\delta H(t)$
of the Hubble parameter as
\begin{align}
\delta H(t)&=
\int_{-\infty}^tdt^\prime\delta \dot{H}(t^\prime)
=\dot{H}_{\rm sr}\int_{t_\ast}^tdt^\prime\kappa(t^\prime)
=-\epsilon_{\rm sr}H_{\rm sr}^2\int_{t_\ast}^tdt^\prime\kappa(t^\prime)\,,
\end{align}
where higher order terms in the slow-roll expansion
are dropped.
More explicitly,
it is given by
\begin{align}
\nonumber
\delta H(t)
&=\alpha^2\epsilon_{\rm sr}H_{\rm sr}\left[
\frac{1}{6}\Big(1-e^{-3H_{\rm sr}(t-t_\ast)}\Big)
-\frac{1}{4\mu } e^{-3H_{\rm sr}(t-t_\ast)}\sin[2H_{\rm sr}\mu (t-t_\ast)]\right.
\\*
&\qquad\qquad\qquad
\left.+\frac{3}{8\mu^2}e^{-3H_{\rm sr}(t-t_\ast)}\Big(1-\cos[2H_{\rm sr}\mu(t-t_\ast)]\Big)
\right]\,.
\end{align}
Notice that
$\delta H(t)$ is first order in $\epsilon_{\rm sr}$
as well as second order in $\alpha$.

\subsection{Perturbations in the unitary gauge}
Let us next consider perturbations
around the background discussed in
the previous subsection.
We first construct the action in the unitary gauge 
in this subsection
and then introduce that for the Nambu-Goldstone boson
in the next subsection.

\subsubsection{Generalities}
In the unitary gauge,
the adiabatic mode is eaten by graviton
so that
the physical degrees of freedom
are three physical modes of graviton
and one massive isocurvature mode.
Corresponding to the degrees of freedom
to change bases of the field space,
there are some ambiguities in the description
of the massive isocurvature  $\sigma$.\footnote{
See also~\cite{Gao:2013zga} for recent discussions
on bases of perturbations in multiple-field inflation.}
In fact,
we can introduce the following family
of definitions of the isocurvature perturbation $\sigma$:
\begin{align}
\phi_1(x)&=\bar{\phi}_1(t)
-\sin\beta(t)\sigma(x)\,,\\
\phi_2(x)&=\bar{\phi}_2(t)+\cos\beta(t)\sigma(x)\,.
\end{align}
Here $\beta(t)$ is an arbitrary function of time $t$
and determines
the direction of $\sigma$ perturbations
in the field space.
We will take one particular choice of $\beta(t)$ later.
Before that,
however,
let us summarize the action in the unitary gauge
for general $\beta(t)$.
First, derivatives of $\phi_i$'s are given by
\begin{align}
\partial_\mu\phi_1(x)&=\delta^0_\mu\left[\dot{\bar{\phi}}_1(t)
-\cos\beta(t)\,\dot{\beta}(t)\sigma(x)\right]
-\sin\beta(t)\partial_\mu\sigma(x)\,,\\
\partial_\mu\phi_2(x)&=\delta^0_\mu\left[\dot{\bar{\phi}}_2(t)
-\sin\beta(t)\dot{\beta}(t)\sigma(x)\right]
+\cos\beta(t)\partial_\mu\sigma(x)\,.
\end{align}
Then, the kinetic term of $\phi_i$'s can be
written in terms of $\bar{\phi}_i$'s and $\sigma$ as
\begin{align}
\nonumber
-\sum_{i=1,2}\frac{1}{2}\partial_\mu\phi_i\partial^\mu\phi_i
&=-\frac{1}{2}g^{00}\left(\dot{\bar{\phi}}_1^2+\dot{\bar{\phi}}_2^2\right)-\frac{1}{2}\partial_\mu\sigma\partial^\mu\sigma
-\frac{1}{2}\dot{\beta}^2g^{00}\sigma^2\\*
\nonumber
&\quad
+\left[\sin\beta\,\dot{\bar{\phi}}_1-\cos\beta\,\dot{\bar{\phi}}_2\right]\partial^0\sigma
+\left[\cos\beta\,\dot{\bar{\phi}}_1+\sin\beta\,\dot{\bar{\phi}}_2\right]\dot{\beta}\,g^{00}\sigma
\\
\nonumber
&=-\frac{1}{2}g^{00}\left(\dot{\bar{\phi}}_1^2+\dot{\bar{\phi}}_2^2\right)-\frac{1}{2}\partial_\mu\sigma\partial^\mu\sigma
+\frac{1}{2}\dot{\beta}^2\sigma^2\\*
\nonumber
&\quad+\left[\cos\beta\,\dot{\bar{\phi}}_1+\sin\beta\,\dot{\bar{\phi}}_2\right]\dot{\beta}\,\delta g^{00}\sigma\\*
\nonumber
&\quad
+\left[\sin\beta\,\dot{\bar{\phi}}_1-\cos\beta\,\dot{\bar{\phi}}_2\right]\partial^0\sigma
-\left[\cos\beta\,\dot{\bar{\phi}}_1+\sin\beta\,\dot{\bar{\phi}}_2\right]\dot{\beta}\,\sigma\\*
\label{unitary_action}
&\quad 
-\frac{1}{2}\dot{\beta}^2\delta g^{00}\sigma^2\,,
\end{align}
where the linear order fluctuation terms
become total derivatives
after taking into account
the Einstein-Hilbert term
and the potential term.
In the framework of
the effective field theory approach~\cite{Noumi:2012vr},
the total action (\ref{total_action}) can be described
in the unitary gauge as
\begin{align}
\nonumber
S&=\int d^4x\sqrt{-g}\Bigg[
\frac{1}{2}M_{\rm Pl}^2R
+M_{\rm Pl}^2\dot{H}g^{00}
-M_{\rm Pl}^2(3H^2+\dot{H})
-\frac{1}{2}\partial_\mu\sigma\partial^\mu\sigma
\\*
\label{unitary_EFT}
&\quad\qquad\qquad\qquad
+\beta_1\delta g^{00}\sigma
+\beta_3(t)\partial^0\sigma
-(\dot{\beta}_3+3H\beta_3)\sigma
+\overline{\gamma}_1\delta g^{00}\sigma^2
-V_\sigma(\sigma;t)
\Bigg]\,,
\end{align}
where the parameters
$H$, $\dot{H}$, $\beta_1$, $\beta_3$, and $\overline{\gamma}_1$
are given by
\begin{align}
\nonumber
M_{\rm Pl}^2H^2
=\frac{1}{3}\left[\frac{1}{2}\left(\dot{\bar{\phi}}_1^2+\dot{\bar{\phi}}_2^2\right)
+V(\bar{\phi}_i)
\right]\,,
\quad
M_{\rm Pl}^2\dot{H}=-\frac{1}{2}\left(\dot{\bar{\phi}}_1^2+\dot{\bar{\phi}}_2^2\right)\,,\\*
\label{in_EFT_general}
\beta_1=\left[\cos\beta\,\dot{\bar{\phi}}_1+\sin\beta\,\dot{\bar{\phi}}_2\right]\dot{\beta}\,,
\quad
\beta_3=\sin\beta\,\dot{\bar{\phi}}_1-\cos\beta\,\dot{\bar{\phi}}_2\,,
\quad
\overline{\gamma}_1=-\frac{1}{2}\dot{\beta}^2\,.
\end{align}
The time-dependent potential $V_\sigma(\sigma;t)$ of $\sigma$
takes the form
\begin{align}
\nonumber
V_\sigma(\sigma;t)&=
\theta(t_\ast-t)\Big[V_{\rm sr}\big(\bar{\phi}_{\rm sr}-\sin(\beta-\alpha)\,\sigma\big)
+V_\bot\big(\cos(\beta-\alpha)\,\sigma\big)
-\frac{1}{2}\dot{\beta}^2\sigma^2
\Big]\\*
\nonumber
&\quad
+\theta(t-t_\ast)
\Big[V_{\rm sr}(\bar{\phi}_1-\sin\beta\,\sigma)
+V_\bot(\bar{\phi}_2+\cos\beta\,\sigma)
-\frac{1}{2}\dot{\beta}^2\sigma^2\Big]\\*
\label{in_EFT_general_2}
&\quad-\text{(zeroth and first order terms in $\sigma$)}\,,
\end{align}
where we used $(\bar{\phi}_1,\bar{\phi}_2)=(\bar{\phi}_{\rm sr},0)$
for $t<t_\ast$.
Note that the background value of
$V_{\rm sr}+V_\bot$
contributes to $-M_{\rm Pl}^2(3H^2+\dot{H})$ term
in (\ref{unitary_EFT})
and the first order fluctuation terms
contribute to the $-(\dot{\beta}_3+3H\beta_3)\sigma$ term.

\subsubsection{Action in the kinetic basis}
\label{subsubsec:action_kin}
From the expression~(\ref{in_EFT_general}),
we notice that the $\beta_3$ coupling vanishes
when $\sigma$ perturbations are orthogonal to the background trajectory.
We define such a kinetic basis in the following way:
\begin{align}
\beta(t)&=\arctan\frac{\dot{\bar{\phi}}_2}{\dot{\bar{\phi}}_1}
=\theta(t_\ast-t)\,\alpha
+\theta(t-t_\ast)\,\gamma(t)\quad
{\rm with}
\quad
\gamma(t)=\arctan\frac{\dot{\varphi}_2}{\dot{\bar{\phi}}_{\rm sr}+\dot{\varphi}_1}\,.
\end{align}
Note that
$\displaystyle\gamma(t)=\frac{\dot{\varphi}_2}{\dot{\bar{\phi}}_{\rm sr}}$
at the leading order in $\alpha$.
The time-derivative of $\beta$ is then given by\footnote{
\label{footnote:kinetic_basis}
Here it should be noticed that
$\dot{\beta}(t)$
is proportional to a delta function
in the bases with $\gamma(t_\ast)\neq\alpha$.
For example, in the potential basis defined by $\gamma(t)=0$,
we have $\dot{\beta}(t)=-\alpha\,\delta(t-t_\ast)$.
In such bases,
the coupling $\overline{\gamma}_1$
in~(\ref{in_EFT_general})
and the term $-\frac{1}{2}\dot{\beta}^2\sigma^2$
in (\ref{in_EFT_general_2})
are proportional to the delta function squared,
which leads to spurious singularities
in the calculation of spectra.
On the other hand,
in the kinetic basis,
$\dot{\beta}(t)$ is regular
and no spurious singularities appear.
This is one of the reasons why we take the kinetic basis
in this paper.}
\begin{align}
\dot{\beta}(t)=\delta(t-t_\ast)\big(\gamma(t_\ast)-\alpha\big)+\theta(t-t_\ast)\,\dot{\gamma}(t)=\theta(t-t_\ast)\,\dot{\gamma}(t)\,,
\end{align}
and therefore, we have
\begin{align}
\beta_1=\theta(t-t_\ast)\,\big|\dot{\bar{\phi}}_i\big|\,\dot{\gamma}\,,\quad
\beta_3=0\,,\quad
\overline{\gamma}_1=-\frac{1}{2}\theta(t-t_\ast)\,\dot{\gamma}^2\,.
\end{align}
Here $\big|\dot{\bar{\phi}}_i\big|^2=(\dot{\bar{\phi}}_1)^2+(\dot{\bar{\phi}}_2)^2=\dot{\phi}_{\rm sr}^2+\mathcal{O}(\alpha^2)$.
At the leading order in $\alpha$ and $\epsilon_{\rm sr}$,
they are given by
\begin{align}
\label{EFT_parameters_kinetic}
\beta_1=\theta(t-t_\ast)\ddot{\varphi}_2\,,\quad
\beta_3=0\,,\quad
\overline{\gamma}_1=-\frac{1}{2}\theta(t-t_\ast)\left(\frac{\ddot{\varphi}_2}{\dot{\phi}_{\rm sr}}\right)^2\,.
\end{align}
In this basis,
the potential $V(\sigma;t)$ can be expressed as
\begin{align}
\nonumber
V_\sigma(\sigma;t)&=
\theta(t_\ast-t)\Big[V_{\rm sr}\big(\bar{\phi}_{\rm sr}\big)
+V_\bot\big(\sigma\big)
\Big]\\*
\nonumber
&\quad
+\theta(t-t_\ast)
\Big[V_{\rm sr}(\bar{\phi}_1-\sin\gamma\,\sigma)
+V_\bot(\bar{\phi}_2+\cos\gamma\,\sigma)
-\frac{1}{2}\dot{\gamma}^2\sigma^2\Big]\\*
\label{V(s;t)_kinetic}
&\quad-\text{(zeroth and first order terms in $\sigma$)}\,.
\end{align}
For the potential~(\ref{massive_potential}),
its concrete form is given by
\begin{align}
\nonumber
V_\sigma(\sigma;t)&=
\theta(t_\ast-t)
\frac{m^2}{2}\sigma^2\\*
\nonumber
&\quad
+\theta(t-t_\ast)
\left[
\frac{1}{2}V_{\rm sr}^{\prime\prime}(\bar{\phi}_1)\gamma^2\sigma^2
-\frac{1}{3!}V_{\rm sr}^{\prime\prime\prime}(\bar{\phi}_1)\gamma^3\sigma^3
+\Big(\frac{m^2}{2}+\frac{\lambda}{4}\varphi_2^2-\frac{1}{2}\dot{\gamma}^2\Big)\sigma^2
+\frac{\lambda}{3!}\varphi_2\sigma^3
\right]\\*
&\quad
+\mathcal{O}(\sigma^4)\,.
\end{align}
To summarize,
in the kinetic basis,
the action in the unitary gauge
is given by
\begin{align}
\nonumber
S&=\int d^4x\sqrt{-g}\Bigg[
\frac{1}{2}M_{\rm Pl}^2R
+M_{\rm Pl}^2\dot{H}g^{00}
-M_{\rm Pl}^2(3H^2+\dot{H})\\
\label{unitary_action_EFT}
&\quad\qquad\qquad\qquad
-\frac{1}{2}\partial_\mu\sigma\partial^\mu\sigma
+\beta_1\delta g^{00}\sigma
+\overline{\gamma}_1\delta g^{00}\sigma^2
-V_\sigma(\sigma;t)
\Bigg]\,,
\end{align}
where $\beta_1$, $\bar{\gamma}_1$, and $V_\sigma(\sigma;t)$
are given in (\ref{EFT_parameters_kinetic}) and (\ref{V(s;t)_kinetic}).
In the rest of this paper,
we take the kinetic basis and
use (\ref{unitary_action_EFT})
as the action in the unitary gauge.

\subsubsection{Perturbativity}
\label{subsubsec:perturbativity}
Our models~(\ref{unitary_action_EFT}) contain
three free parameters
in addition to the Hubble parameter
$H_{\rm sr}$ in the single field slow-roll model:
the turning angle $\alpha$,
the mass $m$ of the massive potential,
and the coupling constant $\lambda$
for the quartic self-interaction
of massive directions $\phi_\bot$.
In the calculation of primordial spectra,
we would like to treat the parameters
$\alpha$ and $\lambda$ as perturbations.
From the viewpoint of the action~(\ref{unitary_action_EFT})
for metric perturbations and massive isocurvatures,
the perturbativity of $\alpha$ and $\lambda$
can be rephrased by
that of
the following quadratic and cubic interactions:
\begin{align}
\Big(\frac{\lambda}{4}\varphi_2^2-\frac{1}{2}\dot{\gamma}^2\Big)\sigma^2
\quad
{\rm and}
\quad
\frac{\lambda}{3!}\varphi_2\sigma^3\,.
\end{align}
For the perturbativity of quadratic interactions,
we require that
they are smaller than the mass term
\begin{align}
\lambda\varphi_2^2\lesssim m^2\,,
\quad
\dot{\gamma}^2\lesssim m^2\,.
\end{align}
On the other hand,
we impose the following condition on the cubic interaction:
\begin{align}
(\text{cubic coupling})\times(\text{propagator})
\times(\text{cubic coupling})
\sim
\frac{(\lambda\varphi_2)^2}{m^2}\lesssim1\,,
\end{align}
which is essentially the same
as the condition that
the four-point amplitudes are not large.
Using the expression~(\ref{vaphi2_in_epsilon})
of $\varphi_2$,
these conditions can be stated as
\begin{align}
\label{perturbativity_condition}
\frac{\lambda\alpha^2}{\mu^4 }\frac{2M_{\rm Pl}^2\epsilon_{\rm sr}}{H_{\rm sr}^2}\lesssim 1\,,
\quad
\alpha\lesssim
1
\,,
\quad
\frac{\lambda^2\alpha^2}{\mu^4 }\frac{2M_{\rm Pl}^2\epsilon_{\rm sr}}{H_{\rm sr}^2}\lesssim 1\,.
\end{align}
If the mass $\mu$ and the turning angle $\alpha$
are specified,
(\ref{perturbativity_condition})
can be though of as conditions on
the coupling $\lambda$ of the quartic interaction.
Using $\displaystyle\frac{H_{\rm sr}^2}{2M_{\rm Pl}^2\epsilon_{\rm sr}}\simeq 4\pi^2P_{\zeta}\sim10^{-8}$,
we have for example
\begin{align}
\lambda\lesssim0.01 
\quad{\rm for}
\quad
(\mu,\alpha)=(10,0.1)
\quad
{\rm and}
\quad
\lambda\lesssim1 
\quad{\rm for}
\quad
(\mu,\alpha)=(10,0.01)\,.
\end{align}

\subsection{Action for the Goldstone boson}
For the calculation of primordial perturbations,
it is convenient to introduce the Goldstone boson $\pi$
by the St\"uckelberg method.
In this subsection
we construct the action for $\pi$
and discuss its relevant terms to
tree-level two point and three point functions
of scalar perturbations $\zeta$.
The action for the Goldstone boson $\pi$
can be obtained by the gauge transformation
\begin{align}
\label{unitary_to_pi}
t\to \tilde{t}\,,\quad
x^i\to\tilde{x}^i
\quad{\rm with}
\quad
\tilde{t} +\tilde{\pi}(\tilde{t},\tilde{x})=t\,,\quad
\tilde{x}^{i}=x^i\,,
\end{align}
which is realized by the following replacement~\cite{Cheung:2007st}:
\begin{align}
\nonumber
&\delta^0_\mu\to \delta^0_\mu+\partial_\mu\pi\,,\quad
f(t)\to f(t+\pi)\,,
\quad
\int d^4x \sqrt{-g}\to\int d^4x \sqrt{-g}\,,\\
\label{replacement}
&\nabla_\mu\to\nabla_\mu\,,
\quad g_{\mu\nu}\to g_{\mu\nu}\,,
\quad g^{\mu\nu}\to g^{\mu\nu}\,,
\quad R_{\mu\nu\rho\sigma}\to R_{\mu\nu\rho\sigma}\,,
\end{align}
where we dropped the tilde for simplicity.
The action corresponding to~(\ref{unitary_action_EFT})
is then given by~\cite{Noumi:2012vr}
\begin{align}
\nonumber
S&=\int d^4x\sqrt{-g}\Bigg[
\frac{1}{2}M_{\rm Pl}^2R
+M_{\rm Pl}^2\dot{H}(t+\pi)(g^{00}+2\partial^0\pi+\partial_\mu\pi\partial^\mu\pi)
-M_{\rm Pl}^2(3H^2+\dot{H})(t+\pi)\\*
\nonumber
&\quad\qquad\qquad\qquad
-\frac{1}{2}\partial_\mu\sigma\partial^\mu\sigma
+\beta_1(t+\pi)(\delta g^{00}+2\partial^0\pi+\partial_\mu\pi\partial^\mu\pi)\sigma\\*
\label{before_decouple}
&\quad\qquad\qquad\qquad
+\overline{\gamma}_1(t+\pi)(\delta g^{00}+2\partial^0\pi+\partial_\mu\pi\partial^\mu\pi)\sigma^2
-V_\sigma(\sigma;t+\pi)
\Bigg]\,.
\end{align}
We next discuss which terms are
relevant to
tree-level three point functions of $\pi$.
For this purpose, let us take
the spatially flat gauge
and rewrite the action~(\ref{before_decouple})
in terms of the ADM decomposition:
\begin{equation}
ds^2=-(N^2-N_iN^i)dt^2+2N_i dx^i dt+a^2(e^{\gamma})_{ij}\,dx^idx^j
\quad
{\rm with}
\quad
\gamma_{ii}=\partial_i\gamma_{ij}=0
\,.
\end{equation}
Here and in what follows we use the spatial metric $h_{ij}=a^2
(e^\gamma)_{ij}$ and its inverse $h^{ij}=a^{-2} (e^{-\gamma})_{ij}$ to
raise or lower the indices of $N^i$.
In this gauge, there are no second order mixing terms of $\pi$ and
$\gamma_{ij}$ because $\gamma_{ij}$ has two spatial indices and
is transverse-traceless. Then, the tensor fluctuation $\gamma_{ij}$ does
not contribute to tree-level three point functions of $\pi$.
Therefore, possible contributions of metric perturbations come only
from the auxiliary fields $\delta N=N-1$ and $N^i$.
For the calculation of three-point functions,
it is sufficient to solve the constraints up to first order~\cite{Maldacena:2002vr}.
For the action (\ref{before_decouple}),
it is performed as follows~\cite{Noumi:2012vr}:
\begin{align}
\label{ADM_constraints}
\delta N&=-\frac{\dot{H}}{H}\pi\,,\quad
N^i=a^{-2}\partial_i\psi
\quad
{\rm with}
\quad
\psi=a^2\partial^{-2}
\Big(\frac{\dot{H}}{H^2}(\dot{H}\pi+H\dot{\pi})
+\frac{\beta_1}{M_{\rm Pl}^2H}\sigma\Big)\,.
\end{align}
In the following discussions,
we decompose the kinetic term of $\pi$
as
\begin{align}
M_{\rm Pl}^2\dot{H}\partial_\mu\pi\partial^\mu\pi
=M_{\rm Pl}^2\dot{H}_{\rm sr}\partial_\mu\pi\partial^\mu\pi
+M_{\rm Pl}^2\delta\dot{H}\partial_\mu\pi\partial^\mu\pi\,,
\end{align}
and treat the second term as an interaction term.
In such a case,
the canonical
normalization
is given by
\begin{align}
\pi_c\sim M_{\rm Pl}(-\dot{H}_{\rm sr})^{1/2}\pi\,,
\quad
\sigma_c=\sigma\,,
\quad
\delta N_c\sim M_{\rm Pl}\delta N\,,
\quad
N^i_c\sim M_{\rm Pl}N^i\,,
\end{align}
and we redefine the coupling constants $\beta_1$ correspondingly as
\begin{align}
\beta^c_1\sim\frac{1}{M_{\rm Pl}(-\dot{H}_{\rm sr})^{1/2}}\beta_1\,.
\end{align}
Then, the constraints~(\ref{ADM_constraints}) can be written as
\begin{align}
\delta N_c\sim \epsilon^{1/2}_{\rm sr}(1+\kappa)\pi_c\,,
\quad
N^i_c\sim\epsilon_{\rm sr}^{1/2}\frac{\partial_i}{\partial^2}
\Big(-\dot{\pi}_c+\beta^c_1\sigma_c
\Big)\,,
\end{align}
where higher order terms in $\epsilon_{\rm sr}$ and $\eta_{\rm sr}$
are dropped.
It is
manifest that $\delta N_c$ and $N_c^i$ are suppressed
by the slow-roll parameter
$\epsilon_{\rm sr}^{1/2}$ and contributions from $\delta N_c$ and $N_c^i$
are expected to be irrelevant when the slow-roll direction
of the potential satisfies the slow-roll conditions.
In fact, we can explicitly show that
they are irrelevant in our calculations.
We then drop these contributions
to obtain the following action
up to cubic order perturbations:
\begin{align}
\nonumber
S&=\int d^4x \,a^3\Big[
-M_{\rm Pl}^2\dot{H}\Big(\dot{\pi}^2-\frac{(\partial_i\pi)^2}{a^2}\Big)
-3M_{\rm Pl}^2\dot{H}^2\pi^2
-M_{\rm Pl}^2\ddot{H}\pi\Big(\dot{\pi}^2-\frac{(\partial_i\pi)^2}{a^2}\Big)
- 3 M_{\rm Pl}^2 \dot{H} \ddot{H} \pi^3\\*
\nonumber
&\qquad\qquad\qquad
+\frac{1}{2}\Big(\dot{\sigma}^2-\frac{(\partial_i\sigma)^2}{a^2}\Big)
-2\beta_1\dot{\pi}\sigma
-\beta_1\Big(\dot{\pi}^2-\frac{(\partial_i\pi)^2}{a^2}\Big)\sigma
-2\dot{\beta}_1\pi\dot{\pi}\sigma 
\\*
&\qquad\qquad\qquad
-2\overline{\gamma}_1\dot{\pi}\sigma^2
-V(\sigma;t+\pi)
\Big]\,.
\end{align}
By further dropping the sub-leading terms in the slow-roll expansion,
we finally obtain
\begin{align}
\nonumber
S&=\int d^4x \,a^3\Big[
-M_{\rm Pl}^2\dot{H}\Big(\dot{\pi}^2-\frac{(\partial_i\pi)^2}{a^2}\Big)
-M_{\rm Pl}^2\delta\ddot{H}\pi\Big(\dot{\pi}^2-\frac{(\partial_i\pi)^2}{a^2}\Big)\\*
\nonumber
&\qquad\qquad\qquad
+\frac{1}{2}\Big(\dot{\sigma}^2-\frac{(\partial_i\sigma)^2}{a^2}\Big)
-2\beta_1\dot{\pi}\sigma
-\beta_1\Big(\dot{\pi}^2-\frac{(\partial_i\pi)^2}{a^2}\Big)\sigma
-2\dot{\beta}_1\pi\dot{\pi}\sigma 
\\*
&\qquad\qquad\qquad
-2\overline{\gamma}_1\dot{\pi}\sigma^2
-V(\sigma;t+\pi)
\Big]\,.
\end{align}
Note that
$\delta\ddot{H}$ and $\dot{\beta}_1$
can be written in terms of $\kappa$ and $\gamma$
as
\begin{align}
\nonumber
\delta\ddot{H}&=
\theta(t-t_\ast)\big(\ddot{H}_{\rm sr}\kappa
+\dot{H}_{\rm sr}\dot{\kappa}\big)
+\delta(t-t_\ast)\dot{H}_{\rm sr}\kappa\\
\nonumber
&\simeq
\theta(t-t_\ast)\dot{H}_{\rm sr}\dot{\kappa}
+\delta(t-t_\ast)\dot{H}_{\rm sr}\kappa\\
&=\theta(t-t_\ast)\dot{H}_{\rm sr}\dot{\kappa}\,,
\\
\nonumber
\dot{\beta}_1&=\theta(t-t_\ast)(\ddot{\bar{\phi}}_{\rm sr}\dot{\gamma}+\dot{\bar{\phi}}_{\rm sr}\ddot{\gamma})
+\delta(t-t_\ast)\dot{\gamma}\\
&\simeq
\theta(t-t_\ast)\dot{\bar{\phi}}_{\rm sr}\ddot{\gamma}
+\delta(t-t_\ast)\dot{\gamma}\,,
\end{align}
where we used $\kappa(t_\ast)=0$
and dropped higher order terms
in $\epsilon_{\rm sr}$ and $\eta_{\rm sr}$.

\subsection{Hamiltonian in the interaction picture}
In the following two sections
the primordial power spectrum
and bispectrum
of scalar perturbations $\zeta$
are calculated
using the in-in formalism.
For this purpose,
let us introduce the Hamiltonian in the interaction
picture up to the cubic order fluctuations.
Since we would like to
discuss the deviation
from single field slow-roll inflation,
we choose the free part and the interaction part
of the Hamiltonian
in the following way:
\begin{align}
H_{\rm free}&=\int d^3x\,a^3\left[-M_{\rm Pl}^2\dot{H}_{\rm sr}\Big(\dot{\pi}^2+\frac{(\partial_i\pi)^2}{a^2}\Big)
+\frac{1}{2}\Big(\dot{\sigma}^2+\frac{(\partial_i\sigma)^2}{a^2}+m^2\sigma^2\Big)
\right]\,,\\*
H_{\rm int}&=H_{\rm int}^{(2)}+H_{\rm int}^{(3)}\,,\\*
H_{\rm int}^{(2)}&=\int d^3x\,a^3\left[M_{\rm Pl}^2\delta\dot{H}\Big(\dot{\pi}^2-\frac{(\partial_i\pi)^2}{a^2}\Big)
+2\beta_1\dot{\pi}\sigma\right]\,,\\*
H_{\rm int}^{(3)}&=\int d^3x\,a^3\left[M_{\rm Pl}^2\delta\ddot{H}\pi\Big(\dot{\pi}^2-\frac{(\partial_i\pi)^2}{a^2}\Big)
+\beta_1\Big(\dot{\pi}^2-\frac{(\partial_i\pi)^2}{a^2}\Big)\sigma
+2\dot{\beta}_1\pi\dot{\pi}\sigma 
+\theta(t-t_\ast)\frac{\lambda\varphi_2}{6}\sigma^3\right]\,,
\end{align}
where we dropped terms irrelevant to
the spectra up to the leading order in the turning angle~$\alpha$.\footnote{
Note that contributions from the $2\overline{\gamma}_1\dot{\pi}\sigma^2$ term are sub-leading in the tree-level three-point functions of $\pi$.}
The quantum fields $\pi$ and $\sigma$ in the interaction picture
are expanded in the Fourier space as
\begin{align}
\pi_{\bf k}=u_{k}\, a_{\bf k}+u^\ast_{k}\, a^\dagger_{-{\bf k}}\,,
\quad
\sigma_{\bf k}=v_{k}\, b_{\bf k}+v^\ast_{k}\, b^\dagger_{-{\bf k}}
\end{align}
with the standard commutation relations
\begin{align}
\label{creation_annihilation}
[a_{\bf k},a^\dagger_{\bf k^\prime}]=(2\pi)^3\delta^{(3)}({\bf k}-{\bf k}^\prime)\,,
\quad
[b_{\bf k},b^\dagger_{\bf k^\prime}]=(2\pi)^3\delta^{(3)}({\bf k}-{\bf k}^\prime)\,.
\end{align}
Here the mode functions $u_{k}$ and $v_{k}$ satisfy the equations of
motion in the free theory and depend on $k=|{\bf k}|$:
\begin{align}
\label{eom_no_app}
\ddot{u}_{ k}+3H_{\rm sr}\dot{u}_{ k}+\frac{k^2}{a^2}u_{ k}=0\,,
\quad
\ddot{v}_{ k}+3H_{\rm sr}\dot{v}_{k}+\Big(m^2+\frac{k^2}{a^2}\Big)v_{ k}=0\,,
\end{align}
where
higher order terms in $\epsilon_{\rm sr}$ and $\eta_{\rm sr}$ are dropped.
In terms of the conformal time $d\tau=a^{-1}dt$,
they can be written as
\begin{align}
\label{eom_app}
u_{k}^{\prime\prime}-\frac{2}{\tau}u_{k}^\prime+k^2u_{ k}=0\,,
\qquad
v_{k}^{\prime\prime}-\frac{2}{\tau}v_{k}^\prime+k^2v_{ k}+\frac{m^2}{H_{\rm sr}^2\tau^{2}}v_k=0\,,
\end{align}
where
the primes denote derivatives with respect to $\tau$
and we used $\tau=-1/(aH_{\rm sr})$ at the leading order
in $\epsilon_{\rm sr}$ and $\eta_{\rm sr}$.
The normalizations of the mode functions follow from
\begin{align}
\label{normalization_mode}
2M_{\rm Pl}^2H_{\rm sr}^2\epsilon_{\rm sr}\,a^3\left(u_{k}\,\dot{u}_{k}^\ast-\dot{u}_{k}\,u^\ast_{k}\right)=i\,,
\quad
a^3\left(v_{k}\,\dot{v}_{k}^\ast-\dot{v}_{k}\,v^\ast_{k}\right)=i\,.
\end{align}
Assuming that
$\pi$ and $\sigma$
are in the Bunch-Davies vacuum
before the turn,
the mode functions $u_k$ and $v_k$
are given by
\begin{align}
\label{mode_u_expression}
u_k&=\frac{1}{2M_{\rm Pl}\epsilon_{\rm sr}^{1/2} k^{3/2}}(1+i k\tau)e^{-i k\tau}
=\frac{1}{2M_{\rm Pl}\epsilon_{\rm sr}^{1/2} k^{3/2}}(1-i x)e^{ ix}\,,\\*
\label{mode_v_expression}
v_k&=-ie^{-\frac{\pi}{2}\mu+\frac{i}{4}\pi}\frac{\sqrt{\pi}H_{\rm sr}}{2}
(-\tau)^{3/2}H_{i\mu}^{(1)}(- k\tau)
=-ie^{-\frac{\pi}{2}\mu+\frac{i}{4}\pi}\frac{\sqrt{\pi}H_{\rm sr}}{2k^{3/2}}
x^{3/2}H_{i\mu}^{(1)}(x)\,,
\end{align}
where $x=-k\tau$ and $H_\nu^{(1)}=J_\nu+iY_\nu$ is the Hankel function.
Note that the time derivative of $u_k$
is given by
\begin{align}
\dot{u}_k&=-H_{\rm sr}\tau u_k^\prime=-\frac{H_{\rm sr}}{2M_{\rm Pl}\epsilon_{\rm sr}^{1/2} k^{3/2}} x^2e^{i x}\,.
\end{align}
It is also convenient to
express $\kappa$ and $\varphi_2$
in terms of the conformal time:
\begin{align}
\nonumber
\kappa(t)&=-\alpha^2(\tau/\tau_\ast)^3\left[\Big(\frac{1}{2}-\frac{9}{8\mu^2}\Big)(1-\cos[2\mu\ln(\tau/\tau_\ast)])
-\frac{3}{2\mu}\sin[2\mu\ln(\tau/\tau_\ast)]
\right]\\*
&=-\alpha^2\left[\Big(\frac{1}{2}-\frac{9}{8\mu^2}\Big)(\tau/\tau_\ast)^3
+\frac{(3+2i\mu)^2}{16\mu^2}(\tau/\tau_\ast)^{3+2i\mu}
+\frac{(3-2i\mu)^2}{16\mu^2}(\tau/\tau_\ast)^{3-2i\mu}
\right]
\,,\\
\label{Hamiltonian_varphi2}
\varphi_2(t)&=-\frac{\alpha\bar{\dot{\phi}}_{\rm sr}}{\mu H_{\rm sr}}(\tau/\tau_\ast)^{3/2} \sin[\mu\ln(\tau/\tau_\ast)]
=i\frac{\alpha\bar{\dot{\phi}}_{\rm sr}}{2\mu H_{\rm sr}}
\left[(\tau/\tau_\ast)^{\frac{3}{2}+i\mu}
-(\tau/\tau_\ast)^{\frac{3}{2}-i\mu}
\right]\,,
\end{align}
where $\tau_\ast$ is the conformal time
corresponding to the turning time $t_\ast$.
In the following,
we use the above setup for the calculation of
primordial spectra.

\section{Primordial power spectrum}
\label{sec:PS}
\begin{figure}[t]
\begin{center}
\includegraphics[width=160mm, bb=0 0 580 137]{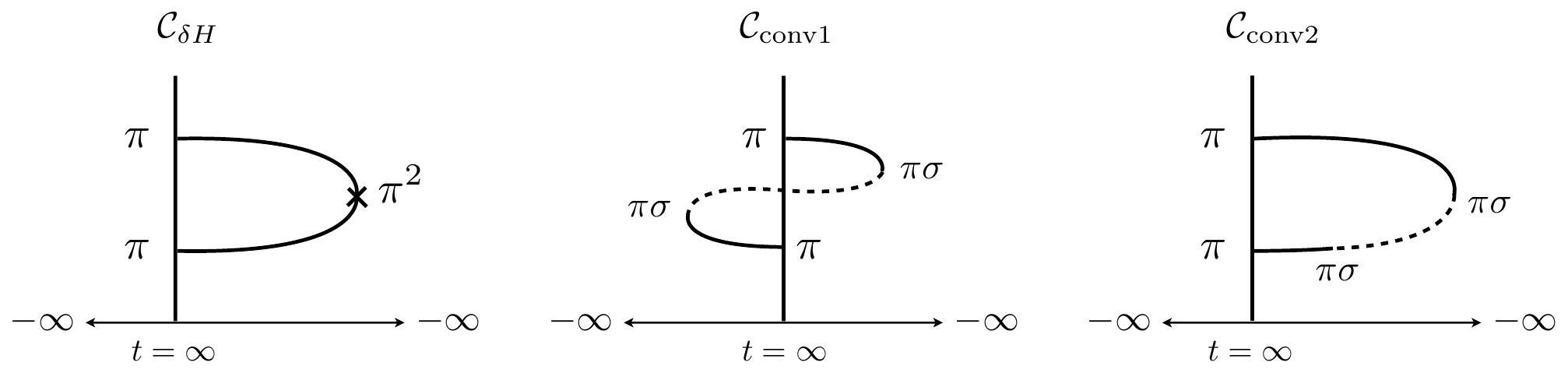}
 \end{center}
\vspace{-5mm}
\caption{
Diagrams corresponding to $\mathcal{C}_{\delta H}$, $\mathcal{C}_{\rm conv1}$,
and $\mathcal{C}_{\rm conv2}$
(depicted from the left).
The solid and dotted lines represent
the propagation of $\pi$ and $\sigma$, respectively.}
\label{fig:PS_diagram}
\end{figure}

\setcounter{equation}{0}
Let us then calculate the primordial power spectrum.
Since the scalar perturbation
$\zeta$ is given at the linear order by $\zeta=-H\pi$,
we calculate the expectation value of $\pi_{\bf k}(t)\pi_{\bf k^\prime}(t)$.
Using the in-in formalism,
it is calculated as
\begin{align}
\nonumber
\langle\pi_{\bf k}(t)\pi_{\bf k^\prime}(t)\rangle
&=\langle0|
\left[\bar{T}\exp\Big(i\int_{t_0}^tdt^\prime H_{\rm int}(t^\prime)\Big)\right]
\pi_{\bf k}(t)\pi_{\bf k^\prime}(t)
\left[T\exp\Big(-i\int_{t_0}^tdt^\prime H_{\rm int}(t^\prime)\Big)\right]
|0\rangle\\
\nonumber
&=\langle0|
\pi_{\bf k}(t)\pi_{\bf k^\prime}(t)
|0\rangle
+2{\rm Re}\left[-i\int_{t_0}^tdt_1\langle0|
\pi_{\bf k}(t)\pi_{\bf k^\prime}(t)H_{\rm int}(t_1)
|0\rangle\right]\\
\nonumber
&\quad+\int_{t_0}^td\tilde{t}_1\int_{t_0}^tdt_1\langle0|
 H_{\rm int}(\tilde{t}_1)
\pi_{\bf k}(t)\pi_{\bf k^\prime}(t)
H_{\rm int}(t_1)
|0\rangle\\
\label{ps_H}
&\quad-2{\rm Re}\left[\int_{t_0}^tdt_1\int_{t_0}^{t_1}dt_2\langle0|
\pi_{\bf k}(t)\pi_{\bf k^\prime}(t)
H_{\rm int}(t_1)H_{\rm int}(t_2)|0\rangle\right]
+\ldots\,,
\end{align}
where the dots stand for higher order terms
in the couplings.
In the following,
we set $t=\infty$ and $t_0=-\infty$
and drop the indication of the time~$t$ in correlators.
In terms of the mode functions and couplings,
the two point function~(\ref{ps_H})
can be expressed up to the leading order corrections
from $\kappa$ and $\beta_1$
as
\begin{align}
\label{pi-pi-C}
\langle\pi_{\bf k}\pi_{\bf k^\prime}\rangle&=(2\pi)^3\delta^{(3)}({\bf k}+{\bf k}^\prime)\frac{1}{4M_{\rm Pl}^2\epsilon_{\rm sr}k^3}
\Big[1+\mathcal{C}_{\delta H}+\mathcal{C}_{\rm conv}\Big]\,,
\end{align}
where $\mathcal{C}_{\delta H}$ and $\mathcal{C}_{\rm conv}$ are defined by
(see Fig.~\ref{fig:PS_diagram} for the corresponding diagrams)
\begin{align}
\mathcal{C}_{\delta H}&=2{\rm Re}\Bigg[
i\int_{-\infty}^\infty dt_1\,a^3\,
\big(2M_{\rm Pl}^2\dot{H}_{\rm sr}\kappa(t_1)\big)
\Big(\dot{u}_k^2
(t_1)-\frac{k^2}{a^2}u_k^2(t_1)\Big)\Bigg]\,,\\
\label{C_conv_start}
\mathcal{C}_{\rm conv}&=\mathcal{C}_{\rm conv1}+\mathcal{C}_{\rm conv2}\,,\\
\label{C_conv_start1}
\mathcal{C}_{\rm conv1}&=8\left|\int_{-\infty}^\infty dt_1\,a^3(t_1)
\beta_1(t_1)\dot{u}_k(t_1)v_k
(t_1)\right|^2\,,\\
\label{C_conv_start2}
\mathcal{C}_{\rm conv2}&=-16{\rm Re}\Bigg[
\int_{-\infty}^\infty dt_1\,a^3(t_1)
\beta_1(t_1)\dot{u}^\ast_k(t_1)v_k(t_1)
\int_{-\infty}^{t_1}dt_2
\,a^3(t_2)
\beta_1(t_2)\dot{u}^\ast_k(t_2)v^\ast_k(t_2)
\Bigg]\,.
\end{align}
Note that
$\mathcal{C}_{\delta H}$
describes the effects of the deformation~$\delta H$
of the background Hubble parameter
and
$\mathcal{C}_{\rm conv}$ describes
the conversion effects between
adiabatic and isocurvature perturbations.
Also notice that
they are second order in the turning angle $\alpha$
because $\kappa$ and $\beta_1$
are second and first order in $\alpha$, respectively.
The two-point functions of scalar perturbations
are then given by
\begin{align}
\langle\zeta_{\bf k}(t)\zeta_{\bf k^\prime}(t)\rangle&=(2\pi)^3\delta^{(3)}
({\bf k}+{\bf k}^\prime)
\frac{2\pi^2}{k^3}
\mathcal{P}_\zeta(k)\,,
\end{align}
where the power spectrum $\mathcal{P}_\zeta(k)$ is
\begin{align}
\label{C_to_P}
\mathcal{P}_\zeta(k)&=\frac{H_{\rm sr}^2}{8\pi^2M_{\rm sr}^2\epsilon_{\rm sr}}
\left(1+\mathcal{C}_{\delta H}+\mathcal{C}_{\rm conv}\right)\,.
\end{align}
In the rest of this section,
we evaluate $\mathcal{C}_{\delta H}$ and $\mathcal{C}_{\rm conv}$,
which can be thought of as deviation factors from the single field slow-roll model.

\subsection{Hubble deformation effects}
\label{subsec:Hubble}
\begin{figure}[t]
\begin{center}
\includegraphics[width=160mm, bb=0 0 580 212]{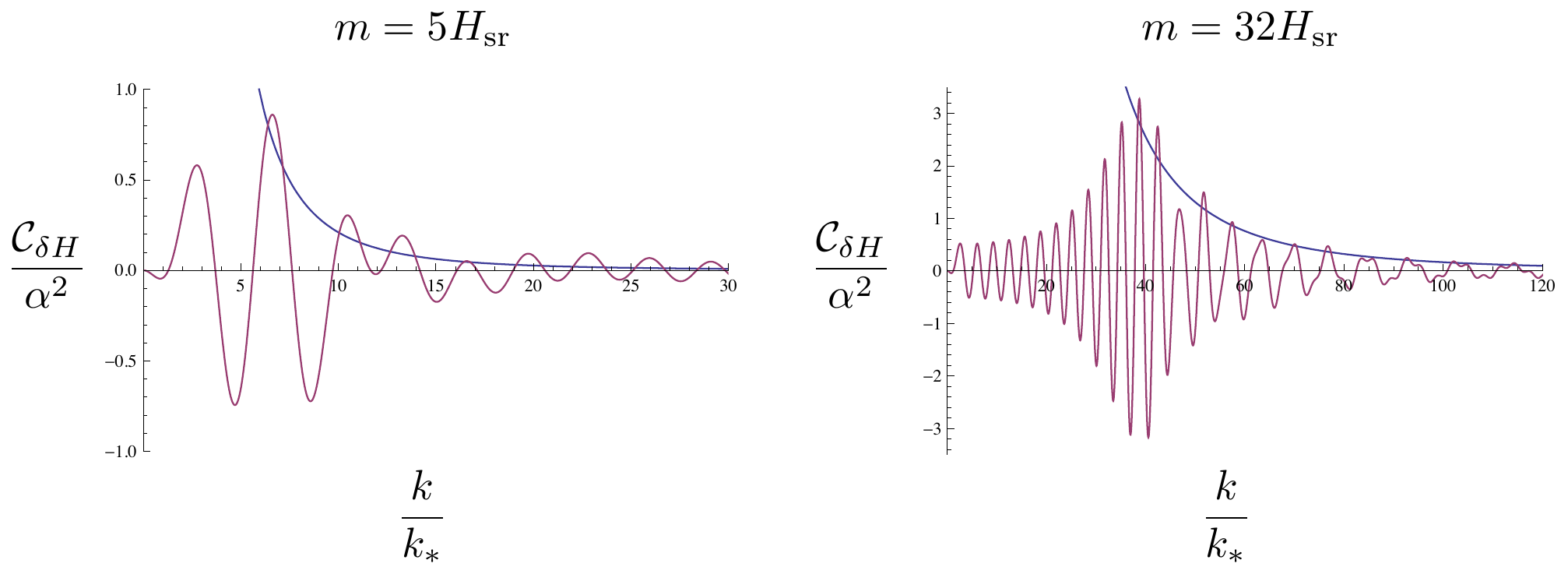}
 \end{center}
\vspace{-5mm}
\caption{Scale-dependence of $\mathcal{C}_{\delta H}$
for $m=5H_{\rm sr}$ (left figure)
and $m=32H_{\rm sr}$ (right figure).
The red curve is the exact expression (\ref{CdH_exact})
for $\mathcal{C}_{\delta H}$.
The blue curve is the resonance size
estimated via the parametric resonance arguments:
$\mathcal{C}_{\delta H}=
\frac{\sqrt{\pi}}{2}\alpha^2 \mu^{1/2}\big(k/(\mu k_\ast)\big)^{-3}$,
where the numerical coefficient $\sqrt{\pi}/2$
is determined by the stationary phase approximation.
As shown in the above figures,
the parametric resonance arguments
well explain and capture our exact results.}
\label{fig:CdH}
\end{figure}

Using the expression~(\ref{mode_u_expression})
of the mode functions,
$\mathcal{C}_{\delta H}$ can be expressed as
\begin{align}
\label{C_delta_H_x_int}
\mathcal{C}_{\delta H}&=
-{\rm Im}\left[\int_0^{x_\ast}dx\,\kappa\,(2-2ix^{-1}-x^{-2})e^{-2ix}\right]\,,
\end{align}
where $x=-k\tau$ and $x_\ast=-k\tau_\ast$.
It is convenient to introduce
the mode $k_\ast=-1/\tau_\ast$,
which crosses the horizon
at the time $t=t_\ast$ of turning.
In terms of $k_\ast$,
the parameter $x_\ast$ can be expressed
as $x_\ast=k/k_\ast$.
Using $x$ and $x_\ast$,
the coupling $\kappa$ can be written as
\begin{align}
\kappa
&=-\alpha^2\left[\Big(\frac{1}{2}-\frac{9}{8\mu^2}\Big)(x/x_\ast)^3
+\frac{(3+2i\mu)^2}{16\mu^2}(x/x_\ast)^{3+2i\mu}
+\frac{(3-2i\mu)^2}{16\mu^2}(x/x_\ast)^{3-2i\mu}
\right]
\,.
\end{align}
Then,
the integral~(\ref{C_delta_H_x_int}) reduces to
integrals of the form
\begin{align}
\mathcal{I}(\delta,n,x)
&=\int_0^x d\tilde{x}\,(\tilde{x}/x)^{3+\delta}\tilde
{x}^ne^{-i\tilde{x}}\,,
\end{align}
whose analytic expression can be obtained as follows:
\begin{align}
\label{calI_analytic}
\mathcal{I}(\delta,n,x)&=
i^{-1-n}(1/ix)^{3+\delta}
\Big[\Gamma(4+n+\delta)
-\Gamma(4+n+\delta,ix)\Big]\,.
\end{align}
Here $\Gamma(z)$ and $\Gamma(a,z)$
are the gamma function and
the incomplete gamma function, respectively.
In terms of $\mathcal{I}(\delta,n,x)$,
the integral~(\ref{C_delta_H_x_int})
can be expressed as
\begin{align}
\nonumber
\mathcal{C}_{\delta H}(x_\ast)&=
{\rm Im}\Bigg[
\alpha^2\Big(\frac{1}{2}-\frac{9}{8\mu^2}\Big)
\Big[
\mathcal{I}(0,0,2x_\ast)
-2i\mathcal{I}(0,-1,2x_\ast)
-2\mathcal{I}(0,-2,2x_\ast)
\Big]
\\
\nonumber
&\quad\qquad
+\alpha^2\frac{(3+2i\mu)^2}{16\mu^2}
\Big[
\mathcal{I}(2i\mu,0,2x_\ast)
-2i\mathcal{I}(2i\mu,-1,2x_\ast)
-2\mathcal{I}(2i\mu,-2,2x_\ast)
\Big]
\\
\label{CdH_exact}
&\quad\qquad
+\alpha^2\frac{(3-2i\mu)^2}{16\mu^2}
\Big[
\mathcal{I}(-2i\mu,0,2x_\ast)
-2i\mathcal{I}(-2i\mu,-1,2x_\ast)
-2\mathcal{I}(-2i\mu,-2,2x_\ast)
\Big]\Bigg]\,.
\end{align}
As displayed in Fig.~\ref{fig:CdH},
a kind of resonance appears in $\mathcal{C}_{\delta H}$
around the scale $k\sim \frac{m}{H_{\rm sr}}k_\ast\sim \mu k_\ast$~\cite{Chen:2011zf,Chen:2011tu,Saito:2012pd,Battefeld:2013xka,Saito:2013aqa}.
In the rest of this subsection,
we discuss qualitative features of this resonance effect.

\subsubsection{Mathieu equation and the resonance effect}
\label{subsubsec:Mathieu}
Following~\cite{Saito:2012pd},
let us introduce the Mathieu equation
characterizing the above resonance effects
and estimate the scale and size of the resonance.
For our purpose,
it is convenient
to introduce a canonically normalized field $u$
defined by
\begin{align}
u=z_\pi\pi
\quad
{\rm with}
\quad
z_\pi=(-2M_{\rm Pl}^2\dot{H}a^3)^{1/2}=(2M_{\rm Pl}^2\epsilon)^{1/2}Ha^{3/2}\,.
\end{align}
In terms of $u$,
the second order action of $\pi$
can be written as
\begin{align}
S&=\int dtd^3x\,a^3(-M_{\rm Pl}^2\dot{H})\Big(\dot{\pi}^2-\frac{(\partial_i\pi^2)}{a^2}\Big)
=\frac{1}{2}\int dt d^3x\,
\Big(\dot{u}^2-\frac{(\partial_iu)^2}{a^2}
-\frac{\ddot{z}_\pi}{z_\pi}u^2\Big)\,,
\end{align}
and the equation of motion in the momentum space
is given by
\begin{align}
\label{eom_resonance_u}
\ddot{u}+\left(\frac{k^2}{a^2}+\frac{\ddot{z}_\pi}{z_\pi}\right)u=0\,,
\end{align}
where the explicit form of $\ddot{z}_\pi/z_\pi$ is
\begin{align}
\frac{\ddot{z}_\pi}{z_\pi}
&=\frac{9}{4}H_{\rm sr}^2+\frac{3}{2}H_{\rm sr}\dot{\kappa}+\frac{1}{2}\ddot{\kappa}+\mathcal{O}(\epsilon_{\rm sr},\eta_{\rm sr},\kappa^2)
\quad
{\rm with}
\quad
\kappa=\frac{\dot{H}-\dot{H}_{\rm sr}}{\dot{H}_{\rm sr}}=\frac{\delta \dot{H}}{\dot{H}_{\rm sr}}
\,.
\end{align}
To characterize the highly oscillating background,
let us rewrite $\kappa$ as
\begin{align}
\kappa=A(t)\sin (2\omega t+\theta)\,,
\end{align}
where
we assume that $\omega\gg H_{\rm sr}$
and
the time-dependence of the normalization factor~$A(t)$
is negligible compared to $\omega$.
At the leading order in $H_{\rm sr}/\omega$,
the equation of motion~(\ref{eom_resonance_u})
can be written as
\begin{align}
\ddot{u}+\left(\frac{k^2}{a^2}+\frac{9}{4}H_{\rm sr}^2-2\omega^2A(t)\sin(2\omega t+\theta)\right)u=0\,,
\end{align}
which can be regarded as the Mathieu equation.
From this expression,
it is obvious that
the mode~$k$ feels
resonance effects around the time $t_k$ given by
\begin{align}
\frac{k}{a(t_k)}=\omega\,.
\end{align}
Here we dropped higher order terms in $\omega/H_{\rm sr}$.
From general discussions on the parametric resonance,
it follows that $u(t)$ is changed around $t=t_k$ as
\begin{align}
\left|\frac{u(t)-u(t_k)}{u(t_k)}\right|\sim \big|A(t_k)\big|\,\omega \,(t-t_k)\,
\end{align}
and
the duration $\Delta t$ of the resonance can be estimated as
\begin{align}
\Delta t\sim (H_{\rm sr}\omega)^{-1/2}\,.
\end{align}
We then conclude that
the mode $k$ is changed after experiencing the parametric resonance
as
\begin{align}
\left|\frac{u_{\rm after}-u_{\rm before}}{u_{\rm before}}\right|\sim \left|A(t_k)\right|\sqrt{\frac{\omega}{H_{\rm sr}}}\,.
\end{align}
In our models,
the parameters $A(t)$ and $\omega$ are
given by
\begin{align}
A(t)\sim \frac{\alpha^2}{2} e^{-3H(t-t_\ast)}\,,
\quad
\omega=\mu H_{\rm sr}\,,
\end{align}
and the mode $k$ feels resonance effects
around the time $t_k$ given by
\begin{align}
\frac{k}{a(t_k)}=\mu H_{\rm sr}
\quad
\leftrightarrow
\quad
\frac{a(t_k)}{a(t_\ast)}=\frac{1}{\mu}\frac{k}{k_\ast}\,.
\end{align}
Since the background oscillation starts at $t=t_\ast$,
the mode $k \lesssim\mu k_\ast$ does not experience
the parametric resonance
and the resonance effect appears at the scale $k\gtrsim \mu k_\ast$.
The size of $\mathcal{C}_{\delta H}$ for $k\gtrsim \mu k_\ast$
can be also estimated as
\begin{align}
\label{CdH_estimated}
\big|\mathcal{C}_{\delta H}\big|\sim \alpha^2 \mu^{1/2}e^{-3H(t_k-t_\ast)}
\sim \alpha^2 \mu^{1/2}\left(\frac{k}{\mu k_\ast}\right)^{-3}\,.
\end{align}
As depicted in Fig.~\ref{fig:CdH},
the above estimation well explains
our exact results.

\medskip It would be also notable that the resonance effects can be also
estimated via the stationary phase approximation,\footnote{ See
e.g. Refs.~\cite{Battefeld:2013xka,Saito:2013aqa} for the use of the
stationary phase approximation in the context of heavy field
oscillations.}  which is useful to evaluate integrals with oscillations.
As discussed in Appendix~\ref{app:mathcalI}, the integral
$\mathcal{I}(2i\mu,n,x)$ in the heavy mass region $\mu\gg1$ can be
evaluated via the stationary phase approximation as
\begin{align}
\mathcal{I}(2i\mu,n,x)
&\simeq\left\{\begin{array}{ccl}
\displaystyle
\left(\frac{2\mu}{x}\right)^{3+2i\mu}
(2\mu)^{n+\frac{1}{2}}\sqrt{2\pi}e^{-2i\mu-\frac{i}{4}\pi}&{\rm for} & x\gtrsim2\mu\,,
\\[2mm]
0&{\rm for} & x\lesssim2\mu\,.
\end{array}\right.
\end{align}
Applying this approximation to~(\ref{CdH_exact}),
the resonance effect can be again expected
to appear at the scale $x_\ast\gtrsim\mu$
and its size can be again estimated
as
\begin{align}
\big|\mathcal{C}_{\delta H}\big|&\simeq
\Bigg|{\rm Im}\left[\frac{\alpha^2}{4}\mathcal{I}(2i\mu,0,2x_\ast)\right]\Bigg|
\simeq
\left|\alpha^2\frac{\sqrt{\pi}}{2}\left(\frac{\mu}{x_\ast}\right)^{3+2i\mu}
\mu^{1/2}e^{-2i\mu-\frac{i}{4}\pi}\right|
\simeq \frac{\sqrt{\pi}}{2}\alpha^2 \mu^{1/2}\left(\frac{k}{\mu k_\ast}\right)^{-3}
\,,
\end{align}
which coincides with Eq.~(\ref{CdH_estimated}). Here note that resonance
effects do not appear in $\mathcal{I}(0,n,x)$ and
$\mathcal{I}(-2i\mu,n,x)$, which are therefore irrelevant in the heavy
mass region.

\subsection{Conversion effects}
\begin{figure}[t]
\begin{center}
\includegraphics[width=160mm, bb=0 0 580 412]{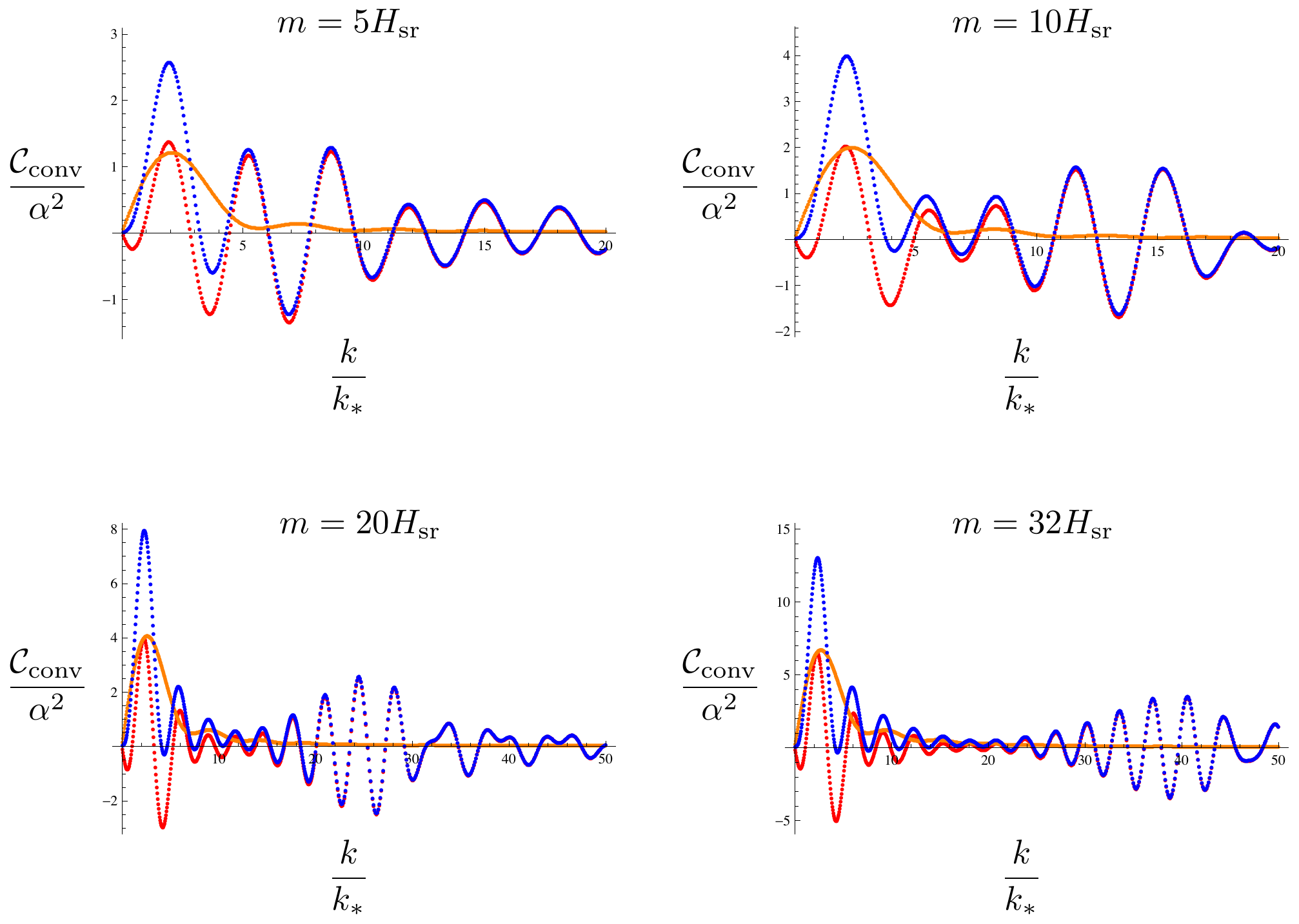}
 \end{center}
\vspace{-5mm}
\caption{Scale-dependence of $\mathcal{C}_{\rm conv}$
for
$m/H_{\rm sr}=5$ (upper left), $10$ (upper right), $20$ (lower left), and $32$ (lower right).
The 
blue/red dots are
numerical results for $\mathcal{C_{\rm conv}}$/$\mathcal{C_{\rm conv2}}$
and the orange curve is the analytic result
for $\mathcal{C_{\rm conv1}}$.
The total conversion effect
has a peak $\mathcal{C}_{\rm conve}\sim 0.5 \mu\alpha^2$
at $k\sim 2k_\ast$
and the resonance at $k\gtrsim \mu k_\ast$.
It is also observed that
the resonance effects arise only from~$\mathcal{C}_{\rm conv2}$.}
\label{fig:conv}
\end{figure}

Let us next discuss the conversion effects $\mathcal{C}_{\rm conv}$.
Substituting the expressions~(\ref{mode_u_expression})
and (\ref{mode_v_expression})
of the mode functions
into (\ref{C_conv_start})-(\ref{C_conv_start2}),
$\mathcal{C}_{\rm conv}$ can be written as
\begin{align}
\mathcal{C}_{\rm conv}&=\mathcal{C}_{\rm conv1}+\mathcal{C}_{\rm conv2}\,,\\
\mathcal{C}_{\rm conv1}&=
\pi e^{-\mu\pi}
\left|\int_{0}^{x_\ast}dx
\frac{\dot{\gamma}}{H_{\rm sr}}\,x^{-1/2}e^{ix}H_{i\mu}^{(1)}(x)\right|^2\,,\\
\mathcal{C}_{\rm conv2}&=
-2\pi e^{-\mu\pi}{\rm Re}\left[
\int_0^{x_\ast}dx_1
\frac{\dot{\gamma}}{H_{\rm sr}}\,x_1^{-1/2}e^{ix_1}H_{-i\mu}^{(2)}(x_1)
\int_{x_1}^{x_\ast}dx_2
\frac{\dot{\gamma}}{H_{\rm sr}}\,x_2^{-1/2}e^{ix_2}H_{i\mu}^{(1)}(x_2)\right]\,,
\end{align}
where $\dot{\gamma}/H_{\rm sr}$ is given by
\begin{align}
\label{gammadot/H}
\frac{\dot{\gamma}}{H_{\rm sr}}
&=
i\alpha 
\left[\frac{(\frac{3}{2}+i\mu)^2}{2\mu}(x/x_\ast)^{3/2+i\mu}
-\frac{(\frac{3}{2}-i\mu)^2}{2\mu}(x/x_\ast)^{3/2-i\mu}\right]\,.
\end{align}
It is convenient to
introduce the indefinite integral
$\mathcal{D}_+(\ell,\mu,x)$ defined by
\begin{align}
\mathcal{D}_+(\ell,\mu,x)
&=\int dx\,x^{-\frac{1}{2}+\ell}e^{ix}H^{(1)}_{i\mu}(x)\,,
\end{align}
whose analytic expression is~\cite{Chen:2012ge,Noumi:2012vr}
\begin{align}
\nonumber
\mathcal{D}_+(\ell,\mu,x)
&=
\frac{2^{i\mu} x^{\frac{1}{2}+\ell-i\mu}\Gamma(i\mu)}{i\pi(\frac{1}{2}+\ell-i\mu)}
{}_2F_2\Big(\frac{1}{2}-i\mu,\frac{1}{2}+\ell-i\mu;\frac{3}{2}+\ell-i\mu,1-2i\mu; 2ix\Big)\\
\label{Dpanalytic}
&\quad+e^{\pi\mu}\frac{2^{-i\mu} x^{\frac{1}{2}+\ell+i\mu}\Gamma(-i\mu)}{i\pi(\frac{1}{2}+\ell+i\mu)}
{}_2F_2\Big(\frac{1}{2}+i\mu,\frac{1}{2}+\ell+i\mu;\frac{3}{2}+\ell+i\mu,1+2i\mu; 2ix\Big)\,.
\end{align}
Using this function $\mathcal{D}_+(\ell,\mu,x)$,
we obtain the following useful relation:
\begin{align}
\nonumber
&\int dx
\frac{\dot{\gamma}}{H_{\rm sr}}\,x^{-1/2}e^{ix}H_{i\mu}^{(1)}(x)
\\*
&\quad
=
i\alpha\left[\frac{(\frac{3}{2}+i\mu)^2}{2\mu}x_\ast^{-(\frac{3}{2}+i\mu)}\mathcal{D}_+(\tfrac{3}{2}+i\mu,\mu,x)
-\frac{(\frac{3}{2}-i\mu)^2}{2\mu}x_\ast^{-(\frac{3}{2}-i\mu)}\mathcal{D}_+(\tfrac{3}{2}-i\mu,\mu,x)
\right]\,,
\end{align}
which
enables us to evaluate $\mathcal{C}_{\rm conv1}$ analytically
and reduces
the calculation of $\mathcal{C}_{\rm conv2}$
to one integral over $x_1$.
Since it seems difficult to perform the $x_1$-integral in~$\mathcal{C}_{\rm conv2}$ analytically,
we evaluated $\mathcal{C}_{\rm conv2}$ numerically.
The obtained $\mathcal{C}_{\rm conv}$, $\mathcal{C}_{\rm conv1}$, and $\mathcal{C}_{\rm conv2}$
are summarized in Fig.~\ref{fig:conv}.
We first notice that
there is a peak at $k\sim 2k_\ast$,
which has been discussed in recent literatures~\cite{Gao:2012uq,Shiu:2011qw,Cespedes:2012hu,Noumi:2012vr}.
In addition,
resonance-like features
can be observed
at the scale $k\gtrsim \mu k_\ast$.
In the rest of this subsection,
we discuss qualitative features of these two effects.

\subsubsection{Peak at the turning scale}
\label{subsubsec:peak}
\begin{figure}[t]
\begin{center}
\includegraphics[width=160mm, bb=0 0 580 412]{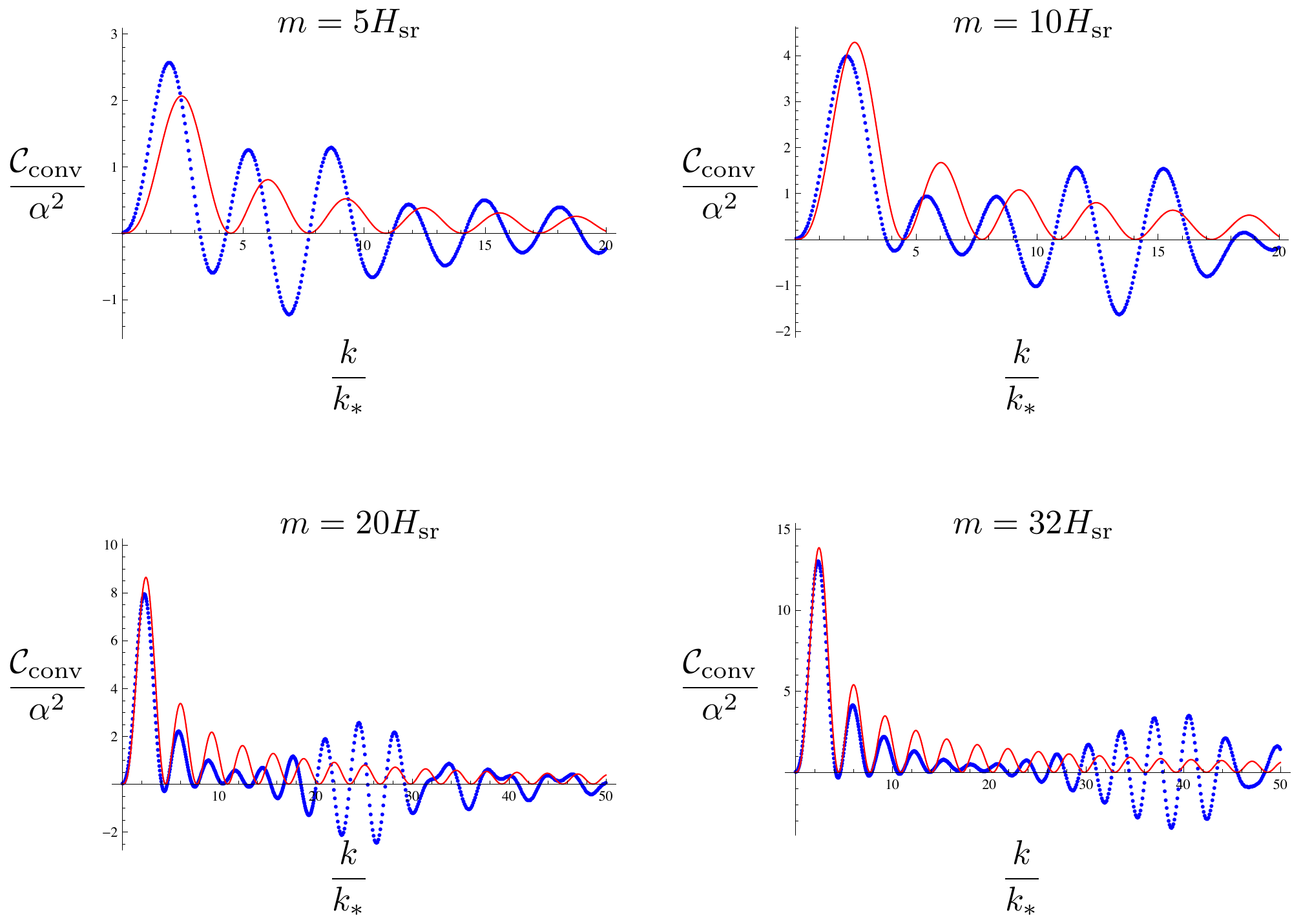}
 \end{center}
\vspace{-5mm}
\caption{
Heavy mass approximation for
$\mathcal{C}_{\rm conv}$ around the turning scale $k\sim k_\ast$.
The red curve is the analytic expression
(\ref{C_conv_heavy}) in the heavy mass approximation
and the blue dots are the numerical results without approximations.
While the behavior around $k\sim k_\ast$
is well explained by the analytic expression,
the approximation becomes worth at $k\gtrsim \mu k_\ast$.}
\label{fig:Cconv_2}
\end{figure}

As depicted in Fig.~\ref{fig:conv},
$\mathcal{C}_{\rm conv}$ has a peak around the turning scale.
In particular,
it is notable that its size becomes larger as the massive direction gets heavier.
Since effects of highly oscillating interactions
and those of heavy fields are suppressed by the high frequency or the heavy mass in general,
this kind of behavior might seem to be curious.
In the following,
we explain how the conversion effects
become relevant around the turning scale
and
analytically calculate $\mathcal{C}_{\rm conv}$ in the heavy mass region.
Using the heavy mass approximation, $m\gg H_{\rm sr}$,
the mode function for $k\sim k_\ast$ can be written as
\begin{align}
v_k(t)\simeq \frac{1}{\sqrt{2ma^3}}e^{-im(t-t_\ast)}
\end{align}
up to a phase factor
irrelevant to the computation of $\mathcal{C}_{\rm conv}$.
With this expression,
$\mathcal{C}_{\rm conv1}$ can be written for example as
\begin{align}
\nonumber
\mathcal{C}_{\rm conv1}(k)&=8\left|\int_{-\infty}^\infty dt_1\,a^3(t_1)
\beta_1(t_1)\dot{u}_k(t_1)v_k
(t_1)\right|^2\\*
\label{peak_C_conv}
&=
\frac{8\alpha^2m^2\dot{\bar{\phi}}_{\rm sr}^2}{2m}\left|\int_{t_\ast}^\infty dt\,a^{3/2}
e^{-\frac{3}{2}H_{\rm sr}(t-t_\ast)}
\sin [m(t-t_\ast)]e^{-im(t-t_\ast)}\dot{u}_k(t)\right|^2\,.
\end{align}
An important observation is that
the coupling $\beta_1$
and the mode function $v_k$ have the same frequency~$m$
so that $\beta_1v_k$ has a non-oscillating component:
\begin{align}
\sin [m(t-t_\ast)]e^{-im(t-t_\ast)}&=\frac{1-e^{-2im(t-t_\ast)}}{2i}\,.
\end{align}
Then,
this non-oscillating component becomes relevant
to the integral in~(\ref{peak_C_conv})
and $\mathcal{C}_{\rm conv1}$ can be evaluated as follows:
\begin{align}
\mathcal{C}_{\rm conv1}(k)&\simeq
\alpha^2m\,\dot{\bar{\phi}}_{\rm sr}^2
\left|\int_{t_\ast}^\infty dt\,a^{3/2}
e^{-\frac{3}{2}H_{\rm sr}(t-t_\ast)}
\dot{u}_k(t)\right|^2
=\frac{\mu\alpha^2}{2x_\ast^3}\Big|e^{ix_\ast}(1-ix_\ast)-1\Big|^2\,,
\end{align}
where $x_\ast=k/k_\ast$.
Notice that
the contributions from the non-oscillating part
are proportional to $\mu\simeq m/H_{\rm sr}$
while those from oscillating components
are suppressed by the factor $1/\mu$.
Similarly, $\mathcal{C}_{\rm conv2}$ is calculated as
\begin{align}
\nonumber
\mathcal{C}_{\rm conv2}(k)
&\simeq - \frac{\mu\alpha^2}{2x_\ast^3}{\rm Re}\left[
\big(e^{-ix_\ast}(1+ix_\ast)-1\big)^2\right]\,,
\end{align}
and we have the following analytic expression
for $\mathcal{C}_{\rm conv}$ in the heavy mass approximation:
\begin{align}
\label{C_conv_heavy}
\mathcal{C}_{\rm conv}(k)
&\simeq\mu\alpha^2\frac{(\sin x_\ast-x_\ast\cos x_\ast)^2}{x_\ast^3}\,.
\end{align}
Note that (\ref{C_conv_heavy})
takes its maximum value $\mathcal{C}_{\rm conv}\sim 0.43\,\mu\alpha^2$ at $x_\ast\sim2.46$.
As depicted in Fig.~\ref{fig:Cconv_2},
the analytic expression (\ref{C_conv_heavy})
well explains the behavior around the turning scale
$k\sim k_\ast$,
but
the approximation becomes worse
in the region $k\gtrsim \mu k_\ast$
essentially because of resonance effects
discussed below.
To summarize,
the peak around the turning scale
arises because of the coincidence between the frequency of
the oscillating coupling $\beta_1$
and the mass $m$ of massive isocurvature perturbations.
Using the heavy mass approximation,
the scale of the peak
and its size can be estimated as
$k\sim2.46k_\ast$ and $\mathcal{C}_{\rm conv}\sim 0.43\,\mu\alpha^2$.

\subsubsection{Resonance effects}
\label{subsubsec:resonance}
Let us next discuss the resonance-like
effects at the scale $k\gtrsim\mu k_\ast$
using the heavy mass approximation $m\gg H_{\rm sr}$.
In this approximation, 
the mode function for $k\sim \mu k_\ast$ is given
up to an irrelevant phase factor as
\begin{align}
v_k(t)\simeq \frac{1}{\sqrt{2Ea^3}}e^{-iE(t-t_\ast)}
\quad
{\rm with}
\quad
E=\sqrt{m^2+\frac{k^2}{a^2}}\,.
\end{align}
Using this expression,
we first estimate the size of $\mathcal{C}_{\rm conv1}$:
\begin{align}
\nonumber
\mathcal{C}_{\rm conv1}(k)&=8\left|\int_{-\infty}^\infty dt_1\,a^3(t_1)
\beta_1(t_1)\dot{u}_k(t_1)v_k
(t_1)\right|^2\\*
\label{C_vonv_resonance}
&\simeq
8\alpha^2m^2\dot{\bar{\phi}}_{\rm sr}^2\left|\int_{t_\ast}^\infty dt\,\frac{a^{3/2}}{\sqrt{2E}}
e^{-\frac{3}{2}H_{\rm sr}(t-t_\ast)}
\sin [m(t-t_\ast)]e^{-iE(t-t_\ast)}\dot{u}_k(t)\right|^2\,.
\end{align}
The phase factor of the integrand
takes the form
\begin{align}
\sim \exp\left[ i\Big(\pm m-E-\frac{k}{a}\Big)t\right]=\exp \left[i\Big(\pm m-\sqrt{m^2+\frac{k^2}{a^2}}-\frac{k}{a}\Big)t\right]\,,
\end{align}
which always oscillates with a frequency of the order $m$.
Therefore,
the integral in (\ref{C_vonv_resonance})
is suppressed by $1/(m\,E^{1/2})\sim m^{-3/2}$ and
it is found that
$\mathcal{C}_{\rm conv1}\sim \alpha^2/\mu$,
which is irrelevant in the heavy mass region.

\medskip
Let us then consider  $\mathcal{C}_{\rm conv2}$
in the heavy mass approximation:
\begin{align}
\nonumber
\mathcal{C}_{\rm conv2}&=-16{\rm Re}\Bigg[
\int_{-\infty}^\infty dt_1\,a^3(t_1)
\beta_1(t_1)\dot{u}^\ast_k(t_1)v_k(t_1)
\int_{-\infty}^{t_1}dt_2
\,a^3(t_2)
\beta_1(t_2)\dot{u}^\ast_k(t_2)v^\ast_k(t_2)
\Bigg]\\
\nonumber
&=-16\alpha^2m^2\dot{\bar{\phi}}_{\rm sr}^2\,{\rm Re}\Bigg[
\int_{t_{\ast}}^\infty dt_1\frac{a^{3/2}}{\sqrt{2E}}\,
e^{-\frac{3}{2}H_{\rm sr}(t_1-t_\ast)}
\sin [m(t_1-t_\ast)]e^{-iE(t_1-t_\ast)}\dot{u}^\ast_k(t_1)\\
\label{C_conv2_heavy}
&\qquad\qquad\qquad\qquad\quad
\times
\int_{t_{\ast}}^{t_1}dt_2\frac{a^{3/2}}{\sqrt{2E}}\,
e^{-\frac{3}{2}H_{\rm sr}(t_2-t_\ast)}
\sin [m(t_2-t_\ast)]e^{iE(t_2-t_\ast)}\dot{u}^\ast_k(t_2)
\Bigg]\,.
\end{align}
Apparently,
(\ref{C_conv2_heavy}) may seem to be suppressed
by the factor $1/\mu$ like $\mathcal{C}_{\rm conv2}$.
However,
it turns out that
non-trivial contributions arise
around the time $t_1=t_k$
determined by $\displaystyle\frac{k}{a(t_k)}\sim m$,
which is the time when resonance effects
from the Hubble deformation appear.
Around the time $t_2=t_k$,
$\sin[m(t_2-t_\ast)]\dot{u}^\ast_k(t_2)$
has non-oscillating components
(whose time-dependence is negligible
compared to $E$)
so that the $t_2$-integral in~(\ref{C_conv2_heavy})
can be performed as
\begin{align}
\nonumber
&\int_{t_{\ast}}^{t_1}dt_2
\,\frac{a^{3/2}}{\sqrt{2E}}\,
e^{-\frac{3}{2}H_{\rm sr}(t_2-t_\ast)}
\sin [m(t_2-t_\ast)]e^{iE(t_2-t_\ast)}\dot{u}^\ast_k(t_2)\\
&\qquad
\simeq
\frac{1}{iE}
\frac{a^{3/2}}{\sqrt{2E}}\,
e^{-\frac{3}{2}H_{\rm sr}(t_1-t_\ast)}
\sin [m(t_2-t_\ast)]e^{iE(t_1-t_\ast)}\dot{u}^\ast_k(t_1)\,,
\end{align}
which implies that
adiabatic and isocurvature perturbations
around $t\sim t_k$
are converted to each other instantaneously.
We therefore obtain the following expression for~(\ref{C_conv2_heavy}):
\begin{align}
\nonumber
\mathcal{C}_{\rm conv2}
&\simeq
-16\alpha^2m^2\dot{\bar{\phi}}_{\rm sr}^2\,{\rm Re}\Bigg[
\int_{t_\ast}^\infty dt_1\,\frac{a^{3}}{2iE^2}
e^{-3H_{\rm sr}(t_1-t_\ast)}
\sin^2 [m(t_1-t_\ast)]\big(\dot{u}^\ast_k(t_1)\big)^2\Bigg]\\*
\label{C_vonv2_resonance_2}
&=8\alpha^2M_{\rm Pl}^2\dot{H}_{\rm sr}\,{\rm Re}\Bigg[
i\int_{t_\ast}^\infty dt_1\,a^{3}\frac{m^2}{E^2}
e^{-3H_{\rm sr}(t_1-t_\ast)}
\big(1-\cos[2m(t_1-t_\ast)]\big)\dot{u}^2_k(t_1)\Bigg]\,.
\end{align}
When $t_1\sim t_k$,
non-oscillating components appear
in $\cos[2m(t_1-t_\ast)]\dot{u}_k^2(t_1)$
so that the $t_1$-integral has a non-trivial contribution
from $t_1\sim t_k$,
which can be observed as the resonance in Fig.~\ref{fig:conv}.
It is also notable that
(\ref{C_vonv2_resonance_2})
can be written up to higher order terms in $1/\mu$
as
\begin{align}
\mathcal{C}_{\rm conv2}
&=-2M_{\rm Pl}^2\dot{H}_{\rm sr}\,{\rm Re}\Bigg[
i\int_{t_\ast}^\infty dt_1\,a^{3}\frac{8m^2}{E^2}
\kappa(t_1)
\dot{u}^2_k(t_1)\Bigg]\,,
\end{align}
which takes a similar form
as $\mathcal{C}_{\delta H}$.
Since the leading contribution
in the $t_1$-integral
arises from~$t_1\sim t_k$,
we can further rewrite $\mathcal{C}_{\rm conv2}$
in the following way:
\begin{align}
\nonumber
\mathcal{C}_{\rm conv2}
&\simeq-2M_{\rm Pl}^2\dot{H}_{\rm sr}\,{\rm Re}\Bigg[
i\int_{t_\ast}^\infty dt_1\,a^{3}\frac{4m^2}{E^2}
\kappa(t_1)
\Big(\dot{u}^2_k(t_1)-\frac{(\partial_iu_k)^2}{a^2}\Big)\Bigg]\\&\simeq-2\,{\rm Re}\Bigg[
i\int_{t_\ast}^\infty dt_1\,a^{3}
(2M_{\rm Pl}^2\dot{H}_{\rm sr}\kappa(t_1))
\Big(\dot{u}^2_k(t_1)-\frac{(\partial_iu_k)^2}{a^2}\Big)\Bigg]\,,
\end{align}
where we used
the relation
$\displaystyle\dot{u}_k^2=-\frac{(\partial_iu_k)^2}{a^2}$
in the heavy mass limit
and the relation $E^2(t_k)=2m^2$.
Surprisingly,
we obtained the relation
$\mathcal{C}_{\rm conv2}=-\mathcal{C}_{\delta H}$
in the heavy mass limit,
which suggests that
the resonance effects from the Hubble deformation
and the conversion interaction
cancel each other out.

\medskip
To summarize,
at the scale $k\gtrsim \mu k_\ast$,
the leading contribution from the conversion interaction
arises from $t\sim t_k$
and
resonance effects are induced.
In particular,
it is suggested that
the resonance in $\mathcal{C}_{\delta H}$
and $\mathcal{C}_{\rm conv}$
cancels each other out in the heavy mass limit.
In the next subsection,
we confirm this cancellation by numerical calculations.
At the end of this section,
it is also discussed
from the effective interaction perspective
that this kind of cancellation generically occurs in a wider class of models with heavy field oscillations.

\subsection{Total power spectrum}
\label{subsec:totalPS}
\begin{figure}[h]
\begin{center}
\includegraphics[width=160mm, bb=0 0 580 418]{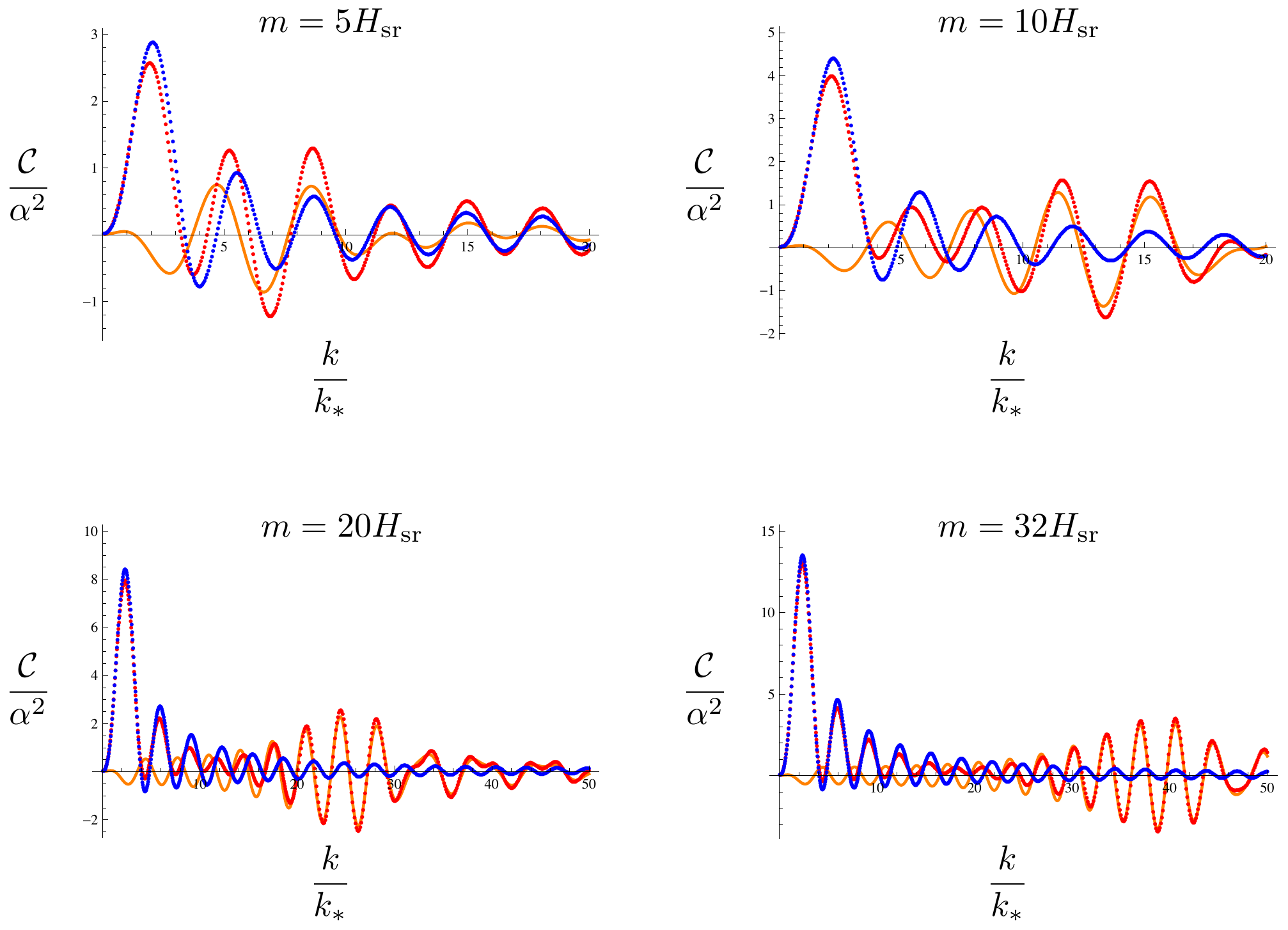}
 \end{center}
\vspace{-5mm}
\caption{Total deviation factor $\mathcal{C}$.
The blue/red dots are
numerical results for
$\mathcal{C}$/$\mathcal{C}_{\rm conv}$.
The orange curve is the analytic result
for $-\mathcal{C}_{\delta H}$.
It is observed that
the resonances in $\mathcal{C}_{\rm conv}$ and $-\mathcal{C}_{\delta H}$
well coincide in the heavy mass region
so that these resonance effects disappear
in the total deviation factor $\mathcal{C}$.}
\label{fig:Ctotal}
\end{figure}
\begin{figure}[t]
\begin{center}
\includegraphics[width=160mm, bb=0 0 580 418]{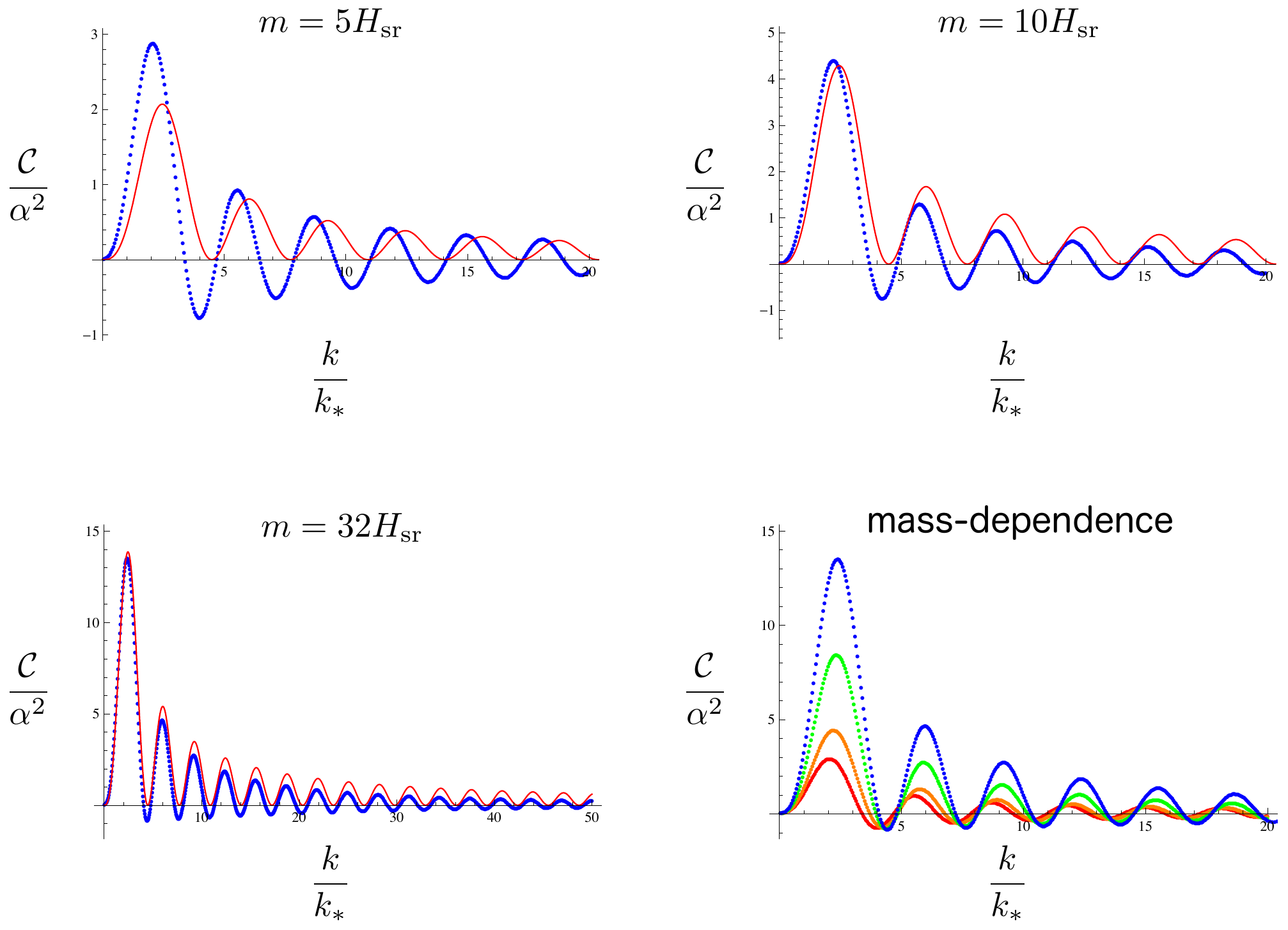}
 \end{center}
\vspace{-5mm}
\caption{Heavy mass approximation
for the total deviation factor $\mathcal{C}$
(upper figures and lower left figure).
The analytic expression~(\ref{C_total_app})
denoted by the red curve
provides a good approximation
for the full total deviation factor~$\mathcal{C}$ (blue dots).
As is suggested from the expression~(\ref{C_total_app}),
the shape of $\mathcal{C}$ does not depend on the mass $m$
up to a normalization factor proportional to~$\mu\alpha^2$ (lower right figure; the red/orange/green/blue dotes
are numerical results of $\mathcal{C}$ for $m=5H_{\rm sr}$/$10H_{\rm sr}$/$20H_{\rm sr}$/$32H_{\rm sr}$).}
\label{fig:Ctotal_2}
\end{figure}

Now we combine the results in the previous
two subsections and evaluate the total deviation factor
$\mathcal{C}=\mathcal{C}_{\delta H}+\mathcal{C}_{\rm conv}$.
As depicted in Fig.~\ref{fig:Ctotal},
we can observe the resonance cancellation
discussed in the previous subsection
and no resonance 
appears in the power spectrum
after taking into account
both of the Hubble deformation effects
and the conversion effects appropriately.
An important consequence of this cancellation
is that the peak at the turning scale becomes clearer
and
the analytic expression~(\ref{C_conv_heavy})
describing the peak around the turning scale
becomes applicable to
the total deviation factor
$\mathcal{C}$ in the whole scale:\footnote{
\label{footnote:comments_on_GLM}
Interestingly,
our analytic expression~(\ref{C_total_app}) for the total deviation factor
$\mathcal{C}$
exactly coincides with
the sudden turning limit of
$\mathcal{F}_{h}$ in~\cite{Gao:2012uq},
where the power spectra in
a wider class of models with turning potentials
were investigated via the potential basis.
Our deviation factor $\mathcal{C}$
corresponds to $\Delta\mathcal{P}/\mathcal{P}_0=\mathcal{F}_{l}+\mathcal{F}_{h}+\mathcal{F}_{lh}$ there
and the sudden turning limit is realized by $\mu\to\infty$
in their language.
As we have mentioned in the introduction and 
footnote~\ref{footnote:kinetic_basis} in Sec.~\ref{subsubsec:action_kin},
the calculation in the potential basis suffers from
spurious singularities.
Correspondingly,
$\mathcal{F}_{l}$ and $\mathcal{F}_{lh}$
diverges in the limit $\mu\to\infty$.
Based on our results,
we expect that
$\mathcal{F}_{l}$ and $\mathcal{F}_{lh}$
will cancel each other out
and the total deviation factor there
will reduce to the form $\Delta\mathcal{P}/\mathcal{P}_0=\mathcal{F}_{h}$
in the sudden turning limit.}
\begin{align}
\label{C_total_app}
\mathcal{C}\simeq\mu\alpha^2\frac{(\sin x_\ast-x_\ast\cos x_\ast)^2}{x_\ast^3}\,.
\end{align}
As shown in Fig.~\ref{fig:Ctotal_2},
this analytic expression
provides a good approximation.

\subsection{On resonance cancellation for more general settings}
\label{subsec:cancellation}
In this section
we have discussed the effects of
sudden turning potentials
on the primordial power spectrum.
In particular,
we found a non-trivial resonance cancellation
between
the Hubble deformation effects
and the conversion effects.
Before closing this section,
we would like to
point out that
this kind of resonance cancellation
generically occurs
in more general two field models with
canonical kinetic terms.

\medskip
With this class of models in mind,
let us first consider the following action:
\begin{align}
\label{on_resonance_action}
S&=\int d^4x \,a^3\Big[
-M_{\rm Pl}^2\dot{H}\Big(\dot{\pi}^2-\frac{(\partial_i\pi)^2}{a^2}\Big)
+\frac{1}{2}\Big(\dot{\sigma}^2-\frac{(\partial_i\sigma)^2}{a^2}
-m^2\sigma^2\Big)
-2\beta_1\dot{\pi}\sigma
\Big]\,,
\end{align}
where $\pi$ and $\sigma$
correspond to the Goldstone boson
and the massive isocurvature perturbation
in the kinetic basis, respectively.
Suppose that
the Hubble parameter can be 
decomposed into the slow-roll part $H_{\rm sr}$ and
the oscillating part $\delta H$,
\begin{align}
H=H_{\rm sr}+\delta H\,,
\end{align}
and $H_{\rm sr}$ satisfies the slow-roll conditions
$\epsilon_{\rm sr}=-\dot{H}_{\rm sr}/H_{\rm sr}^2\ll1$
and
$\eta_{\rm sr}=\dot{\epsilon}_{\rm sr}/(H_{\rm sr}\epsilon_{\rm sr})\ll1$.
We also assume that
$\delta H$ can be treated as perturbations:
$\displaystyle\kappa=\frac{\delta \dot{H}}{\dot{H}_{\rm sr}}\ll1$.
As discussed in Sec.~\ref{subsubsec:Mathieu},
when $\kappa$ oscillates with a high frequency $2\omega\gg H_{\rm sr}$,
the mode $k$ experiences the parametric resonance
around the time $t_k$ determined
by $\displaystyle\frac{k}{a(t_k)}=\omega$
due to the Hubble deformations.

\medskip
We then consider the conversion effects.
Suppose that the coupling $\beta_1$
oscillates with a high frequency $\widetilde{\omega}\gg H_{\rm sr}$
and the mass $m$ of $\sigma$ is
also heavy $m\gg H_{\rm sr}$.
In this setting,
we would like to integrate out
the heavy field $\sigma$
and
to construct the effective action for $\pi$.
In general,
when the time-dependence of $\beta_1\dot{\pi}$
is comparable to the mass of $\sigma$,
it is not possible to integrate out~$\sigma$
in a simple manner
and
it is required
to consider its full dynamics.
As discussed in Sec.~\ref{subsubsec:resonance},
however,
the time-dependence of $\beta_1 \dot{\pi}$
becomes negligible
when $\displaystyle\frac{k}{a}\sim \widetilde{\omega}$.
In such a case,
$\sigma$ can be integrated out
and the conversion effects around the time
$t= \tilde{t}_k$
defined by $\displaystyle\frac{k}{a(\tilde{t}_k)}=\widetilde{\omega}$
can be simplified in the following way
(see e.g.~\cite{Noumi:2012vr} for a detailed discussion):
\begin{align}
\nonumber
&\int d^4x \,a^3\Big[
\frac{1}{2}\Big(\dot{\sigma}^2-\frac{(\partial_i\sigma)^2}{a^2}
-m^2\sigma^2\Big)
-2\beta_1\dot{\pi}\sigma
\Big]\\*
\nonumber
&\simeq\int dt\frac{d^3k}{(2\pi)^3}\,a^3\left[-\frac{1}{2}\Big(\frac{k^2}{a^2}+m^2\Big)\sigma^2
-2\beta_1\dot{\pi}\sigma
\right]\\*
&=\int dt\frac{d^3k}{(2\pi)^3}\,a^3\left[-\frac{m^2+\widetilde{\omega}^2}{2}\Big(\sigma+\beta_1\dot{\pi}\Big)^2+\frac{2\beta_1^2}{m^2+\widetilde{\omega}^2}\dot{\pi}^2
\right]
\to\int d^4x\,a^3\frac{2\beta_1^2}{m^2+\widetilde{\omega}^2}\dot{\pi}^2\,,
\end{align}
where
we used $\displaystyle \frac{\omega}{a(t_k)}=\widetilde{\omega}$
and $\sigma$ was integrated out at the last arrow.
Then,
the dynamics of $\pi$ at $t\sim\tilde{t}_k$
can be described by the effective action
\begin{align}
\label{on_resonance_effective}
S_{\rm eff}&=\int d^4x \,a^3\Big[
-M_{\rm Pl}^2\dot{H}\Big(\dot{\pi}^2-\frac{(\partial_i\pi)^2}{a^2}\Big)
+\frac{2\beta_1^2}{m^2+\widetilde{\omega}^2}\dot{\pi}^2
\Big]\,.
\end{align}
Here notice that
the time $t_k$ of resonances from Hubble deformations
and the time $\tilde{t}_k$ of those from conversion interactions
coincide if and only if $\omega=\widetilde{\omega}$.

\medskip
Finally, let us apply the above discussions
to two-field models with canonical kinetic terms.
Suppose that
a heavy field $\phi_\bot$ starts oscillating
at some time $t=t_\ast$
and the action for $t>t_\ast$
is given by
\begin{align}
\label{on_resonance_general_action}
S=\int d^4x\sqrt{-g}\left[\frac{1}{2}M_{\rm Pl}^2R-\sum_{i=\parallel,\bot}\frac{1}{2}\partial_\mu\phi_i\partial^\mu\phi_i
-V_{\rm sr}(\phi_\parallel)-V_\bot(\phi_\bot)\right]\,.
\end{align}
Here $V_{\rm sr}$ is a slow-roll potential and $\phi_\bot$ is defined
such that the massive potential $V_\bot$ takes its minimum value at
$\phi_\bot=0$.  Let us define the deviations $\varphi_i$ of the
background trajectory from the single field slow-roll model as
\begin{align}
\bar{\phi}_\parallel(t)=\bar{\phi}_{\rm sr}(t)+\varphi_1\,,
\quad
\bar{\phi}_\bot(t)=\varphi_2\,.
\end{align}
Then,
applying our general discussions in Sec.~\ref{sec:setup},
the action for the Goldstone boson in the kinetic basis
is given by (\ref{on_resonance_action})
with the parameters
\begin{align}
\kappa=2\frac{\dot{\varphi}_1}{\dot{\bar{\phi}}_{\rm sr}}+\left(\frac{\dot{\varphi}_2}{\dot{\bar{\phi}}_{\rm sr}}\right)^2\,,\quad
\beta_1=
\ddot{\varphi}_2\,.
\end{align}
Since the typical time-dependence of $\varphi_1$ is of order $H_{\rm sr}$,
$\varphi_1$ is irrelevant to the resonance effects.
We therefore concentrate on the effects of $\varphi_2$ 
and introduce the frequency $\omega$ characterizing the highly oscillating background $\varphi_2$
as
$\varphi_2=\alpha\sin(\omega t+\theta)$,
where the time-dependence of the normalization factor $\alpha$ and the phase factor $\theta$ is negligible
compared to $\omega$.
Notice that
the couplings $\kappa$ and $\beta_1$
oscillate with the frequency $2\omega$ and $\omega$,
respectively,
so that the resonance effects
from the Hubble deformation
and the conversion effect appear at the same time.
The effective action~(\ref{on_resonance_effective}) for $\pi$ at $t\sim t_k$
can be then written as
\begin{align}
\nonumber
S_{\rm eff}
&=\int d^4x\,a^3\left[-M_{\rm Pl}^2\dot{H}_{\rm sr}\left(\dot{\pi}^2-\frac{(\partial_i\pi)^2}{a^2}\right)
+\frac{1}{2}\dot{\varphi}_2^2\left(\dot{\pi}^2-\frac{(\partial_i\pi)^2}{a^2}\right)
+\frac{\ddot{\varphi}_2^2}{m^2+\omega^2}\dot{\pi}^2
\right]\\*[1mm]
\nonumber
&=\int d^4x\,a^3\left[-M_{\rm Pl}^2\dot{H}_{\rm sr}\left(\dot{\pi}^2-\frac{(\partial_i\pi)^2}{a^2}\right)\right.\\
\label{on_resonance_eff_final}
&\qquad\qquad\qquad
\left.
+\frac{\alpha^2\omega^2}{2}\cos^2(\omega t+\theta)\left(\dot{\pi}^2-\frac{(\partial_i\pi)^2}{a^2}\right)
+\frac{\alpha^2\omega^4}{m^2+\omega^2}\sin^2(\omega t+\theta)\dot{\pi}^2
\right]\,.
\end{align}
As discussed in Sec.~\ref{subsubsec:resonance},
the $\displaystyle\frac{(\partial_i\pi)^2}{a^2}$
interaction gives the same contribution
as $-\dot{\pi}^2$
when $\displaystyle \frac{k}{a}\gg H_{\rm sr}$.
Then, the interaction terms
in~(\ref{on_resonance_eff_final})
can be written effectively as
\begin{align}
\dot{\varphi}_2^2\dot{\pi}^2
+\frac{2\ddot{\varphi}^2_2}{m^2+\omega^2}\dot{\pi}^2
&=\alpha^2\omega^2\left(\cos^2(\omega t+\theta)+\frac{2\omega^2}{m^2+\omega^2}\sin^2(\omega t+\theta)\right)\dot{\pi}^2\,.
\end{align}
In particular,
when the mass of the heavy isocurvature $\sigma$
coincides with
the frequency of the highly oscillating background $\varphi_2$, $m=\omega$,
the oscillating features of the two interactions in the action
cancel each other out, which causes no resonance in the power spectrum:
\begin{align}
\dot{\varphi}_2^2\dot{\pi}^2
+\frac{2\ddot{\varphi}^2_2}{m^2+\omega^2}\dot{\pi}^2
=\alpha^2m^2\dot{\pi}^2\,.
\end{align}
Note that the condition $m=\omega$ is always satisfied
at the leading order in $\varphi_i$'s
in the models described by
the action~(\ref{on_resonance_general_action}).
We then conclude that
no resonance appears
in the power spectrum
so that
the important signature of this class of models
is the peak at the turning scale,
whose analytic expression in the heavy mass approximation can be obtained straightforwardly
by extending the discussion
in Sec.~\ref{subsubsec:peak}.

\section{Primordial bispectrum}
\label{sec:bispectrum}
\setcounter{equation}{0}
Let us next calculate the primordial bispectrum
induced by the sudden turning potentials.
Using the relation $\zeta=-H\pi$ at the linear order,
three-point functions of scalar perturbations $\zeta$
can be obtained as
\begin{align}
\langle\zeta_{\bf k_1}\zeta_{\bf k_2}\zeta_{\bf k_3}\rangle
&=-H^3\langle\pi_{\bf k_1}\pi_{\bf k_2}\pi_{\bf k_3}\rangle\,.
\end{align}
We define $\mathcal{B}_\zeta(k_1,k_2,k_3)$
and $\mathcal{B}_\pi(k_1,k_2,k_3)$
by the three point functions without the delta function factor:
\begin{align}
\langle\zeta_{\bf k_1}\zeta_{\bf k_2}\zeta_{\bf k_3}\rangle
&=
(2\pi)^3\delta^{(3)}({\bf k_1+k_2+k_3})\mathcal{B}_\zeta(k_1,k_2,k_3)\,,\\
\langle\pi_{\bf k_1}\pi_{\bf k_2}\pi_{\bf k_3}\rangle
&=
(2\pi)^3\delta^{(3)}({\bf k_1+k_2+k_3})\mathcal{B}_\pi(k_1,k_2,k_3)\,.
\end{align}
It is also convenient to introduce the
conventional shape function $S$ and $f_{NL}$ parameter:
\begin{align}
S(k_1,k_2,k_3)&=\frac{k_1^2k_2^2k_3^2}{(2\pi)^4\mathcal{P}_{\zeta}^2}\mathcal{B}_\zeta(k_1,k_2,k_3)\,,\\
f_{NL}(k_1,k_2,k_3)&=\frac{10}{3}\frac{k_1^3k_2^3k_3^3}{\sum_ik_i^3}
\frac{1}{(2\pi)^4\mathcal{P}_{\zeta}^2}\mathcal{B}_\zeta(k_1,k_2,k_3)
=\frac{10}{3}\frac{k_1k_2k_3}{\sum_i
 k_i^3}S(k_1,k_2,k_3)\,,
\end{align}
where $\mathcal{P}_\zeta$ is given at the leading order in slow-roll parameters as
\begin{align}
\mathcal{P}_\zeta&=\frac{1}{(2\pi)^2}\frac{H_{\rm sr}^2}{2M_{\rm Pl}^2\epsilon_{\rm sr}}\,.
\end{align}
Using this expression,
$S(k_1,k_2,k_3)$ and $f_{NL}(k_1,k_2,k_3)$
can be related to $\mathcal{B}_\pi(k_1,k_2,k_3)$ as
\begin{align}
S(k_1,k_2,k_3)&=-\frac{4M_{\rm Pl}^4\epsilon_{\rm sr}^2}{H_{\rm sr}}
k_1^2k_2^2k_3^2\mathcal{B}_{\pi}(k_1,k_2,k_3)\,,\\
f_{NL}(k_1,k_2,k_3)&=-\frac{10}{3}\frac{4M_{\rm Pl}^4\epsilon_{\rm
 sr}^2}{H_{\rm sr}}\frac{k_1^3k_2^3k_3^3}{\sum_i
 k_i^3}\mathcal{B}_{\pi}(k_1,k_2,k_3)
=\frac{10}{3}\frac{k_1k_2k_3}{\sum_i
 k_i^3}S(k_1,k_2,k_3)\,.
\end{align}
Let us then calculate the three point functions $\mathcal{B}_\pi(k_1,k_2,k_3)$ of $\pi$
using the in-in formalism.
As we introduced earlier,
the interaction Hamiltonian of our model takes the form
\begin{align}
H_{\rm int}^{(2)}&=\int d^3x\,a^3\left[\theta(t-t_\ast)M_{\rm Pl}^2\dot{H}_{\rm sr}\kappa\Big(\dot{\pi}^2-\frac{(\partial_i\pi)^2}{a^2}\Big)
+2\beta_1\dot{\pi}\sigma\right]\,,\\*
\nonumber
H_{\rm int}^{(3)}&=\int d^3x\,a^3\left[\theta(t-t_\ast)M_{\rm Pl}^2\dot{H}_{\rm sr}\dot{\kappa}\pi\Big(\dot{\pi}^2-\frac{(\partial_i\pi)^2}{a^2}\Big)
+\beta_1\Big(\dot{\pi}^2-\frac{(\partial_i\pi)^2}{a^2}\Big)\sigma
+2\dot{\beta}_1\pi\dot{\pi}\sigma \right.\\*
\label{cubic_int_bispectrum}
&
\qquad\qquad\qquad
\left.+\theta(t-t_\ast)\frac{\lambda\varphi_2}{6}\sigma^3\right]\,.
\end{align}
\begin{figure}[t]
\begin{center}
\includegraphics[width=160mm, bb=0 0 580 294]{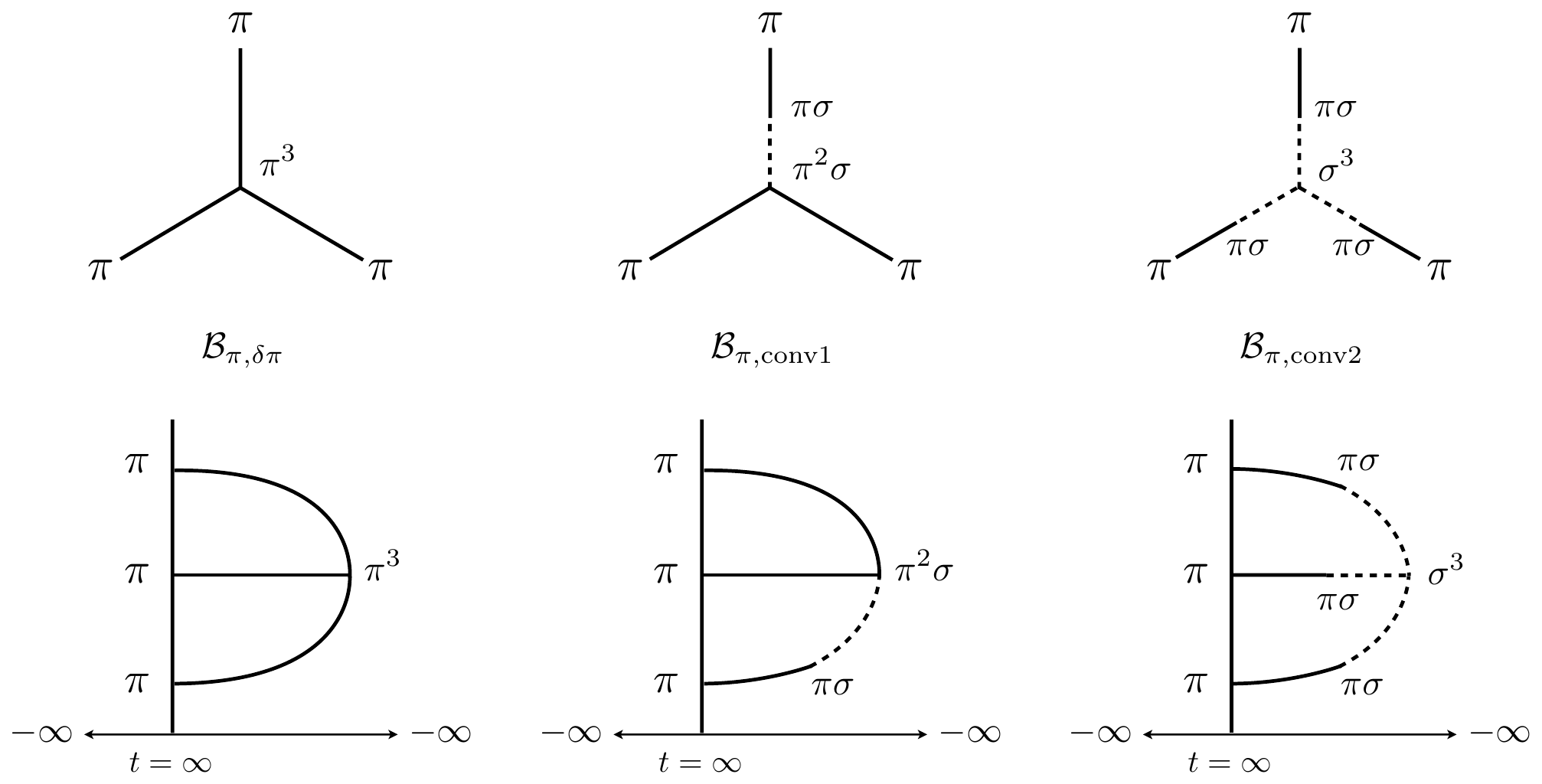}
 \end{center}
\vspace{-5mm}
\caption{Diagrams for the three-point functions of $\pi$
are classified into
$\mathcal{B}_{\pi,\delta H}$,
$\mathcal{B}_{\pi,\delta H}$,
and $\mathcal{B}_{\pi,\delta H}$
with respect to the number of conversion interactions
(upper figures).
Examples of corresponding diagrams
for the in-in calculations are given
in lower figures.}
\label{fig:BS_diagram}
\end{figure}
Corresponding to three types of diagrams depicted in Fig~\ref{fig:BS_diagram},
$\mathcal{B}_{\pi}(k_1,k_2,k_3)$ contains three types of contributions:
\begin{align}
\mathcal{B}_{\pi}(k_1,k_2,k_3)=\mathcal{B}_{\pi,\delta H}(k_1,k_2,k_3)+\mathcal{B}_{\pi,{\rm conv}1}(k_1,k_2,k_3)+\mathcal{B}_{\pi,{\rm conv}2}(k_1,k_2,k_3)\,.
\end{align}
The first contribution $\mathcal{B}_{\pi,\delta H}(k_1,k_2,k_3)$
originates from the first term in the cubic interaction 
(\ref{cubic_int_bispectrum})
and describes the Hubble deformation effects.
It can be expressed in terms of the mode functions and couplings~as
\begin{align}
\nonumber
\mathcal{B}_{\pi,\delta H}(k_1,k_2,k_3)&=
\frac{1}{2M_{\rm Pl}^3\epsilon_{\rm sr}^{3/2}(k_1k_2k_3)^{3/2}}{\rm Re}
\left[-i\int_{t_\ast}^\infty dt\,a^3
M_{\rm Pl}^2\dot{H}_{\rm sr}\dot{\kappa}u_{k_3}^\ast
\Big(\dot{u}_{k_1}^\ast\dot{u}_{k_2}^\ast
+\frac{{\bf k_1}\cdot {\bf k_2}}{a^2}u_{k_1}^\ast u_{k_2}^\ast\Big)\right]
\\*
&\quad
+(\text{$2$ permutations})\,.
\end{align}
The second and the third contributions,
$\mathcal{B}_{\pi,{\rm conv}1}(k_1,k_2,k_3)$
and
$\mathcal{B}_{\pi,{\rm conv}2}(k_1,k_2,k_3)$,
describe the conversion effects.
They originate from the second and third terms
and the last term in (\ref{cubic_int_bispectrum}),
respectively.
In terms of the mode functions and couplings,
they can be written as
\begin{align}
\mathcal{B}_{\pi,{\rm conv}1}(k_1,k_2,k_3)
\nonumber
&=
\frac{1}{4M_{\rm Pl}^3\epsilon_{\rm sr}^{3/2}(k_1k_2k_3)^{3/2}}
{\rm Re}
\left[-i\int_{-\infty}^\infty dt\,a^3
\Big\{2\beta_1\Big(\dot{u}_{k_1}^\ast \dot{u}_{k_2}^\ast 
+\frac{{\bf k_1}\cdot {\bf k_2}}{a^2}u_{k_1}^\ast u_{k_2}^\ast \Big)\mathcal{F}_3(t)\right.\\*
\nonumber
&\qquad\qquad\qquad\qquad\qquad\qquad\qquad\qquad\qquad
\left.+2\dot{\beta}_1(u_{k_1}^\ast\dot{u}_{k_2}^\ast+\dot{u}_{k_1}^\ast u_{k_2}^\ast)
\mathcal{F}_3(t)\Big\}
\right]
\\*
&\quad
+(\text{$2$ permutations})\,,\\
\mathcal{B}_{\pi,{\rm conv}2}(k_1,k_2,k_3)&=
\frac{1}{4M_{\rm Pl}^3\epsilon_{\rm sr}^{3/2}(k_1k_2k_3)^{3/2}}{\rm Re}
\left[-i\int_{-\infty}^\infty dt\,\theta(t-t_\ast)\,a^3
\lambda \varphi_2 
\mathcal{F}_1(t)\mathcal{F}_2(t)\mathcal{F}_3(t)\right]
\,,
\end{align}
where $\mathcal{F}_i(t)$ is a deformed propagator defined by
\begin{align}
\nonumber
\mathcal{F}_i(t)&=
-iv_{k_i}^\ast(t)
\int_t^\infty dt^\prime a^3 2\beta_1\dot{u}_{k_i}^\ast v_{k_i}(t^\prime)
-iv_{k_i}(t)
\int_{t_\ast}^t dt^\prime a^32\beta_1\dot{u}_{k_i}^\ast v_{k_i}^\ast(t^\prime)\\
&\quad
+iv_{k_i}^\ast(t)
\int_{t_\ast}^\infty dt^\prime a^32\beta_1\dot{u}_{k_i} v_{k_i}(t^\prime)\,.
\end{align}
In the rest of this section,
we calculate each contribution
and evaluate non-Gaussianities for our models.

\subsection{Hubble deformation effects}
\label{subsec:SdH}
Let us start from the Hubble deformation effects $\mathcal{B}_{\pi,\delta H}$:
\begin{align}
\nonumber
\mathcal{B}_{\pi,\delta H}(k_1,k_2,k_3)
&=-\frac{H_{\rm sr}}{16M_{\rm Pl}^4\epsilon_{\rm sr}^2}\frac{k_t^3}{k_1^3k_2^3k_3^3}\\*
\nonumber
&\quad
\times{\rm Im}\left[\int_0^{x_\ast} dx\,\frac{\dot{\kappa}}{H_{\rm sr}}
(1+ix_3)\left(\frac{x_1^2x_2^2}{x^4}+\frac{\bf k_1\cdot k_2}{k_t^2}\frac{(1+ix_1)(1+ix_2)}{x^2}\right)e^{-ix}
\right]\\*
\label{B_dH_x}
&\quad
+(\text{$2$ permutations})\,,
\end{align}
where $k_t=k_1+k_2+k_3$, $x_i=-k_i\tau$, $x=-k_t\tau$, and $x_\ast=k_t/k_\ast$.
The corresponding shape function $S_{\delta H}$ is also given by
\begin{align}
\nonumber
S_{\delta H}(k_1,k_2,k_3)
&=\frac{1}{4}\frac{k_t^3}{k_1k_2k_3}\,{\rm Im}\left[\int_0^{x_\ast} dx\,\frac{\dot{\kappa}}{H_{\rm sr}}
(1+ix_3)\left(\frac{x_1^2x_2^2}{x^4}+\frac{\bf k_1\cdot k_2}{k_t^2}\frac{(1+ix_1)(1+ix_2)}{x^2}\right)e^{-ix}
\right]\\
&\quad
+(\text{$2$ permutations})\,.
\end{align}
After some algebraic calculations,
the following expression of $S_{\delta H}$ can be obtained:
\begin{align}
\nonumber
S_{\delta H}(k_1,k_2,k_3)
&=\frac{1}{4}\frac{k_t^3}{k_1k_2k_3}\,{\rm Im}\left[\int_0^{x_\ast} dx\,\frac{\dot{\kappa}}{H_{\rm sr}}
\bigg(\frac{i}{2}\alpha_1\alpha_2\alpha_3x
+\Big(\frac{1}{2}\sum_{i>j}\alpha_i\alpha_j-2\alpha_1\alpha_2\alpha_3\Big)\right.\\*
\nonumber
&\qquad\qquad\qquad\qquad\qquad\qquad\qquad
\left.
-\frac{i}{2}\sum_i\alpha_i^2x^{-1}-\frac{1}{2}\sum_i\alpha_i^2x^{-2}\bigg)e^{-ix}
\right]\\
&=\frac{1}{8}\,{\rm Im}\left[\int_0^{x_\ast} dx\,\frac{\dot{\kappa}}{H_{\rm sr}}
\bigg(ix
+\Big(\sum_{i>j}\frac{\alpha_i\alpha_j}{\alpha_1\alpha_2\alpha_3}-4\Big)
-\frac{\sum_i\alpha_i^2}{\alpha_1\alpha_2\alpha_3}
(ix^{-1}+x^{-2})
\bigg)e^{-ix}\right]\,,
\end{align}
where $\alpha_i=k_i/k_t$.
This integral
reduces to the following integral $\widetilde{I}(n,\mu,x_\ast)$ defined by
\begin{align}
\widetilde{I}(n,\mu,x_\ast)=
\int_0^{x_\ast}dx\frac{\dot{\kappa}}{H_{\rm sr}}x^ne^{-ix}\,,
\end{align}
whose analytic expression can be obtained
using the function $\mathcal{I}(\delta,n,x)$
in (\ref{calI_analytic}) as
\begin{align}
\widetilde{I}(n,\mu,x_\ast)=\alpha^2
\left[
\Big(\frac{3}{2}-\frac{27}{8\mu^2}\Big)\mathcal{I}(0,n,x_\ast)
+\frac{(3+2i\mu)^3}{16\mu^2}\mathcal{I}(2i\mu,n,x_\ast)
+\frac{(3-2i\mu)^3}{16\mu^2}\mathcal{I}(-2i\mu,n,x_\ast)
\right]\,.
\end{align}
In terms of $\widetilde{I}(n;\mu;x_\ast)$,
the shape function $S_{\delta H}(k_1,k_2,k_3)$
is then given by
\begin{align}
\nonumber
S_{\delta H}(k_1,k_2,k_3)
&=\frac{1}{8}\,{\rm Im}\left[i\,
\widetilde{I}(1,\mu,x_\ast)
+\Big(\sum_{i>j}\frac{\alpha_i\alpha_j}{\alpha_1\alpha_2\alpha_3}-4\Big)
\widetilde{I}(0,\mu,x_\ast)\right.\\*
\label{S_dH_tildeI}
&\qquad\qquad\qquad\qquad\quad
\left.
-\sum_i\frac{\alpha_i^2}{\alpha_1\alpha_2\alpha_3}\,
\Big(i\,\widetilde{I}(-1,\mu,x_\ast)
+\widetilde{I}(-2,\mu,x_\ast)\Big)
\right]\,.
\end{align}
From this expression,
we first notice that
there are no contributions of $\mathcal{O}(\mathcal{P}_\zeta^{-1/2})$ and higher order
so that large non-Gaussianities
do not arise from the Hubble deformation effects
unless $\widetilde{\mathcal{I}}$ is large.
Also notice that
the shape function depends
on $\alpha_i$ only through the coefficients
in front of $\widetilde{\mathcal{I}}$.
In the rest of this subsection,
we discuss the scale-dependence and shape
of $S_{\delta H}$.

\subsubsection{Scale-dependence of equilateral type non-Gaussianities}
\label{subsubsec:SdH_eq}
\begin{figure}[t]
\begin{center}
\includegraphics[width=160mm, bb=0 0 580 214]{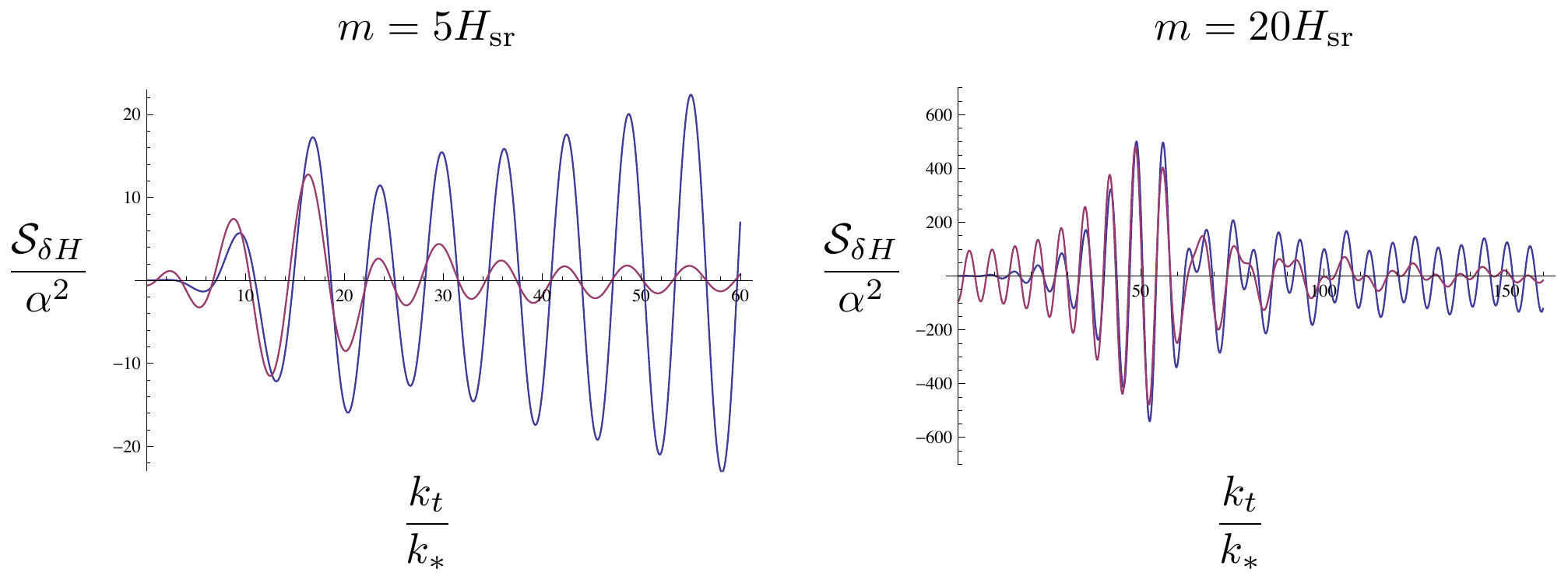}
 \end{center}
\vspace{-5mm}
\caption{Scale-dependence of $S_{\delta H}$ for the equilateral momentum configurations
and singular behaviors deep inside the horizon.
The exact results (\ref{S_dH_tildeI}) for $S_{\delta H}$
(blue curves)
oscillate and linearly diverge
in the region $k_t\gg k_\ast$,
where the mode $k$ is deep inside the horizon.
Based on the assumption that
the Bunch-Davies vacuum is realized deep inside the horizon,
we eliminate this kind of singular behaviors
via the prescription discussed in Appendix~\ref{app:mathcalI}:
We interpolate the original function (blue curves) and
the regularized function obtained by subtracting singular behaviors (red curves)
at the resonance scale $k\simeq\mu k_t$.}
\label{fig:SdH_eq_non}
\end{figure}
\begin{figure}[t]
\begin{center}
\includegraphics[width=160mm, bb=0 0 580 214]{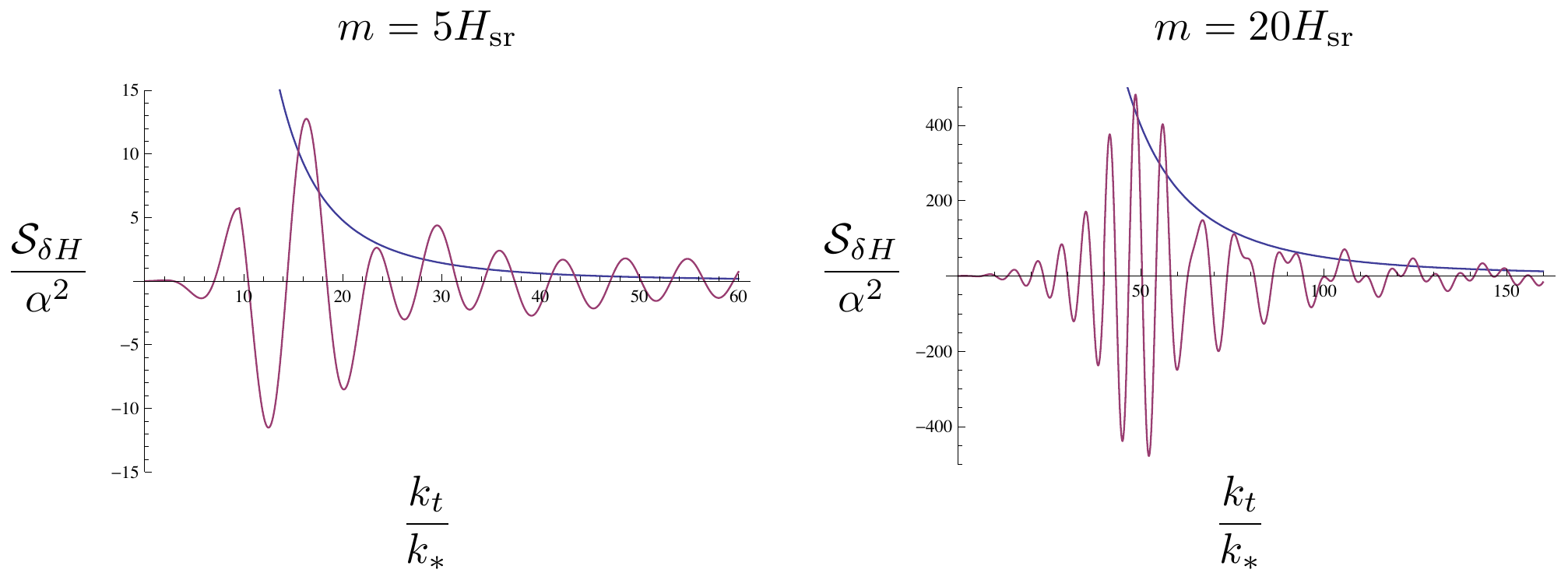}
 \end{center}
\vspace{-5mm}
\caption{Scale-dependence of $S_{\delta H}$ for the equilateral momentum configurations.
The red curve is the exact expression
for $\mathcal{S}_{\delta H}$ with singular behaviors
deep inside the horizon being subtracted.
The blue curve is the estimation based on
the stationary phase approximation:
$S_{\delta H}=
2\sqrt{\pi}\alpha^2 \mu^{5/2}\big(k/(\mu k_\ast)\big)^{-3}$.}
\label{fig:SdH_eq}
\end{figure}

Let us first discuss the scale-dependence of
the shape function $S_{\delta H}$.
As is clear from~(\ref{S_dH_tildeI}),
the scale-dependence of $S_{\delta H}$
is determined by
that of the function $\widetilde{I}$
once the shape of momentum configurations is specified.
For the equilateral momentum configurations $\alpha_i=1/3$,
for example, we have
\begin{align}
\label{SdH_equilateral}
S_{\delta H}(k_t/3,k_t/3,k_t/3)
&=\frac{1}{8}\,{\rm Im}\left[i\,
\widetilde{I}(1,\mu,x_\ast)
+5\,\widetilde{I}(0,\mu,x_\ast)
-9i\,\widetilde{I}(-1,\mu,x_\ast)
-9\,\widetilde{I}(-2,\mu,x_\ast)\Big)
\right]\,.
\end{align}
As we discuss in Appendix~\ref{app:mathcalI},
the integral $\mathcal{I}(\delta,n,x)$ for $n=1,0$,
and therefore, $\widetilde{I}(\delta,n,x)$ and $S_{\delta H}$
suffer from singular behaviors in the region $x\gg1$,
where the mode is deep inside the horizon.
This kind of singular behaviors
deep inside the horizon
are common in the interacting theory
and they are usually eliminated
by assuming that the Bunch-Davies vacuum is realized
deep inside the horizon.
In the bispectrum calculation of this paper,
we subtract these singular behaviors based on a similar assumption (see Appendix~\ref{app:mathcalI}).
The scale-dependence of $S_{\delta H}$ before and after the subtraction is displayed in Fig.~\ref{fig:SdH_eq_non},
where resonance-like effects can be observed
at the scale $k\gtrsim 2\mu k_\ast$.
As discussed in~\cite{Saito:2013aqa},
these resonance effects
can be understood by noticing that
$S_{\delta H}$ contains the same integrals $\mathcal{I}$
as the Hubble deformation effects $\mathcal{C}_{\delta H}$ on power spectra.
Using the stationary phase approximation
as in Sec.~\ref{subsubsec:Mathieu},
(\ref{SdH_equilateral}) can be estimated
in the heavy mass region as
\begin{align}
\big|S_{\delta H}(k_t/3,k_t/3,k_t/3)\big|
&\simeq \left|\frac{\mu}{16}\alpha^2\,{\rm Im}\left[
\mathcal{I}(2i\mu,1,x_\ast)\right]\right|
\sim
\left\{\begin{array}{ccl}
\displaystyle
2\sqrt{\pi}\alpha^2 \mu^{5/2}\left(\frac{k}{\mu k_\ast}\right)^{-3}
&{\rm for}&x_\ast\gtrsim2\mu\,, \\
0&{\rm for}&x_\ast\lesssim2\mu\,.
\end{array}\right.
\end{align}
As displayed in Fig.~\ref{fig:SdH_eq},
the above estimation well agrees
with the exact results.
To summarize,
resonances in $S_{\delta H}(k_t/3,k_t/3,k_t/3)$
appear at the scale $k_t\gtrsim 2\mu k_\ast$
and the size is of the order $\alpha^2\mu^{5/2}$
in the heavy mass approximation.

\subsubsection{Shape of $S_{\delta H}$}
\begin{figure}[t]
\begin{center}
\includegraphics[width=160mm, bb=0 0 580 227]{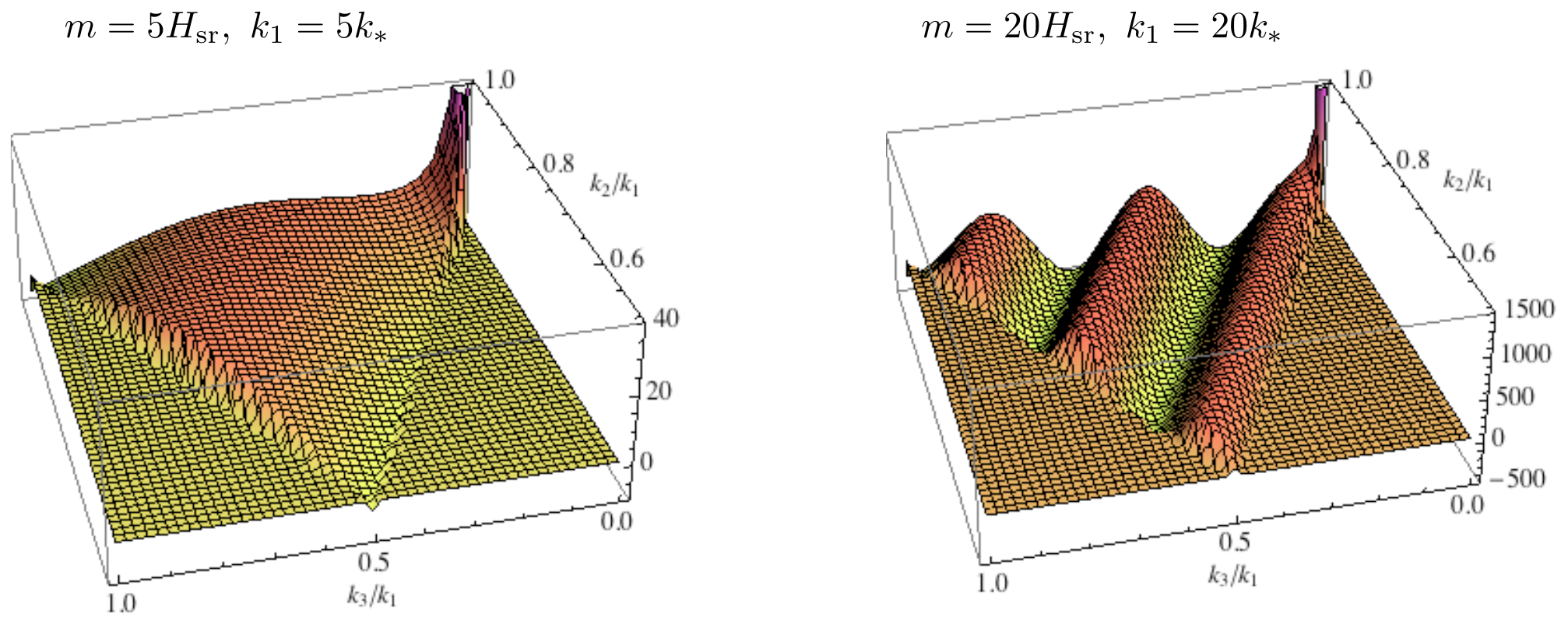}
 \end{center}
\vspace{-5mm}
\caption{Shape of $S_{\delta H}$
around the resonance scale
with fixed maximum momentum.
In the left figure,
$-S_{\delta H}/\alpha^2$ with $m=5H_{\rm sr}$ and $k_1=5k_\ast$
is plotted as a function of $k_2/k_1$ and $k_3/k_1$.
In the right figure,
$S_{\delta H}/\alpha^2$ with $m=20H_{\rm sr}$ and $k_1=20k_\ast$
is plotted as a function of $k_2/k_1$ and $k_3/k_1$.
In this type of conventional $3$D plots,
$S_{\delta H}$ takes a wavy form
and has a peak
at the squeezed momentum configuration
$k_1=k_2\gg k_3$.}
\label{fig:SdH_kmax}
\end{figure}
We next discuss the shape
of $S_{\delta H}(k_1,k_2,k_3)$.
Although shape functions are scale-invariant $S(k_1,k_2,k_3)=S(\lambda k_1,\lambda  k_2,\lambda  k_3)$
in inflationary models with scale-invariance,
those in our models
are not scale-invariant
because
the sudden turning potentials
and associated background oscillations
break the scale-invariance.
We therefore
need to specify both of the shape of momentum configurations
and the maximum momentum
in order to draw the conventional $3$D plot
of bispectra.
In Fig.~\ref{fig:SdH_kmax},
we plotted $S_{\delta H}$ around the resonance scale
as a function of $k_2/k_1$ and $k_3/k_1$,
where $k_1>k_2>k_3$
and the maximum momentum $k_1$ is fixed.
In this standard $3$D plot,
the shape function takes a wavy form
and it is almost flat along the direction $k_2+k_3={\rm constant}$~\cite{Saito:2013aqa}.
We can also find a peak at the squeezed configuration
$k_1=k_2\gg k_3$.
These features can be understood using
the analytic expression~(\ref{S_dH_tildeI})
as follows:
First,
just as discussions in Sec.~\ref{subsubsec:SdH_eq},
the integral $\widetilde{I}$ can be estimated
via the stationary phase approximation
as
\begin{align}
\widetilde{I}(n,\mu,x)\simeq-\frac{i}{2}\alpha^2\mu\,\mathcal{I}(2i\mu,n,x)
\sim 2^{3+n}\sqrt{\pi}\,\alpha^2\left(\frac{\mu}{x}\right)^{3}
\mu^{n+\frac{3}{2}}\,.
\end{align}
Then,
the contribution from $\widetilde{I}(1,\mu,x)$
dominates in~(\ref{S_dH_tildeI})
unless the prefactors $\displaystyle\sum_{i>j}\frac{\alpha_i\alpha_j}{\alpha_1\alpha_2\alpha_3}$
and/or $\displaystyle\sum_{i}\frac{\alpha_i^2}{\alpha_1\alpha_2\alpha_3}$
are large.
In other words,
$\widetilde{I}(1,\mu,x)$
dominates unless we take the squeezed limit:
\begin{align}
\label{SdH_non_squeezed}
S_{\delta H}(k_1,k_2,k_3)
\simeq
\frac{1}{8}{\rm Im}\left[i\,\widetilde{I}\,(1,\mu,k_t/k_\ast)\right]
\quad
{\rm for}
\quad
\alpha_3\gtrsim\frac{1}{\mu}\,,
\end{align}
which depends only on the total momentum
$k_t=k_1+k_2+k_3$.
Because of this property,
$S_{\delta H}$ is almost flat
along the direction $k_2+k_3={\rm constant}$.
In the squeezed limit $k_1=k_2\gg k_3$,
on the other hand,
the prefactors
$\displaystyle\sum_{i>j}\frac{\alpha_i\alpha_j}{\alpha_1\alpha_2\alpha_3}$
and $\displaystyle\sum_{i}\frac{\alpha_i^2}{\alpha_1\alpha_2\alpha_3}$ diverge
so that a peak appears
at the squeezed point.
Also notice that
the peak becomes sharp for the large mass
because $\displaystyle\frac{\widetilde{I}\,(1,\mu,k_t/k_\ast)}{\widetilde{I}\,(0,\mu,k_t/k_\ast)}\sim \mu$
and therefore the region in which $\widetilde{I}\,(1,\mu,k_t/k_\ast)$
dominates
becomes wider for larger $\mu$.
\begin{figure}[t]
\begin{center}
\includegraphics[width=135mm, bb=0 0 580 237]{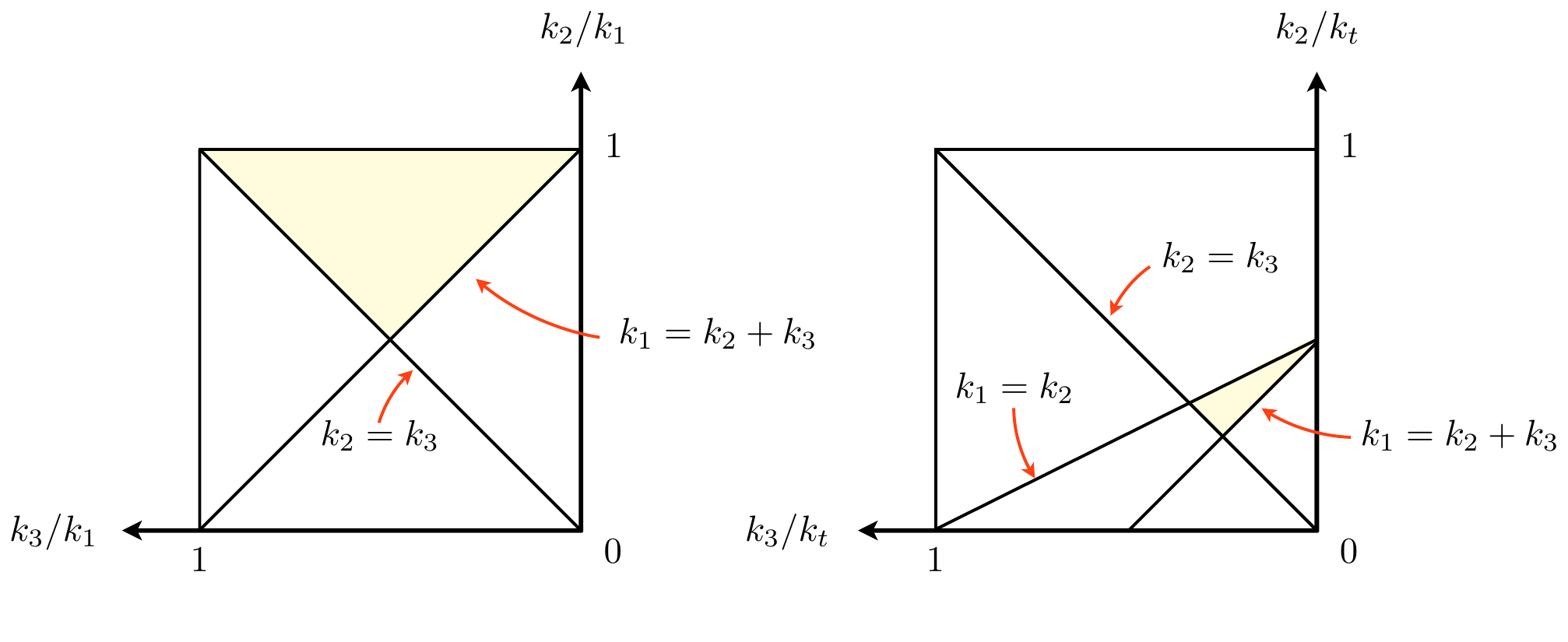}
\end{center}
\vspace{-10mm}
 \caption{In the conventional $3$D plots (the left figure),
shape functions are plotted
in the region satisfying
$1>k_2/k_1>k_3/k_1$ and $1>k_2/k_1+k_3/k_1$.
On the other hand, in the $3$D plots
with fixed total momenta
(the right figure),
they are plotted
in the region satisfying
$k_1/k_t>k_2/k_t>k_3/k_t$ and $k_1/k_t>k_2/k_t+k_3/k_t$,
which can be rephrased in terms of $\alpha_2$ and $\alpha_3$
as~(\ref{3Dplot_conditions_alpha}).
}
\label{fig:conv_vs_ktotal_1}
\begin{center}
\includegraphics[width=150mm, bb=0 0 580 222]{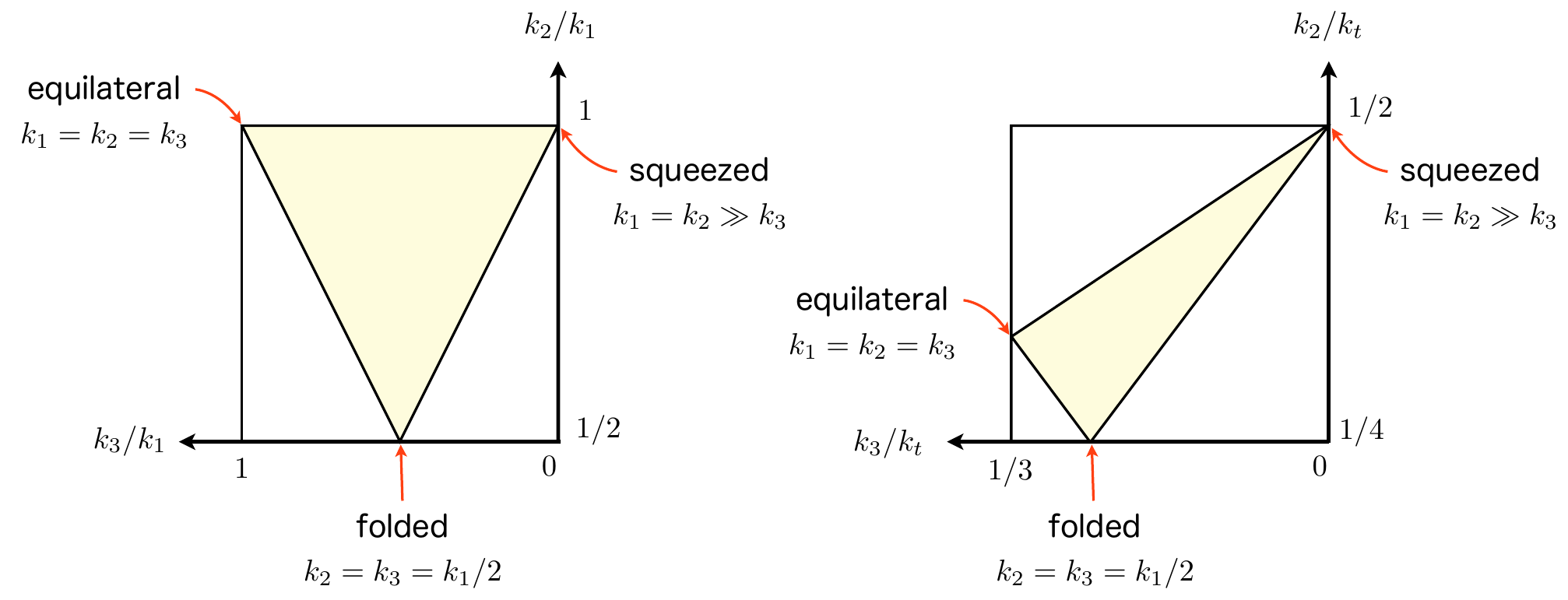}
 \end{center}
\vspace{-5mm}
\caption{
Closeup of Fig.~\ref{fig:conv_vs_ktotal_1}
with squeezed/equilateral/folded configurations depicted.}
\label{fig:conv_vs_ktotal_2}
\end{figure}

\medskip
In order to make the above properties manifest,
it may be convenient to introduce a different type
of $3$D plots, where the total momentum $k_t$
is fixed instead of the maximum momentum $k_1$
and the shape function is plotted as a function
of $\alpha_2=k_2/k_t$ and $\alpha_3=k_3/k_t$.
As in the case of the conventional $3$D plot,
we can take $k_i$ such that $k_1>k_2>k_3$ without loss of generality
and the momentum conservation implies
$k_1<k_2+k_3$.
In terms of $\alpha_2$ and $\alpha_3$,
these conditions can be rephrased as
\begin{align}
\nonumber
&k_1>k_2
\quad\leftrightarrow\quad
\alpha_1>\alpha_2
\quad\leftrightarrow\quad
1>2\alpha_2+\alpha_3\,,\\
\nonumber
&k_2>k_3
\quad\leftrightarrow\quad
\alpha_2>\alpha_3\,,\\
\label{3Dplot_conditions_alpha}
&k_1<k_2+k_3
\quad\leftrightarrow\quad
\alpha_1<\alpha_2+\alpha_3
\quad\leftrightarrow\quad
\frac{1}{2}<\alpha_2+\alpha_3\,,
\end{align}
where we used $\alpha_1+\alpha_2+\alpha_3=1$.
The region satisfying the above three conditions
is depicted in Fig.~\ref{fig:conv_vs_ktotal_1}
and a close up of this region
is given in Fig.~\ref{fig:conv_vs_ktotal_2},
where
the points corresponding
to the squeezed, equilateral, and folded
momentum configurations are also described.
\begin{figure}[t]
\begin{center}
\includegraphics[width=160mm, bb=0 0 580 230]{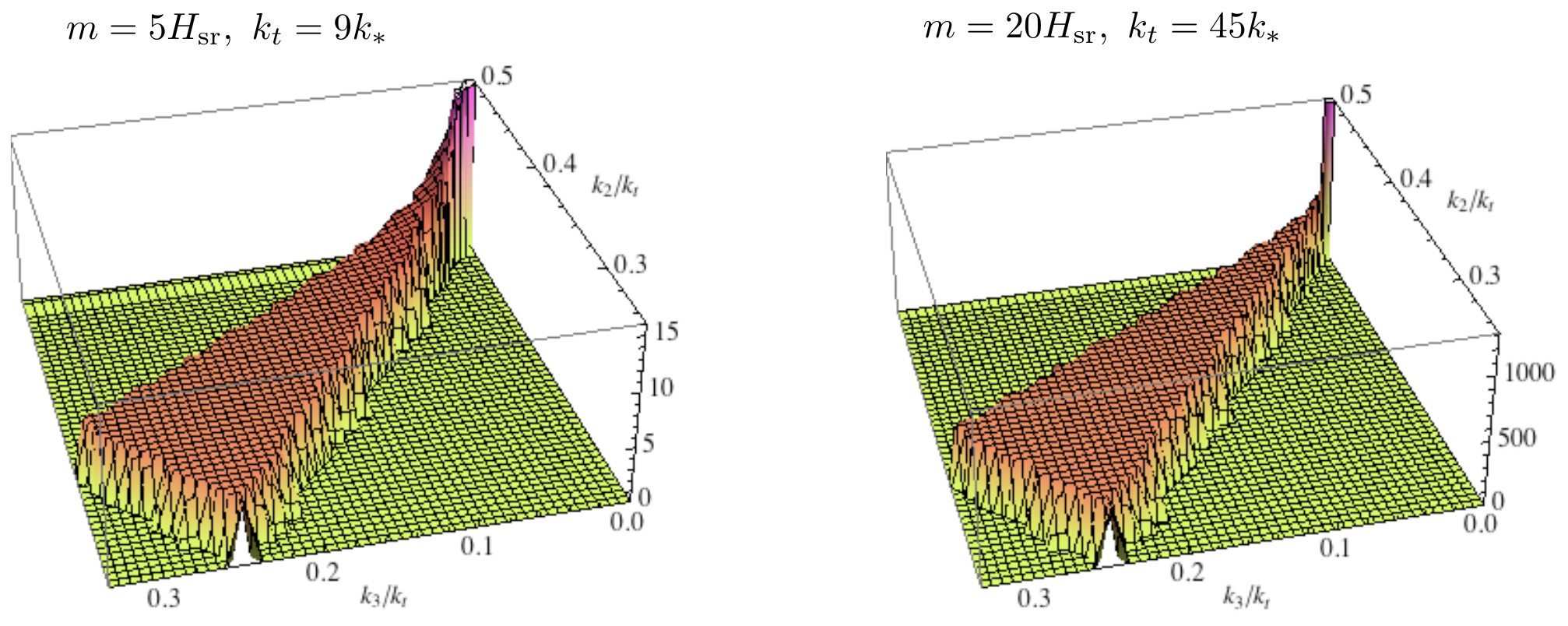}
 \end{center}
\vspace{-5mm}
\caption{
Shape of $S_{\delta H}$
around the resonance scale
with fixed total momenta.
In the left and right figures,
$S_{\delta H}/\alpha^2$ with $m=5H_{\rm sr}$ and
$-S_{\delta H}/\alpha^2$ with $m=20H_{\rm sr}$ are plotted.
The total momenta are taken as
$k_t=9k_\ast$ and $k_t=45k_\ast$, respectively.
In this type of $3$D plots,
it is clear
that
the region where $\widetilde{I}\,(1,\mu,k_t/k_\ast)$ dominates becomes wide
and the peak at the squeezed configurations
become sharp
for large~$\mu$.}
\label{fig:SdH_ktotal}
\end{figure}
As described in Fig.~\ref{fig:SdH_ktotal},
in this type of $3$D plots,
the shape function $S_{\delta H}$
takes a flat form
in the region where $\widetilde{I}\,(1,\mu,k_t/k_\ast)$ dominates
and
it is clear that the peak at the squeezed configuration
becomes sharp for large mass.
We will use both types of the $3$D plots in the rest of this section.

\subsection{Conversion effects}
We next consider the conversion effects. 
After discussing properties of the deformed
propagator $\mathcal{F}_i(t)$,
shape functions are evaluated.

\subsubsection{Deformed propagators}
\label{subsubsec:deformed}
Let us first introduce
an analytic expression for the deformed propagator $\mathcal{F}_i(t)$:
\begin{align}
\nonumber
\mathcal{F}_i(t)=
\frac{1}{2M_{\rm Pl}\epsilon_{\rm sr}^{1/2}k_i^{3/2}}
\frac{i\pi e^{-\pi\mu}}{2}
&\left[
x_i^{3/2}H_{-i\mu}^{(2)}(x_i)
\int_0^{x_i}dx_i^\prime\frac{\beta_1}{H_{\rm sr}}
x_i^\prime\,^{-1/2}e^{-ix_i^\prime}H_{i\mu}^{(1)}(x_i^\prime)\right.\\*
\nonumber
&\quad
+x_i^{3/2}H_{i\mu}^{(1)}(x_i)
\int_{x_i}^{x_{i\ast}}dx_i^\prime\frac{\beta_1}{H_{\rm sr}}
x_i^\prime\,^{-1/2}e^{-ix_i^\prime}H_{-i\mu}^{(2)}(x_i^\prime)\\*
&\quad
\left.
-x_i^{3/2}H_{-i\mu}^{(2)}(x_i)
\int_0^{x_{i\ast}}dx_i^\prime\frac{\beta_1}{H_{\rm sr}}
x_i^\prime\,^{-1/2}e^{+ix_i^\prime}H_{i\mu}^{(1)}(x_i^\prime)
\right]\,,
\end{align}
where $x_{i\ast}=\alpha_ix_\ast$.
Using the relation
\begin{align}
\frac{\beta_1}{H_{\rm sr}}&=\theta(t-t_{\ast})\dot{\bar{\phi}}_{\rm sr}\frac{\dot{\gamma}}{H_{\rm sr}}
=\theta(t-t_{\ast})\sqrt{2}\epsilon_{\rm sr}^{1/2}M_{\rm Pl}H_{\rm sr}
\frac{\dot{\gamma}}{H_{\rm sr}}\,,
\end{align}
it is rewritten as
\begin{align}
\nonumber
\mathcal{F}_i(t)=
\frac{H_{\rm sr}}{\sqrt{2}k_i^{3/2}}
\frac{i\pi e^{-\pi\mu}}{2}
&\left[
x_i^{3/2}H_{-i\mu}^{(2)}(x_i)
\int_0^{x_i}dx_i^\prime\frac{\dot{\gamma}}{H_{\rm sr}}
x_i^\prime\,^{-1/2}e^{-ix_i^\prime}H_{i\mu}^{(1)}(x_i^\prime)\right.\\*
\nonumber
&\quad
+x_i^{3/2}H_{i\mu}^{(1)}(x_i)
\int_{x_i}^{x_{i\ast}}dx_i^\prime\frac{\dot{\gamma}}{H_{\rm sr}}
x_i^\prime\,^{-1/2}e^{-ix_i^\prime}H_{-i\mu}^{(2)}(x_i^\prime)\\*
&\quad
\left.
-x_i^{3/2}H_{-i\mu}^{(2)}(x_i)
\int_0^{x_{i\ast}}dx_i^\prime\frac{\dot{\gamma}}{H_{\rm sr}}
x_i^\prime\,^{-1/2}e^{+ix_i^\prime}H_{i\mu}^{(1)}(x_i^\prime)
\right]\,.
\end{align}
It is useful to introduce
the following indefinite integral $\mathcal{D}_-(\ell,\mu,x)$
analogously to
$\mathcal{D}_+(\ell,\mu,x)$ in (\ref{Dpanalytic}):
\begin{align}
\mathcal{D}_-(\ell,\mu,x)
&=\int dx\,x^{-\frac{1}{2}+\ell}e^{-ix}H^{(1)}_{i\mu}(x)\,,
\end{align}
whose analytic expression is given by
\begin{align}
\nonumber
\mathcal{D}_-(\ell,\mu,x)
&=
\frac{2^{i\mu} x^{\frac{1}{2}+\ell-i\mu}\Gamma(i\mu)}{i\pi(\frac{1}{2}+\ell-i\mu)}
{}_2F_2\Big(\frac{1}{2}-i\mu,\frac{1}{2}+\ell-i\mu;\frac{3}{2}+\ell-i\mu,1-2i\mu; -2ix\Big)\\*
\label{Dmanalytic}
&\quad+e^{\pi\mu}\frac{2^{-i\mu} x^{\frac{1}{2}+\ell+i\mu}\Gamma(-i\mu)}{i\pi(\frac{1}{2}+\ell+i\mu)}
{}_2F_2\Big(\frac{1}{2}+i\mu,\frac{1}{2}+\ell+i\mu;\frac{3}{2}+\ell+i\mu,1+2i\mu; -2ix\Big)\,.
\end{align}
Then,
the deformed propagator $\mathcal{F}_i$
can be written
in terms of $\mathcal{D}_\pm(\ell,\mu,x)$
as
\begin{align}
\mathcal{F}_i(t)&=\frac{H_{\rm sr}}{\sqrt{2}k_i^{3/2}}
\widetilde{\mathcal{F}}_i(x_i)\,,\\
\nonumber
\widetilde{\mathcal{F}}_i(x_i)&=-\alpha\frac{\pi e^{-\pi\mu}}{2}
\left[
x_i^{3/2}H_{-i\mu}^{(2)}(x_i)
\sum_\pm
(-1)^\pm \frac{(\frac{3}{2}\pm i\mu)^2}{2\mu}x_{i\ast}^{-(\frac{3}{2}\pm i\mu)}
\mathcal{D}_-\Big(\frac{3}{2}\pm i\mu,\mu,x_i\Big)\right.\\
\nonumber
&\quad
+x_i^{3/2}H_{i\mu}^{(1)}(x_i)
\sum_\pm
(-1)^\pm \frac{(\frac{3}{2}\pm i\mu)^2}{2\mu}x_{i\ast}^{-(\frac{3}{2}\pm i\mu)}
\bigg(\mathcal{D}_+\Big(\frac{3}{2}\mp i\mu,\mu,x_{i\ast}\Big)-\mathcal{D}_+\Big(\frac{3}{2}\mp i\mu,\mu,x_{i}\Big)\bigg)^\ast\\
\label{tilde_mathcal_Fi}
&\quad
\left.
-x_i^{3/2}H_{-i\mu}^{(2)}(x_i)
\sum_\pm
(-1)^\pm \frac{(\frac{3}{2}\pm i\mu)^2}{2\mu}x_{i\ast}^{-(\frac{3}{2}\pm i\mu)}
\mathcal{D}_+\Big(\frac{3}{2}\pm i\mu,\mu,x_{i\ast}\Big)
\right]\,,
\end{align}
which reduces the calculation of conversion effects
to one integral.
In order to clarify qualitative features
of the deformed propagators $\mathcal{F}_i(t)$
and $\widetilde{\mathcal{F}}_i(x_i)$,
let us discuss their behaviors
using the heavy mass approximation.
In particular,
we consider the following two regimes:
$\displaystyle\frac{k_i}{a(t)}\sim H_{\rm sr}$
and $\displaystyle\frac{k_i}{a(t)}\sim m\gg H_{\rm sr}$,
or in other words,
$x_i\sim1$
and $x_i\sim\mu\gg1$.
Here we assume
that $t>t_\ast$, or equally, $x_i<x_{i\ast}$.
\begin{enumerate}
\item
$\widetilde{\mathcal{F}}_i(x_i)$
in the regime $\displaystyle\frac{k_i}{a(t)}\sim H_{\rm sr}
\leftrightarrow x_i\sim1$.

In this region,
the mode function $v_{k_i}$ of
the massive isocurvature
can be expressed as
\begin{align}
v_{k_i}\simeq \frac{H_{\rm sr}}{\sqrt{2\mu k_\ast^3}}
\Big(\frac{x_i}{x_{i\ast}}\Big)^{\frac{3}{2}+i\mu}
\end{align}
up to a phase factor irrelevant to our discussions.
Using this expression,
$\widetilde{\mathcal{F}}_i$ can be written~as
\begin{align}
\nonumber
\widetilde{\mathcal{F}}_i(x_i)
&\simeq\frac{i}{\mu}\Bigg[
x_i^{3/2}
\Big(\frac{x_i}{x_{i\ast}}\Big)^{-i\mu}
\int_0^{x_i}dx_i^\prime
\frac{\dot{\gamma}}{H_{\rm sr}}
x_i^{\prime \,-1/2}e^{-ix_i^\prime}
\Big(\frac{x_i}{x_{i\ast}}\Big)^{i\mu}\\*
\nonumber
&\qquad\quad
+x_i^{3/2}
\Big(\frac{x_i}{x_{i\ast}}\Big)^{i\mu}
\int_{x_i}^{x_{i\ast}}dx_i^\prime
\frac{\dot{\gamma}}{H_{\rm sr}}
x_i^{\prime \,-1/2}e^{-ix_i^\prime}
\Big(\frac{x_i}{x_{i\ast}}\Big)^{-i\mu}\\*
&\qquad\quad
-x_i^{3/2}
\Big(\frac{x_i}{x_{i\ast}}\Big)^{-i\mu}
\int_0^{x_{i\ast}}dx_i^\prime
\frac{\dot{\gamma}}{H_{\rm sr}}
x_i^{\prime \,-1/2}e^{ix_i^\prime}
\Big(\frac{x_i}{x_{i\ast}}\Big)^{i\mu}\Bigg]\,.
\end{align}
Just as in the case of power spectra,
$\displaystyle\frac{\dot{\gamma}}{H_{\rm sr}}\Big(\frac{x_i}{x_{i\ast}}\Big)^{\pm i\mu}$
contains a non-oscillating component,
\begin{align}
\frac{\dot{\gamma}}{H_{\rm sr}}\Big(\frac{x_i}{x_{i\ast}}\Big)^{\pm i\mu}
\simeq \pm\frac{i}{2}\alpha\mu
\left[\Big(\frac{x_i}{x_{i\ast}}\Big)^{\frac{3}{2}}-\Big(\frac{x_i}{x_{i\ast}}\Big)^{\frac{3}{2}\pm 2i\mu}\right]\,,
\end{align}
and therefore,
the leading contribution to
$\widetilde{\mathcal{F}}_i$
in the heavy mass approximation
is given~by
\begin{align}
\nonumber
\widetilde{\mathcal{F}}_i(x_i)
&\simeq\frac{\alpha}{2}
\Big(\frac{x_i}{x_{i\ast}}\Big)^{3/2}
\Bigg[
-\Big(\frac{x_i}{x_{i\ast}}\Big)^{-i\mu}
\int_0^{x_i}dx_i^\prime
\,x_i^{\prime}e^{-ix_i^\prime}\\
\nonumber
&\qquad\qquad\qquad\quad
+\Big(\frac{x_i}{x_{i\ast}}\Big)^{i\mu}
\int_{x_i}^{x_{i\ast}}dx_i^\prime
\,x_i^{\prime}e^{-ix_i^\prime}
+\Big(\frac{x_i}{x_{i\ast}}\Big)^{-i\mu}
\int_0^{x_{i\ast}}dx_i^\prime
\,x_i^{\prime}e^{ix_i^\prime}\Bigg]\\
\nonumber
&=\frac{\alpha}{2}
\Bigg[
\Big(\frac{x}{x_{\ast}}\Big)^{\frac{3}{2}+i\mu}
\Big[(1+i\alpha_ix_\ast)e^{-i\alpha_i x_\ast}
-(1+i\alpha_ix)e^{-i\alpha_i x}\Big]
\\
\label{deformed_app_1}
&\qquad\quad
-\Big(\frac{x}{x_{\ast}}\Big)^{\frac{3}{2}-i\mu}
\Big[(1-i\alpha_ix_\ast)e^{i\alpha_i x_\ast}
-(1+i\alpha_ix)e^{-i\alpha_i x}\Big]
\Bigg]\,.
\end{align}
Notice that
(\ref{deformed_app_1}) is of the order $\alpha$
and oscillates with a frequency $\sim\mu$.
Also note that
it is of the order $\alpha_i^2$
in the limit $\alpha_i=k_i/k_t\to0$.

\item
$\mathcal{F}_i(t)$ in the regime
$\displaystyle\frac{k}{a(t)}\sim m\gg 1\leftrightarrow\mu\sim x_i\gg 1$.

As we introduced at the beginning of this section,
the deformed propagator $\mathcal{F}_i(t)$
is given~by
\begin{align}
\nonumber
\mathcal{F}_i(t)&=
-iv_{k_i}^\ast(t)
\int_t^\infty dt^\prime a^3 2\beta_1\dot{u}_{k_i}^\ast v_{k_i}(t^\prime)
-iv_{k_i}(t)
\int_{t_\ast}^t dt^\prime a^32\beta_1\dot{u}_{k_i}^\ast v_{k_i}^\ast(t^\prime)\\
&\quad
+iv_{k_i}^\ast(t)
\int_{t_\ast}^\infty dt^\prime a^32\beta_1\dot{u}_{k_i} v_{k_i}(t^\prime)\,,
\end{align}
and the phase factor of each integrand
takes the form
\begin{align}
\sim \exp\left[ i\Big(\pm m\pm E\pm\frac{k}{a}\Big)t\right]=\exp \left[i\Big(\pm m\pm\sqrt{m^2+\frac{k^2}{a^2}}\pm\frac{k}{a}\Big)t\right]\,.
\end{align}
Here $\displaystyle E=\sqrt{m^2+\frac{k^2}{a^2}}$
and the phase factor always oscillates with a frequency of the order~$m$.
Therefore,
the integral
is suppressed by $1/m$
and $\mathcal{F}_i(t)$
can be roughly order estimated as
\begin{align}
\label{order_estimate_Fi}
\mathcal{F}_i(t)\sim
\frac{1}{m}a^3\beta_1|v_{k_i}|^2\dot{u}_{k_i}^\ast\sim \frac{1}{m^2}\beta_1\dot{u}_{k_i}^\ast\,.
\end{align}
\end{enumerate}

\newpage
\subsubsection{Conversion effect $1$}
\label{subsubsec:conv1}
\begin{figure}[t]
\begin{center}
\includegraphics[width=160mm, bb=0 0 580 200]{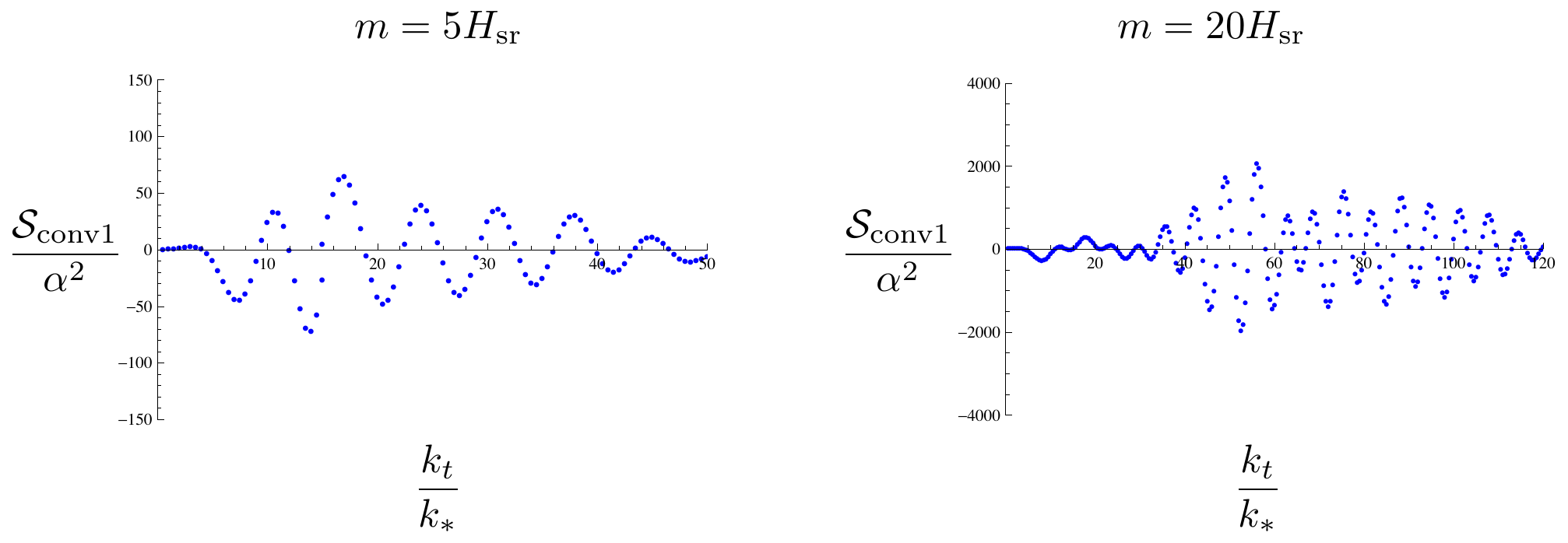}
 \end{center}
\vspace{-5mm}
\caption{Scale-dependence of $S_{\rm conv1}$ for the equilateral momentum configurations.
As in the case of the Hubble deformation effects $S_{\delta H}$,
$S_{\rm conv1}$ becomes singular deep inside the horizon.
Following the assumption that the Bunch-Davies vacuum is realized
deep inside the horizon,
we subtracted such singular behaviors
and the results after the subtraction are plotted in this figure,
where resonances can be observed at the scale $k\gtrsim2\mu$.}
\label{fig:Sconv1_eq}
\end{figure}
\begin{figure}[t]
\begin{center}
\includegraphics[width=140mm, bb=0 0 580 240]{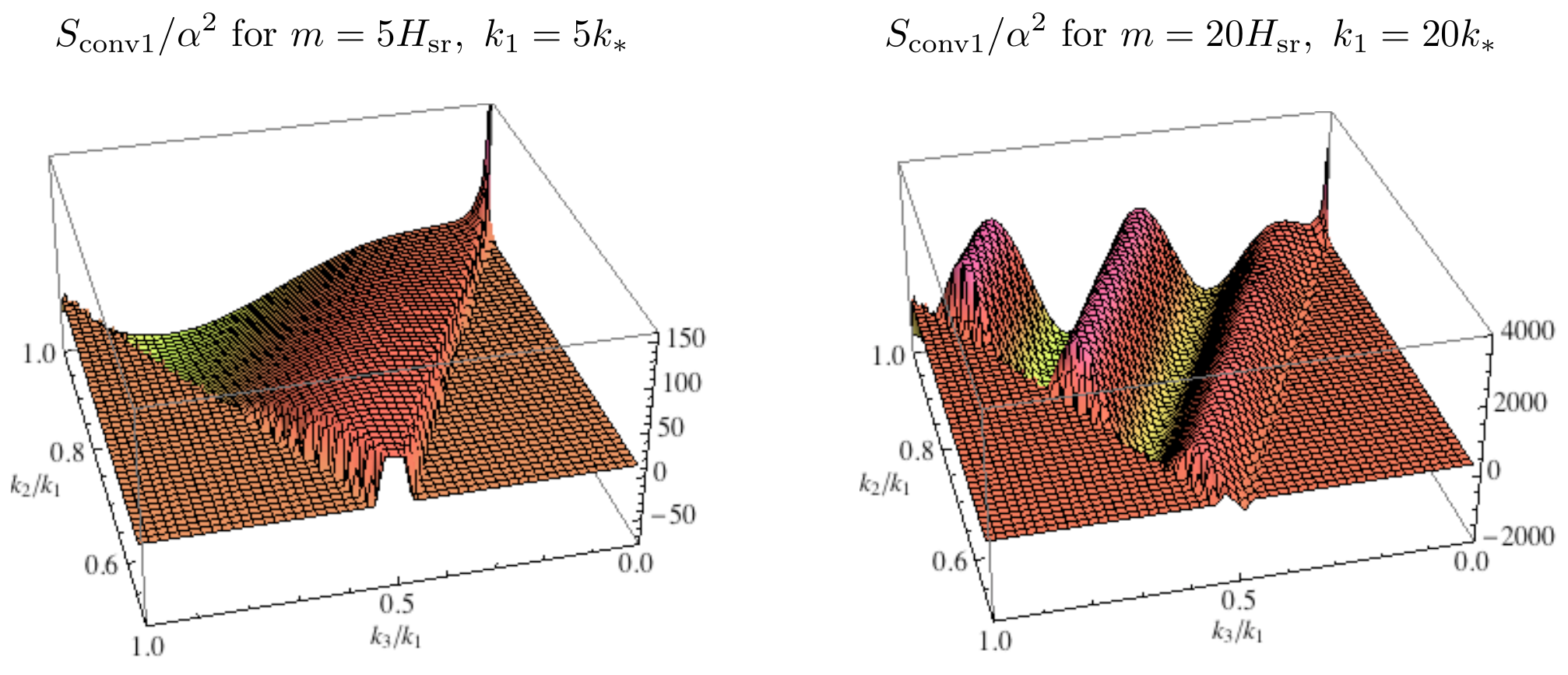}\\[2mm]
\includegraphics[width=140mm, bb=0 0 580 242]{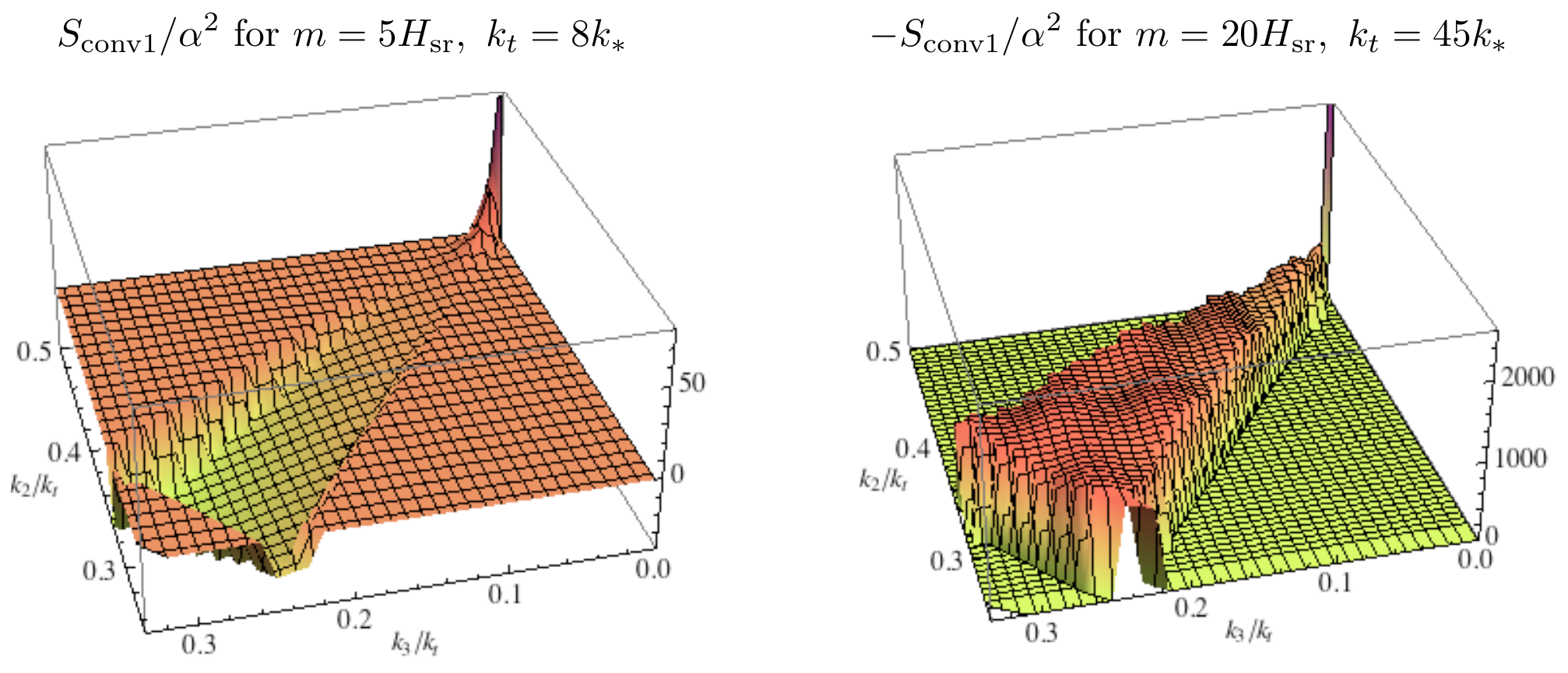}
 \end{center}
\vspace{-5mm}
\caption{Shape of $S_{\rm conv1}$
around the resonance scale.
Similarly to the Hubble deformation effects $S_{\delta H}$,
the shape function takes a wavy form
when plotted with the maximum momentum $k_1$
being fixed (upper figures)
and has a peak at the squeezed configuration.
In $3$D plots with fixed total momenta $k_t$ (lower figures),
the flat region becomes large and the peak at the squeezed configuration becomes sharp
when the mass is heavy.}
\label{fig:Sconv1_3D}
\end{figure}

Now we evaluate the shape function $\mathcal{S}_{\rm conv1}$
for the first type of conversion effects.
Using the expressions
(\ref{mode_u_expression}) and (\ref{mode_v_expression})
of the mode functions,
the three-point function $\mathcal{B}_{\rm conv1}$
and the corresponding shape function
$\mathcal{S}_{\rm conv1}$
can be written as
\begin{align}
\nonumber
\mathcal{B}_{\pi,{\rm conv}1}(k_1,k_2,k_3)
&=
\frac{H_{\rm sr}}{8M_{\rm Pl}^4\epsilon_{\rm sr}^2}
\frac{k_t^3}{k_1^3k_2^3k_3^3}\\*
\nonumber
&\quad\times{\rm Im}
\bigg[\int_0^{\infty} dx\,
\left\{
\theta(x_\ast-x)
\frac{\dot{\gamma}}{H_{\rm sr}}
\left(\frac{x_1^2x_2^2}{x^4}+\frac{\bf k_1\cdot k_2}{k_t^2}\frac{(1+ix_1)(1+ix_2)}{x^2}\right)\right.\\*
\nonumber
&\qquad\qquad\qquad
\left.+\Big(\theta(x_\ast-x)\frac{\ddot{\gamma}}{H_{\rm sr}^2}+\delta(x-x_\ast)x\frac{\dot{\gamma}}{H_{\rm sr}}\Big)\frac{x_1^2+x_2^2+ix_1x_2(x_1+x_2)}{x^4}
\right\}\\*
&\qquad\qquad\qquad\qquad\qquad
\times e^{-i(x_1+x_2)}
\widetilde{\mathcal{F}}_3(x_3)\bigg]+(\text{$2$ permutations})\,,\\
\nonumber
\mathcal{S}_{{\rm conv}1}(k_1,k_2,k_3)
&=
-\frac{1}{2}
\frac{k_t^3}{k_1k_2k_3}\\*
\nonumber
&\quad\times{\rm Im}
\bigg[\int_0^{\infty} dx\,
\left\{
\theta(x_\ast-x)
\frac{\dot{\gamma}}{H_{\rm sr}}
\left(\frac{x_1^2x_2^2}{x^4}+\frac{\bf k_1\cdot k_2}{k_t^2}\frac{(1+ix_1)(1+ix_2)}{x^2}\right)\right.\\*
\nonumber
&\qquad\qquad\qquad
\left.+\Big(\theta(x_\ast-x)\frac{\ddot{\gamma}}{H_{\rm sr}^2}+\delta(x-x_\ast)x\frac{\dot{\gamma}}{H_{\rm sr}}\Big)\frac{x_1^2+x_2^2+ix_1x_2(x_1+x_2)}{x^4}
\right\}\\*
&\qquad\qquad\qquad\qquad\qquad
\times e^{-i(x_1+x_2)}
\widetilde{\mathcal{F}}_3(x_3)\bigg]+(\text{$2$ permutations})\,.
\end{align}
From this expression,
we can see that
there are no contributions of $\mathcal{O}(\mathcal{P}_\zeta^{-1/2})$ and higher order
so that large non-Gaussianities
do not arise
unless the above integral is large.
Using the analytic expression~(\ref{tilde_mathcal_Fi}),
the calculation of $S_{\rm conv1}$
reduces to one integration over $x$.
We performed this integral numerically
and the scale-dependence of the shape function
$S_{\rm conv1}$ for the equilateral momentum configurations
are for example given in Fig.~\ref{fig:Sconv1_eq},
where we find resonances
around the scale $k_t\gtrsim 2\mu k_\ast$
similarly to the case of Hubble deformation effects $S_{\delta H}$.
Let us then order estimate these resonances
by using the heavy mass approximation
and comparing with the Hubble deformation
effects.
As an example,
we discuss the following contribution
to the bispectrum $\mathcal{B}_{\rm conv1}$:
\begin{align}
\mathcal{B}_{\rm conv1}&\ni\frac{1}{4M_{\rm Pl}^3\epsilon_{\rm sr}^{3/2}(k_1k_2k_3)^{3/2}}
{\rm Re}
\left[-i\int_{-\infty}^\infty dt\,a^3
2\beta_1\dot{u}_{k_1}^\ast \dot{u}_{k_2}^\ast 
\mathcal{F}_3(t)\right]\,.
\end{align}
Using~(\ref{order_estimate_Fi}),
it can be roughly estimated as
\begin{align}
\nonumber
&\sim
\frac{1}{2M_{\rm Pl}^3\epsilon_{\rm sr}^{3/2}(k_1k_2k_3)^{3/2}}
\int_{-\infty}^\infty dt\,a^3
\frac{\beta_1^2}{m^2}\dot{u}_{k_1}^\ast \dot{u}_{k_2}^\ast \dot{u}_{k_3}^\ast\\
\label{conv1_estimate}
&\sim\frac{1}{2M_{\rm Pl}^3\epsilon_{\rm sr}^{3/2}(k_1k_2k_3)^{3/2}}
\int_{t_\ast}^\infty dt\,a^3
\dot{\varphi}_2^2\,m^3u_{k_1}^\ast u_{k_2}^\ast u_{k_3}^\ast\,,
\end{align}
where we used
$\beta_1\sim\ddot{\varphi}_2\sim m\dot{\varphi}_2$
and $\dot{u}_{k_i}\sim (k_i/a)\,u_{k_i}\sim mu_{k_i}$.
It is not difficult to see that
the integral~(\ref{conv1_estimate})
has a similar structure
as the Hubble deformation effects $\mathcal{B}_{\pi,\delta H}$.
Let us consider the following contribution
for example:
\begin{align}
\label{BdH_estimate_for_conv}
\mathcal{B}_{\pi,\delta H}(k_1,k_2,k_3)&\ni
\frac{1}{2M_{\rm Pl}^3\epsilon_{\rm sr}^{3/2}(k_1k_2k_3)^{3/2}}{\rm Re}
\left[-i\int_{t_\ast}^\infty dt\,a^3
M_{\rm Pl}^2\dot{H}_{\rm sr}\dot{\kappa}\dot{u}_{k_1}^\ast\dot{u}_{k_2}^\ast u_{k_3}^\ast
\right]\,.
\end{align}
Picking up the oscillating component
$\displaystyle-\frac{\dot{\varphi}_2^2}{2M_{\rm Pl}^2\dot{H}_{\rm sr}}\in\kappa$
relevant to the resonance
and using
$\ddot{\varphi}_2\sim m\dot{\varphi}_2$
and $\dot{u}_{k_i}\sim mu_{k_i}$,
(\ref{BdH_estimate_for_conv})
can be estimated as
\begin{align}
\nonumber
&\sim\frac{1}{4M_{\rm Pl}^3\epsilon_{\rm sr}^{3/2}(k_1k_2k_3)^{3/2}}
\int_{t_\ast}^\infty dt\,a^3
(2\dot{\varphi}_2\ddot{\varphi}_2)\dot{u}_{k_1}^\ast\dot{u}_{k_2}^\ast u_{k_3}^\ast\\
\label{SdH_estimate_for_conv}
&\sim
\frac{1}{2M_{\rm Pl}^3\epsilon_{\rm sr}^{3/2}(k_1k_2k_3)^{3/2}}
\int_{t_\ast}^\infty dt\,a^3
\dot{\varphi}_2^2m^3u_{k_1}^\ast u_{k_2}^\ast u_{k_3}^\ast\,,
\end{align}
which takes a similar form
as~(\ref{conv1_estimate}).
We therefore conclude
that resonance effects appear
also in the first type of contribution $S_{\rm conv1}$
from the conversion interaction
and the size of resonances
is the same as those in the Hubble deformation effects:
the order $\alpha^2\mu^{5/2}$
in the heavy mass approximation.
As is expected from the above discussions,
the shape of non-Gaussianities
around the resonance scale $k_t\sim 2\mu k_\ast$
is also similar to that of the Hubble deformation effects $S_{\delta H}$ (see Fig.~\ref{fig:Sconv1_3D}).

\subsubsection{Conversion effect $2$}
\begin{figure}[t]
\begin{center}
\includegraphics[width=160mm, bb=0 0 580 176]{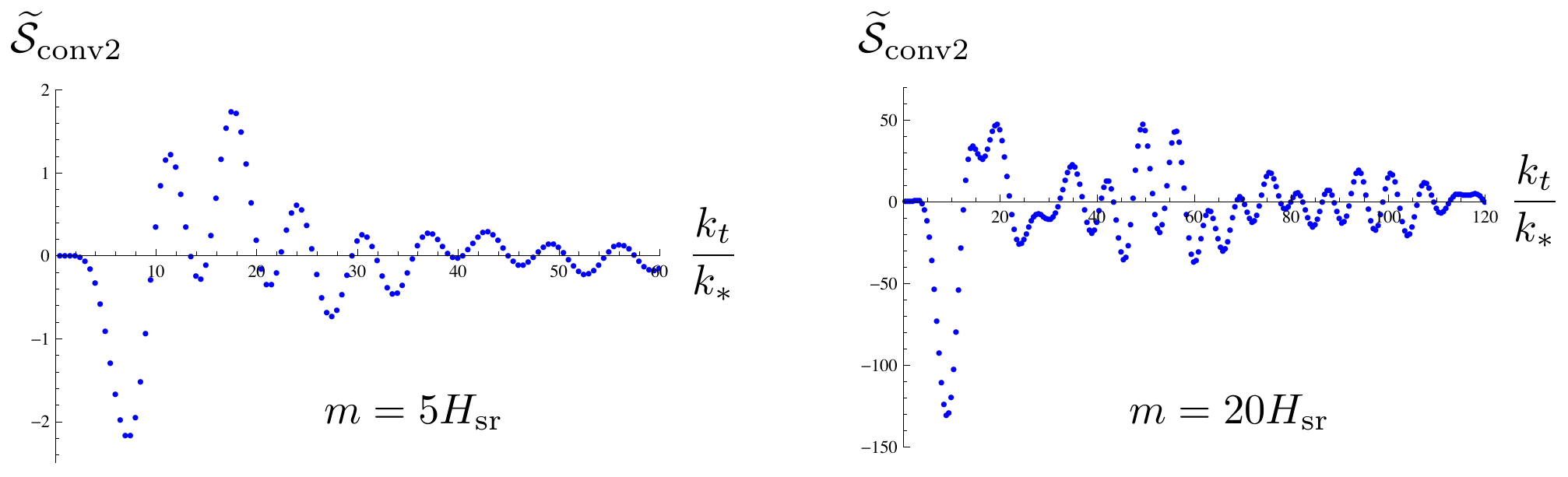}
 \end{center}
\vspace{-5mm}
\caption{Scale-dependence of $S_{\rm conv2}$ for the equilateral momentum configurations.
Here we plotted $\widetilde{S}_{\rm conv2}$
normalized as
$\displaystyle S_{\rm conv2}=\frac{\lambda \alpha^2}{\mu^4}\frac{2M_{\rm Pl}^2\epsilon_{\rm sr}}{H_{\rm sr}^2}\alpha^2\widetilde{S}_{\rm conv2}$.
The shape function $\widetilde{S}_{\rm conv2}$
has a peak at the scale $k\simeq 9 k_\ast$
as well as the resonances
around the scale $k\gtrsim 2\mu k_\ast$ and $k\gtrsim 4\mu k_\ast$.}
\label{fig:Sconv2_eq}
\end{figure}
Let us next evaluate the bispectrum
for the second type of conversion effects.
In terms of $x=-k_t\tau$,
$\mathcal{B}_{\pi,{\rm conv}2}$
and
$S_{\pi,{\rm conv}2}$
can be written as
\begin{align}
\mathcal{B}_{\pi,{\rm conv}2}(k_1,k_2,k_3)&=
\frac{H_{\rm sr}}{8M_{\rm Pl}^4\epsilon_{\rm sr}^{2}}
\frac{k_t^3}{k_1^3k_2^3k_3^3}
\frac{\lambda M_{\rm Pl}^2\epsilon_{\rm sr}}{H_{\rm sr}^2}
{\rm Im}
\left[\int_0^{x_\ast} dx
\frac{\varphi_2}{\dot{\bar{\phi}}_{\rm sr}/H_{\rm sr}} 
\frac{1}{x^4}
\widetilde{\mathcal{F}}_1(x_1)
\widetilde{\mathcal{F}}_2(x_2)
\widetilde{\mathcal{F}}_3(x_3)\right]
\,,\\
S_{{\rm conv}2}(k_1,k_2,k_3)&=
-\frac{1}{2}\frac{k_t^3}{k_1k_2k_3}\frac{\lambda M_{\rm Pl}^2\epsilon_{\rm sr}}{H_{\rm sr}^2}
{\rm Im}
\left[\int_0^{x_\ast} dx
\frac{\varphi_2}{\dot{\bar{\phi}}_{\rm sr}/H_{\rm sr}} 
\frac{1}{x^4}
\widetilde{\mathcal{F}}_1(x_1)
\widetilde{\mathcal{F}}_2(x_2)
\widetilde{\mathcal{F}}_3(x_3)\right]\,.
\end{align}
As discussed in Sec.~\ref{subsubsec:perturbativity},
it is required from the perturbativity
that $\displaystyle\frac{\lambda \alpha^2}{\mu^4}\frac{2M_{\rm Pl}^2\epsilon_{\rm sr}}{H_{\rm sr}^2}\lesssim1$.
With this condition in mind,
we rewrite the shape function as
\begin{align}
S_{{\rm conv}2}(k_1,k_2,k_3)&=
-\frac{1}{4}\frac{k_t^3}{k_1k_2k_3}
\frac{\lambda\alpha^2}{\mu^4}
\frac{2 M_{\rm Pl}^2\epsilon_{\rm sr}}{H_{\rm sr}^2}
{\rm Im}
\left[\int_0^{x_\ast} dx\,
\frac{\mu^4}{\alpha^2}
\frac{\varphi_2}{\dot{\bar{\phi}}_{\rm sr}/H_{\rm sr}} 
\frac{1}{x^4}
\widetilde{\mathcal{F}}_1(x_1)
\widetilde{\mathcal{F}}_2(x_2)
\widetilde{\mathcal{F}}_3(x_3)\right]\,,
\end{align}
from which we can see that
large non-Gaussianities
do not arise unless the integral is large.
Using the analytic expression~(\ref{tilde_mathcal_Fi}),
the calculation of $S_{\rm conv2}$
can be reduced to one integration over $x$
and
we evaluated it numerically.
The scale-dependence of the shape function
$S_{\rm conv2}$ for the equilateral momentum configurations
are for example given by Fig.~\ref{fig:Sconv2_eq},
where
we observe a peak around the turning scale
$k_t\simeq 9k_\ast$
as well as resonances
around $k_t\gtrsim 2\mu k_\ast$ and $k_t\gtrsim 4\mu k_\ast$.

\medskip
We then
estimate the size of the above peak and resonances
using the heavy mass approximation.
Let us first discuss
the resonances around $k_t\gtrsim 2\mu k_\ast$
and $k_t\gtrsim 4\mu k_\ast$.
Extending discussions for the case of $\mathcal{B}_{\rm conv1}$,
the shape function $\mathcal{B}_{\rm conv2}$
in this region
can be estimated as
\begin{align}
\nonumber
\mathcal{B}_{\rm conv2}(k_1,k_2,k_3)&=
\frac{1}{4M_{\rm Pl}^3\epsilon_{\rm sr}^{3/2}(k_1k_2k_3)^{3/2}}{\rm Re}
\left[-i\int_{-\infty}^\infty dt\,a^3
\lambda \varphi_2 
\mathcal{F}_1(t)\mathcal{F}_2(t)\mathcal{F}_3(t)\right]\\
\nonumber
&\sim
\frac{1}{4M_{\rm Pl}^3\epsilon_{\rm sr}^{3/2}(k_1k_2k_3)^{3/2}}
\int_{-\infty}^\infty dt\,a^3
\lambda \varphi_2 \frac{\beta_1^3}{m^6}
\dot{u}_{k_1}^\ast\dot{u}_{k_2}^\ast\dot{u}_{k_3}^\ast\\
\nonumber
&\sim
\frac{1}{4M_{\rm Pl}^3\epsilon_{\rm sr}^{3/2}(k_1k_2k_3)^{3/2}}
\int_{t_\ast}^\infty dt\,a^3
\lambda \frac{\dot{\varphi}_2^4}{m}
u_{k_1}^\ast u_{k_2}^\ast u_{k_3}^\ast\\
\nonumber
&\sim
\frac{1}{4M_{\rm Pl}^3\epsilon_{\rm sr}^{3/2}(k_1k_2k_3)^{3/2}}
\frac{\lambda\alpha^2}{\mu^4}
\frac{2M_{\rm Pl}^2\epsilon_{\rm sr}}{H_{\rm sr}^2}\\*
\label{estimate_conv2_1}
&\qquad
\times
\int_{t_\ast}^\infty dt\,a^3
e^{-3H_{\rm sr}(t-t_\ast)}
\cos^2[m(t-t_\ast)]
\dot{\varphi}_2^2\,m^3
u_{k_1}^\ast u_{k_2}^\ast u_{k_3}^\ast\,,
\end{align}
where we used $\dot{\varphi}_2\simeq
\alpha\, \dot{\bar{\phi}}_{\rm sr}e^{-\frac{3}{2}H_{\rm sr}(t-t_\ast)}\cos[m(t-t_\ast)]$.
We first notice that $\cos^2[m(t-t_\ast)]
\dot{\varphi}_2^2$ in the integrand of~(\ref{estimate_conv2_1})
contains oscillating components with
the frequencies $2m$ and $4m$,
\begin{align}
\cos^2[m(t-t_\ast)]
\dot{\varphi}_2^2=\frac{1}{8}\alpha^2\dot{\bar{\phi}}_{\rm sr}^2e^{-3H_{\rm sr}(t-t_\ast)}
\Big(\cos[4m(t-t_\ast)]+4\cos[2m(t-t_\ast)]+3\Big)\,,
\end{align}
which induce the resonances
around $k_t\gtrsim 2\mu k_\ast$
and $k_t\gtrsim 4\mu k_\ast$.
The size of resonances
can be also estimated
by comparing the Hubble deformation effects.
Since the integral in~(\ref{estimate_conv2_1})
has the same mass-dependence
as the integrals in (\ref{conv1_estimate}) and (\ref{SdH_estimate_for_conv}),
its size can be estimated as $\sim\alpha^2\mu^{5/2}$.
Then, the resonances in the shape function
can be estimated as
\begin{align}
\sim \frac{\lambda\alpha^2}{\mu^4}
\frac{2M_{\rm Pl}^2\epsilon_{\rm sr}}{H_{\rm sr}^2}\alpha^2\mu^{5/2}
\lesssim
\alpha^2\mu^{5/2}\,,
\end{align}
where we used the perturbativity condition
$\displaystyle\frac{\lambda\alpha^2}{\mu^4}
\frac{ 2M_{\rm Pl}^2\epsilon_{\rm sr}}{H_{\rm sr}^2}
\lesssim1$.
This estimation well explains our numerical calculations
in Fig~\ref{fig:Sconv2_eq}.

\begin{figure}[t]
\begin{center}
\includegraphics[width=150mm, bb=0 0 580 241]{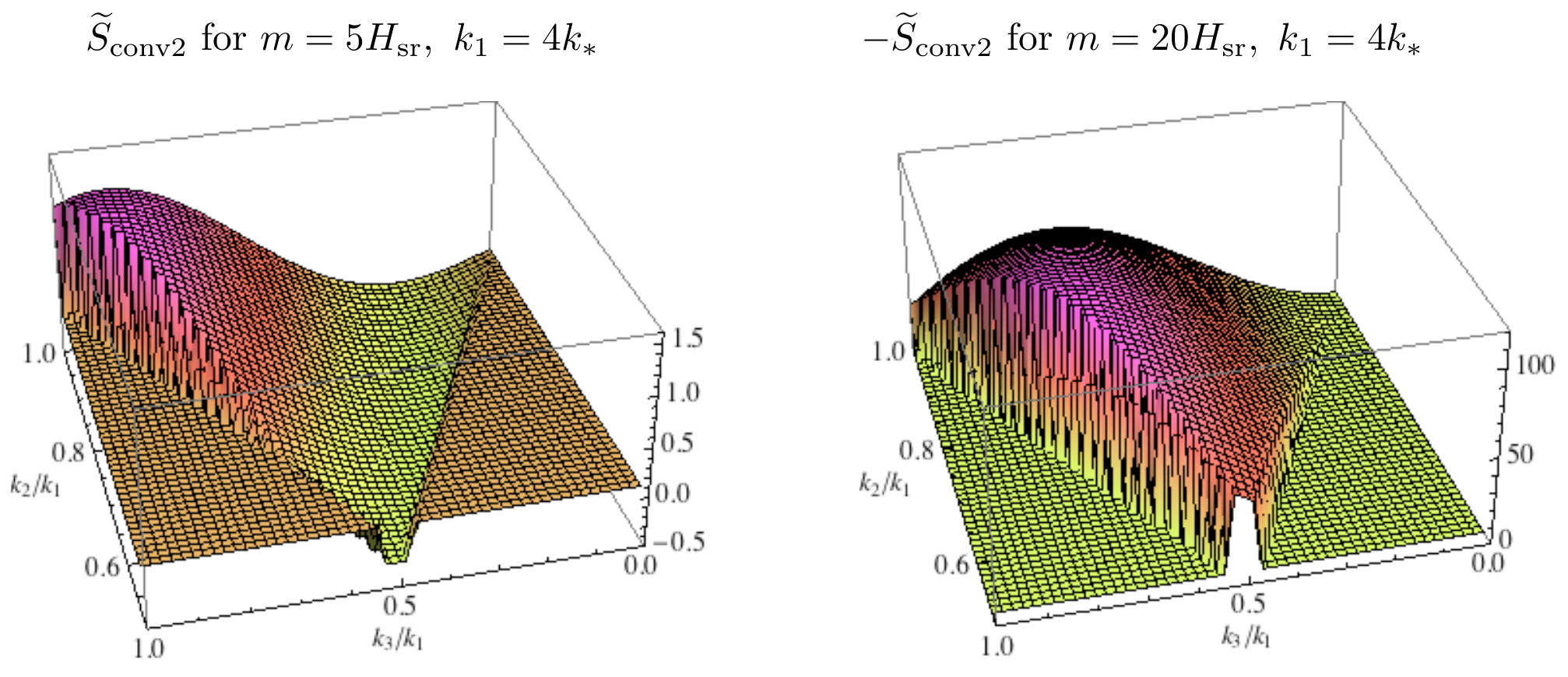}\\[2mm]
\includegraphics[width=150mm, bb=0 0 580 240]{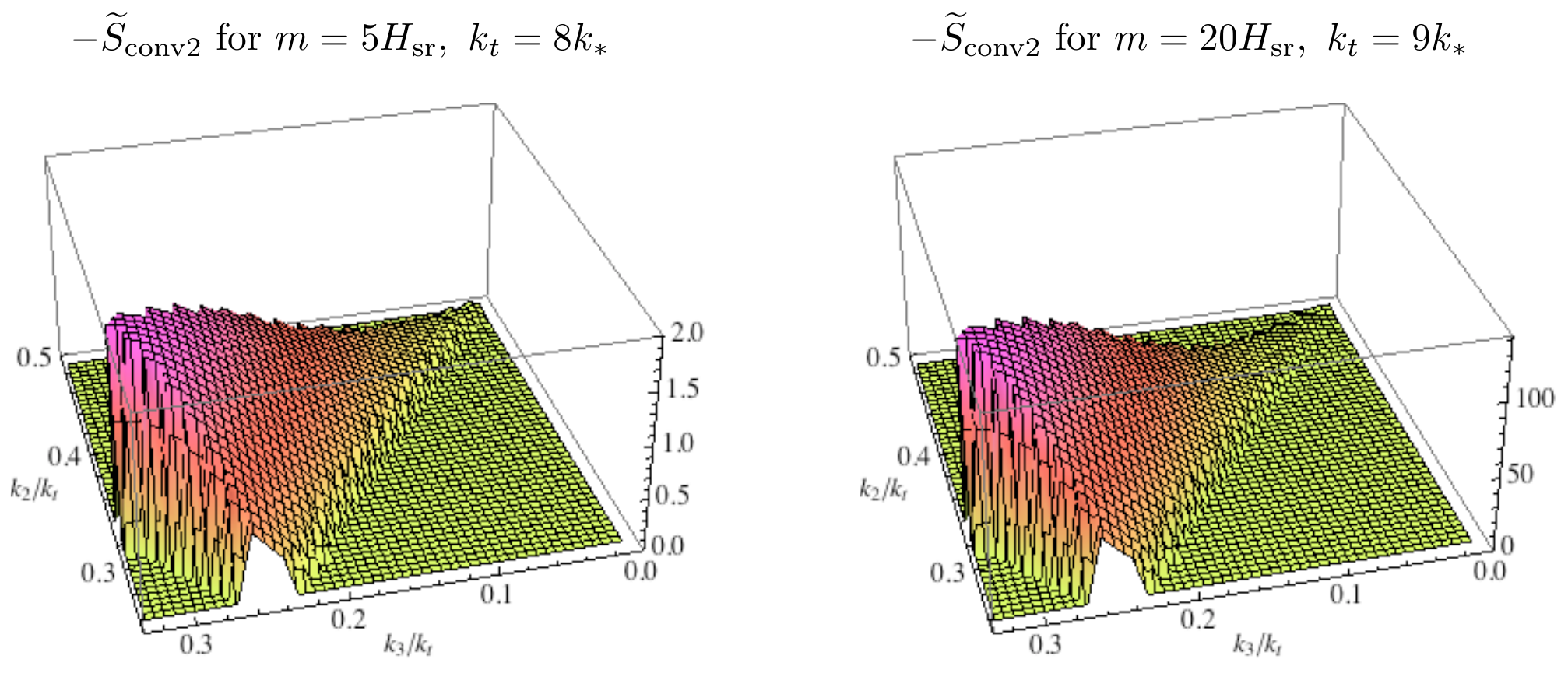}
 \end{center}
\vspace{-5mm}
\caption{Shape of $S_{\rm conv2}$.
In the upper/lower figures,
$\widetilde{S}_{\rm conv2}$ is plotted with maximum/total
momenta being fixed.
We observe that the shape function $S_{\rm conv2}$
around the peak scale
vanishes at the squeezed configuration.}
\label{fig:Sconv2_ktotal}
\end{figure}

\medskip
We next discuss the peak around the turning scale.
The shape function
$S_{{\rm conv}2}$ can be written
using the expression~(\ref{Hamiltonian_varphi2})
as
\begin{align}
\nonumber
S_{{\rm conv}2}(k_1,k_2,k_3)
&=
-\frac{1}{2}\frac{k_t^3}{k_1k_2k_3}
\frac{\lambda M_{\rm Pl}^2\epsilon_{\rm sr}}{H_{\rm sr}^2}
\\*
\label{S_conv2_osci}
&\qquad
\times{\rm Im}
\left[\int_0^{x_\ast} dx\,
\frac{i\alpha}{2\mu}\Big((x/x_\ast)^{\frac{3}{2}+i\mu}-(x/x_\ast)^{\frac{3}{2}-i\mu}\Big)\frac{1}{x^4}
\widetilde{\mathcal{F}}_1(x_1)
\widetilde{\mathcal{F}}_2(x_2)
\widetilde{\mathcal{F}}_3(x_3)\right]\,,
\end{align}
and $\widetilde{\mathcal{F}}_i(x_i)$
around the turning scale
oscillates with high frequency $\sim x^{\pm i\mu}$
as given in~(\ref{deformed_app_1}).
Therefore,
the integrand in~(\ref{S_conv2_osci})
can be seen as
a product of four oscillating functions $\sim x^{\pm i\mu}$.
Since products of
two $x^{i\mu}$'s and two $x^{-i\mu}$'s
lead to non-oscillating components,
the leading contribution to~(\ref{S_conv2_osci})
comes from these non-oscillating components.
Using the previous order estimate $\widetilde{\mathcal{F}}_i\sim\alpha$,
the integral is $\sim\alpha^4\mu^{-1}$
so that the shape function $S_{\rm conv2}$
around the turning scale can be order estimated
as
\begin{align}
S_{\rm conv2}\sim
\frac{\lambda M_{\rm Pl}^2\epsilon_{\rm sr}}{H_{\rm sr}^2}\alpha^4\mu^{-1}
\sim
\frac{\lambda\alpha^2}{\mu^4}
\frac{ 2M_{\rm Pl}^2\epsilon_{\rm sr}}{H_{\rm sr}^2}\alpha^2\mu^{3}
\lesssim \alpha^2\mu^{3}\,,
\end{align}
where the last inequality follows from the perturbativity condition
$\displaystyle\frac{\lambda\alpha^2}{\mu^4}
\frac{ M_{\rm Pl}^2\epsilon_{\rm sr}}{H_{\rm sr}^2}
\lesssim1$.
This estimation well explains our numerical calculations
in Fig.~\ref{fig:Sconv2_eq}.
The shape of $S_{\rm conv2}$
around the tuning scale
can be also discussed in a similar way.
As in Sec.~\ref{subsubsec:deformed},
$\widetilde{\mathcal{F}}_3$ behaves as $\sim \alpha_3^2$
in the squeezed limit $\alpha_3\to0$
so that $S_{\rm conv2}$ vanishes in this limit.
This feature can be confirmed
by our numerical calculations
in Fig.~\ref{fig:Sconv2_ktotal}.

\subsection{Total bispectrum}
At the end of this section,
let us summarize features in the bispectrum.

\paragraph{$\mathcal{O}(\alpha^2)$ contributions}
As we have discussed,
the Hubble deformation effects $S_{\delta H}$
and the first type of conversion effects $S_{\rm conv1}$
are the second order in $\alpha$
and
both of them have resonances
at the scale $k_t\gtrsim 2\mu k_\ast$.
In contrast to the power spectrum,
resonances from the two effects
do not cancel each other out
and the size of total resonances
are estimated as $\mathcal{O}(\alpha^2\mu^{5/2})$.
Around the turning scale,
the shape functions have a peak
at the squeezed momentum configuration
and the peak becomes sharp
as the mass $\mu$ increases.
Using the results of numerical calculations,
the $f_{NL}$ parameter can be
read off as $f_{NL}\sim \alpha^2\mu^{5/2}\times \mathcal{O}(1)$.

\paragraph{$\mathcal{O}(\lambda\alpha^4)$ contributions}
On the other hand,
the second type of conversion effects $S_{\rm conv2}$
are the fourth order in $\alpha$ and
the first order in $\lambda$.
Similarly to the other two contributions $S_{\delta H}$
and $S_{\rm conv1}$,
$S_{\rm conv2}$ also contains resonances.
In this case,
resonances appear around the scale $k\gtrsim 4\mu k_\ast$
as well as $k\gtrsim 2\mu k_\ast$
and their size can be estimated as $\displaystyle\mathcal{O}\Big(\frac{\lambda\alpha^2}{\mu^4}\frac{2M_{\rm Pl}^2\epsilon_{\rm sr}}{H_{\rm sr}^2}\times\alpha^2\mu^{5/2}\Big)$.
In addition to the resonances,
$S_{\rm conv2}$ has a peak around the scale $k\simeq9k_\ast $
and its size is estimated as $\displaystyle\mathcal{O}\Big(\frac{\lambda\alpha^2}{\mu^4}\frac{2M_{\rm Pl}^2\epsilon_{\rm sr}}{H_{\rm sr}^2}\times\alpha^2\mu^{3}\Big)$.
Also discussed that
the shape function $S_{\rm conv2}$
around the peak scale
vanishes in the squeezed limit.
The $f_{NL}$ parameter
around the resonance scale
and at the peak
can be read off
from the numerical results
as
\begin{align}
&f_{NL}\sim \frac{\lambda\alpha^2}{\mu^4}\frac{2M_{\rm Pl}^2\epsilon_{\rm sr}}{H_{\rm sr}^2}\times\alpha^2\mu^{5/2}\times \mathcal{O}(0.1)
\quad
\text{around the resonance scale}
\,,\\
&f_{NL}\sim \frac{\lambda\alpha^2}{\mu^4}\frac{2M_{\rm Pl}^2\epsilon_{\rm sr}}{H_{\rm sr}^2}\times\alpha^2\mu^3\times \mathcal{O}(0.1)
\quad
\text{at the peak}
\,.
\end{align}
Using the perturbativity condition
$\displaystyle\frac{\lambda\alpha^2}{\mu^4}\frac{2M_{\rm Pl}^2\epsilon_{\rm sr}}{H_{\rm sr}^2}\lesssim 1$,
we have
\begin{align}
&f_{NL}\lesssim \alpha^2\mu^{5/2}\times \mathcal{O}(0.1)
\quad
\text{around the resonance scale}
\,,\\
&f_{NL}\lesssim \alpha^2\mu^3\times \mathcal{O}(0.1)
\quad
\text{at the peak}
\,.
\end{align}

\bigskip
To summarize,
the bispectrum in our model has two features:
resonances at $k\gtrsim 2\mu k_\ast$
and a peak at $k\simeq 9k_\ast$.
Because of the perturbativity condition,
the leading contribution to the resonances
are from the Hubble deformation effects $S_{\delta H}$
and the first type of conversion effects $S_{\rm conv1}$.
On the other hand,
the peak arises from the cubic interaction $\sigma^3$
of massive isocurvatures.
Although
a dominant feature changes
depending on the parameter region,
our models are characterized by these two features.

\section{Summary and discussion}
\setcounter{equation}{0}
In this paper,
we discussed effects
of heavy fields on primordial spectra
in inflationary models with sudden turning potentials.
The deviation from single field slow-roll inflation
arises from the deformation of the Hubble parameter
and the conversion interaction between
the adiabatic and the massive isocurvature perturbations,
and those effects
on the power spectrum
and the bispectrum are evaluated.

\medskip
As discussed in Sec.~\ref{sec:PS},
resonance features in the power spectrum
arise from both of the two effects
and their size can be estimated
as $\sim \alpha^2\mu^{1/2}$.
Interestingly,
resonance effects from the two effects
have negative correlations generically
and, in particular,
it was explicitly shown that
such resonance effects cancel each other out
for the case with the canonical kinetic terms.
This resonance cancellation is an important feature in models with canonical kinetic terms,
and therefore, it can be a useful probe to distinguish models with canonical kinetic terms from those with non-canonical ones.
To investigate this possibility,
it would be interesting to extend our discussions
to more general setups such as those with
non-canonical kinetic terms or derivative interactions~\cite{Saito:2012pd,Saito:2013aqa}.
We will discuss this subject elsewhere~\cite{NY}.
The power spectrum
also has a peak at the turning scale.
This feature originates from
the conversion effect
and its size can be estimated
as $\sim \alpha^2\mu$.
In the case with the canonical kinetic terms,
this peak feature becomes clear
because of the resonance cancellation.
In particular,
the total deviation factor
$\displaystyle\mathcal{C}=\frac{\Delta\mathcal{P}_\zeta}{\mathcal{P}_\zeta}$
from single field slow-roll inflation
can be estimated using the heavy mass approximation
as
\begin{align}
\mathcal{C}\simeq
\mathcal{C}_{\rm conv}
\simeq\mu\alpha^2\frac{(\sin x_\ast-x_\ast\cos x_\ast)^2}{x_\ast^3}\,,
\end{align}
which takes the maximum value $\mathcal{C}\sim0.43\,\mu\alpha^2$ at the scale
$x_\ast=k/k_\ast\sim 2.46$.
It is important
that
we can extract the scale heavy field oscillations occurred
from the peak position.
In general,
it is known that
spiky features in primordial power spectra
can induce wavy features on the CMB power spectra
(see e.g.~\cite{Starobinsky:1992ts,Kawasaki:2004pi}).
It would be interesting
to investigate the imprint of this peak feature
on the CMB power spectrum for example:
the remnant of primordial peak features
will tell us when heavy field oscillations occurred during inflation.

\medskip
As discussed in Sec.~\ref{sec:bispectrum},
the bispectrum has resonance and peak features.
In contrast to the case of power spectra,
the resonance cancellation does not
occur in the bispectrum
and both of the resonance and peak features
characterize
this class of models with heavy field oscillations.
The size of the resonance and the peak
can be estimated
as $\lesssim \alpha^2\mu^{5/2}\times\mathcal{O}(1)$
and $\lesssim \alpha^2\mu^{3}\times\mathcal{O}(0.1)$,
respectively.
Although the size of bispectra is not
necessarily large in the perturbative regime,
it would be interesting to
search for this kind of scale-dependent non-Gaussianities
in the observational data.
An important point is that
resonance features in primordial bispectra
generically appear in models with heavy field oscillations
(even in the case of canonical kinetic terms).
Therefore, if the remnant of primordial resonance features
is observed in the bispectra,
it will suggest the existence of heavy field oscillations
during inflation
and, in particular,
we can extract the mass of heavy fields
from the scale of primordial resonances.
Since resonance features do not appear
in the power spectra
for the case with canonical kinetic terms,
we will further obtain details of kinetic terms
by combining the analysis of primordial resonance features
in the power spectra and the bispectra.

\section*{Acknowledgments}
We would like to thank
Xian Gao, David Langlois, and Shuntaro Mizuno
for the correspondence
and valuable discussions.
We are also grateful to
Thorsten Battefeld, Daniel Baumann, and Shi
Pi for useful discussions.
The work of T.N. is supported in part
by Special Postdoctoral Researchers
Program at RIKEN.
The work of M.Y. is supported in part by the
Grant-in-Aid for Scientific Research on Innovative Areas No.~24111706
and the Grant-in-Aid for Scientific Research No.~25287054.\\

\appendix
\section{Sudden limit of turning potentials}
\label{app:sudden}
\setcounter{equation}{0}
\begin{figure}[t]
\begin{center}
\includegraphics[width=120mm, bb=0 0 580 283]{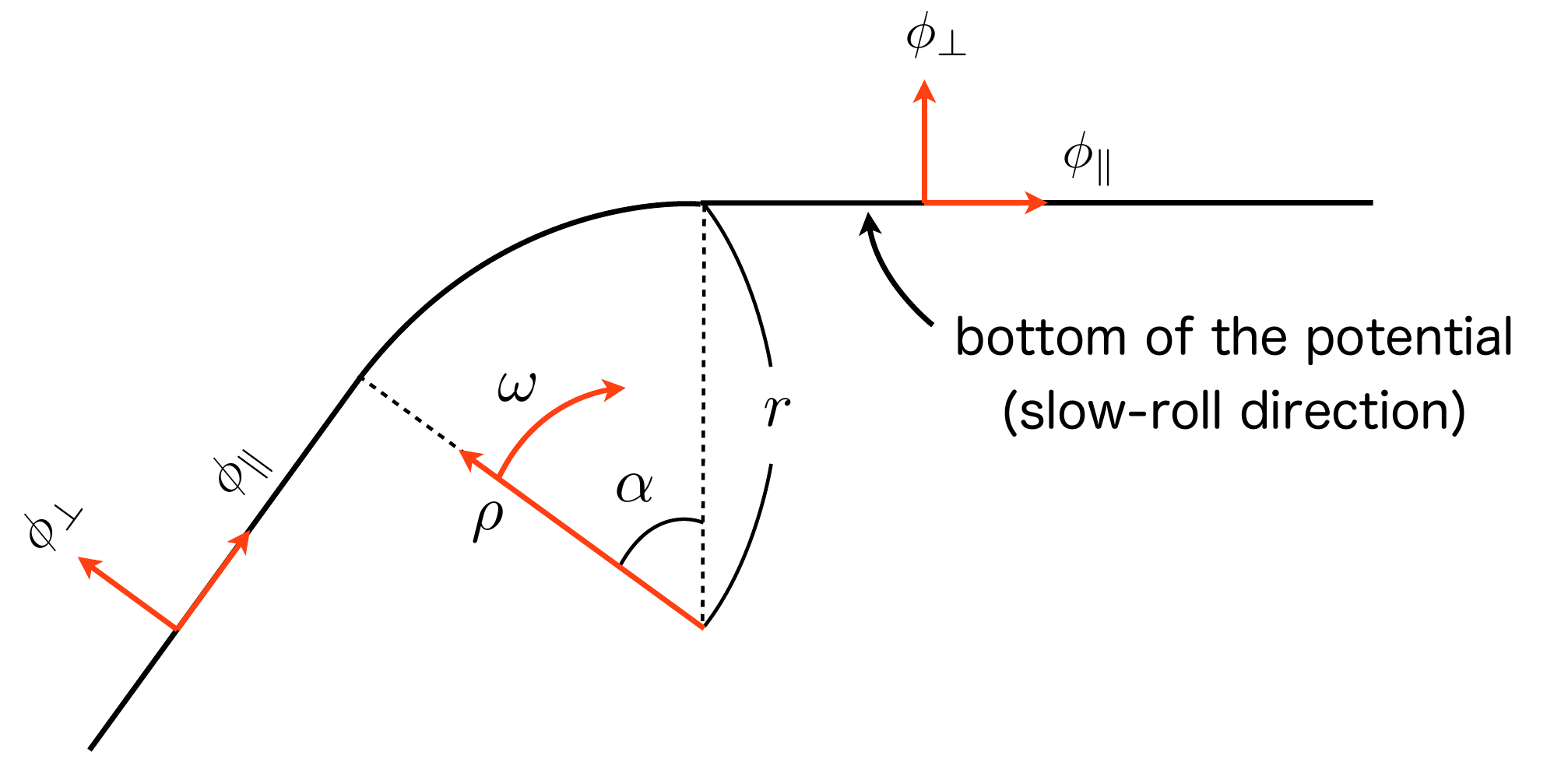}
 \end{center}
\vspace{-5mm}
\caption{Turning potential with turning radius $r$
and turning angle $\alpha$.}
\label{fig:reg_pot}
\end{figure}
In this appendix
we introduce a class
of turning potentials
and discuss its sudden turning limit.
Suppose that the potential
has a slow-roll direction
and its orthogonal direction is massive.
When the slow-roll direction
is along a smooth curve,
we can take
a local coordinate $(\phi_\parallel,\phi_\bot)$
of the field space
such that
$\phi_{\parallel}$
is along the slow-roll direction
and $\phi_{\bot}$
is orthogonal to it.
In such a case,
it is also possible
to define
a separable potential
at least locally:
\begin{align}
V(\phi_i)=V_{\rm sr}(\phi_\parallel)
+V_\bot(\phi_\bot)\,,
\end{align}
where $V_{\rm sr}$ and $V_\bot$
are the slow-roll potential and the massive potential
and $\phi_\bot=0$ is the bottom of the massive potential.
However,
when the slow-roll direction is not smooth
as in the case of sudden turning potentials,
it is not possible
to define such a local coordinate frame
and to introduce a separable potential.
This is nothing but the origin
of the discontinuity in the coordinate frame
in Sec.~\ref{sec:setup}.

\medskip
To clarify this subtlety
and justify our discussion,
let us consider
a class of smoothly turning potentials
and discuss its sudden limit.
As depicted in Fig.~\ref{fig:reg_pot},
we introduce a class of potentials
whose slow-roll direction
turns with a radius $r$ and an angular $\alpha$.
We also assume that the slow-roll direction
is straight before and after the turn.
When the background trajectory
turns along the slow-roll direction
with a velocity $\dot{\bar{\phi}}_{\rm sr}$,
the duration $\Delta t$ of the turn is given by
\begin{align}
\label{def_of_deltat}
\Delta t=\frac{r\alpha}{\dot{\bar{\phi}}_{\rm sr}}\,.
\end{align}
We therefore
would like to define the sudden turning potential
by taking the limit
$\Delta t\to0$ with the turning angle $\alpha$ fixed.
For careful treatments of this limit,
let us introduce the polar coordinate system
$(\rho,\omega)$
of the field space
in the turning regime
such that $(\rho,\omega)=(0,\omega)$
coincides with the center of the turn.
In this coordinate system,
the massive direction coincides with
the radial direction
and
we can define the massive potential $V_\bot$ as
a function of $r$
in the region $r>0$.
Following the setup in Sec.~\ref{sec:setup},
let us define the massive potential as
\begin{align} 
\label{massive_app}
V_\bot=\frac{m^2}{2}(\rho-r)^2+\frac{\lambda}{4!}(\rho-r)^4\,.
\end{align}
In terms of the coordinate $(\phi_\parallel,\phi_\bot)$,
the potential (\ref{massive_app})
can be written as
\begin{align}
\label{massive_app_2}
V_\bot=\frac{m^2}{2}\phi_\bot^2+\frac{\lambda}{4!}\phi_\bot^4\,,
\end{align}
which reproduces the massive potential in Sec.~\ref{sec:setup}.
An important point is that
the description (\ref{massive_app_2})
is valid only in the region $\phi_\bot<r$.
Since quantum fluctuations of massive fields
during inflation can be estimated as
$\displaystyle\sim \frac{H_{\rm sr}^2}{m}$,
it is necessary to assume
$\displaystyle\frac{H_{\rm sr}^2}{m}\lesssim r$
for the use of the expression~(\ref{massive_app_2}).
Therefore,
to define the sudden limit of the turning potential,
we require both of the following two conditions:
\begin{align}
\frac{H_{\rm sr}^2}{m}\lesssim r\,,
\quad
\Delta t\ll \frac{1}{H_{\rm sr}}\,.
\end{align}
Using the definition (\ref{def_of_deltat}) of $\Delta t$,
these conditions can be rephrased as 
\begin{align}
\frac{H_{\rm sr}^2}{m}\lesssim r\ll \frac{\dot{\bar{\phi}}_{\rm sr}}{\alpha H_{\rm sr}}
= \frac{\sqrt{2}M_{\rm Pl}\epsilon_{\rm sr}^{1/2}}{\alpha}
\sim 10^4\times \frac{H_{\rm sr}}{\alpha}\,,
\end{align}
which can be naturally realized in our setup.
We then define the sudden turning potential
by taking the limit $\Delta t\to0$ with $\alpha$ fixed and
$\displaystyle\frac{H_{\rm sr}^2}{m}\lesssim r$ satisfied.
Based on this definition,
the use of the massive potential~(\ref{massive_app_2}) can be validated
and
our discussions in Sec.~\ref{sec:setup}
can be also justified.

\section{Properties of the function $\mathcal{I}(\delta,n,x)$}
\label{app:mathcalI}
\setcounter{equation}{0}
As discussed in Sec.~\ref{subsec:Hubble}
and Sec.~\ref{subsec:SdH},
the Hubble deformation effects $\mathcal{C}_{\delta H}$ and $S_{\delta H}$
on primordial spectra
contain the function
$\mathcal{I}(\delta,n,x)$:
\begin{align}
\nonumber
\mathcal{I}(\delta,n,x)
&=\int_0^x d\tilde{x}\,(\tilde{x}/x)^{3+\delta}\tilde
{x}^ne^{-i\tilde{x}} 
\\
&=i^{-1-n}\left(\frac{1}{ix}\right)^{3+\delta}
\Big[\Gamma(4+n+\delta)-\Gamma(4+n+\delta,ix)\Big]\,.
\end{align}
In this appendix
its asymptotic behavior
in the limits $x\to0$ and
$x\to\infty$,
and its properties relevant to the resonance effects
are discussed.

\subsection{Asymptotic behavior of $\mathcal{I}(\delta,n,x)$}
Let us start from the asymptotic behavior
in the limits $x\to0$ and $x\to\infty$.
Using the asymptotic expansions
of the incomplete gamma function,
\begin{align}
\Gamma(\alpha,z)=\Gamma(\alpha)-z^\alpha
\sum_{k=0}^\infty\frac{(-z)^k}{(\alpha+k)\,k!}\,,
\end{align}
the asymptotic behavior of $\mathcal{I}(\delta,n,x)$
in the limit $x\to0$
can be obtained as
\begin{align}
\mathcal{I}(\delta,n,x)
&=x^{n+1}\sum_{k=0}^\infty\frac{(-ix)^k}{(4+n+\delta+k)\,k!}\\
\label{I_asy_0}
&=\left\{\begin{array}{ccl}
\displaystyle 0&\quad{\rm for}\quad& n=0,1\,,
\\[2mm]
\displaystyle\frac{1}{3+\delta}&\quad{\rm for}\quad& n=-1\,,
\\[2mm]
\displaystyle\frac{1}{(2+\delta)x}-\frac{i}{3+\delta}&\quad{\rm for}\quad&n=-2\,.
\end{array}\right.
\end{align}
As discussed in Sec.~\ref{subsec:Hubble}
and Sec.~\ref{subsec:SdH},
$\mathcal{C}_{\delta H}$ and $S_{\delta H}$
contain $\mathcal{I}(\delta,n,2x_\ast)$ with $n=0,-1,-2$ and $\mathcal{I}(\delta,n,x_\ast)$
with $n=1,0,-1,-2$, respectively.
It may be therefore wondered that
they diverge in the limit $x_\ast\to0$.
However,
using the expression~(\ref{I_asy_0}),
we can show that
they vanish in this limit
so that
they are IR-safe:
\begin{align}
\lim_{x_\ast\to0}\mathcal{C}_{\delta H}(x_\ast)=0\,,
\quad
\lim_{x_\ast\to0}S_{\delta H}(x_\ast)=0\,.
\end{align}
Let us next consider the asymptotic behavior
in the limit $x\to\infty$.
For this purpose,
it is convenient to note the following asymptotic expansions:
\begin{align}
\Gamma(\alpha,z)=e^{-z}z^{\alpha-1}\left(
1-\frac{1-\alpha}{z}+\mathcal{O}(z^{-2})\right)\,.
\end{align}
Using this expansions,
it follows that
\begin{align}
\nonumber
\mathcal{I}(\delta,n,x)
&=-i^{-1-n}e^{-ix}(ix)^n\left(1+\frac{3+n+\delta}{ix}+\mathcal{O}(x^{-2})\right)
\\
\label{I_asy_inf}
&=\left\{\begin{array}{ccl}
\displaystyle 
e^{-ix}(ix+4+\delta)
&\quad{\rm for}\quad& n=1\,,
\\[2mm]
\displaystyle ie^{-ix}&\quad{\rm for}\quad& n=0\,,
\\[2mm]
\displaystyle0&\quad{\rm for}\quad&n=-1,-2\,,
\end{array}\right.
\end{align}
which oscillates and diverges
for $n=0,1$ in the large $x$ region.
As in the case of the $x\to0$ limit,
it follows from~(\ref{I_asy_inf})
that
these singular behaviors cancel out
in $\mathcal{C}_{\delta H}(x_\ast)$
and it vanishes in the $x_\ast\to\infty$ limit:
$\displaystyle\lim_{x_\ast\to\infty}\mathcal{C}_{\delta H}(x_\ast)=0$.
On the other hand,
however,
such cancellation does not occur
in $\mathcal{S_{\delta H}}$
and
it turns out that
the bispectrum $\mathcal{S}_{\delta H}$
is singular in this limit:
$\displaystyle\lim_{x_\ast\to \infty}\mathcal{S}_{\delta H}
\propto x_\ast e^{-ix_\ast}$.
Note that
the mode $k=-x/\tau$ is deep inside the horizon
in the region $x\gg1$.
This kind of oscillating singular behaviors
are common in the interacting theory
and they are usually eliminated by the $i\epsilon$-prescription,
which is based on the assumption
that the Bunch-Davies vacuum is realized
deep inside the horizon.
Following this physical assumptions,
we would like to eliminate
such oscillating parts from $\mathcal{S}_{\delta H}$
and introduce the following regularized functions
$\mathcal{I}_{\,\rm reg}(\delta,1,x)$
and
$\mathcal{I}_{\,\rm reg}(\delta,0,x)$:
\begin{align}
\mathcal{I}_{\,\rm reg}(\delta,1,x)
&=
\mathcal{I}(\delta,1,x)-e^{-ix}(ix+4+\delta)\,,\\*
\mathcal{I}_{\,\rm reg}(\delta,0,x)
&=
\mathcal{I}(\delta,0,x)-ie^{-ix}\,,
\end{align}
which vanish in the $x\to\infty$ limit.
\begin{figure}[t]
\begin{center}
\includegraphics[width=160mm, bb=0 0 580 422]{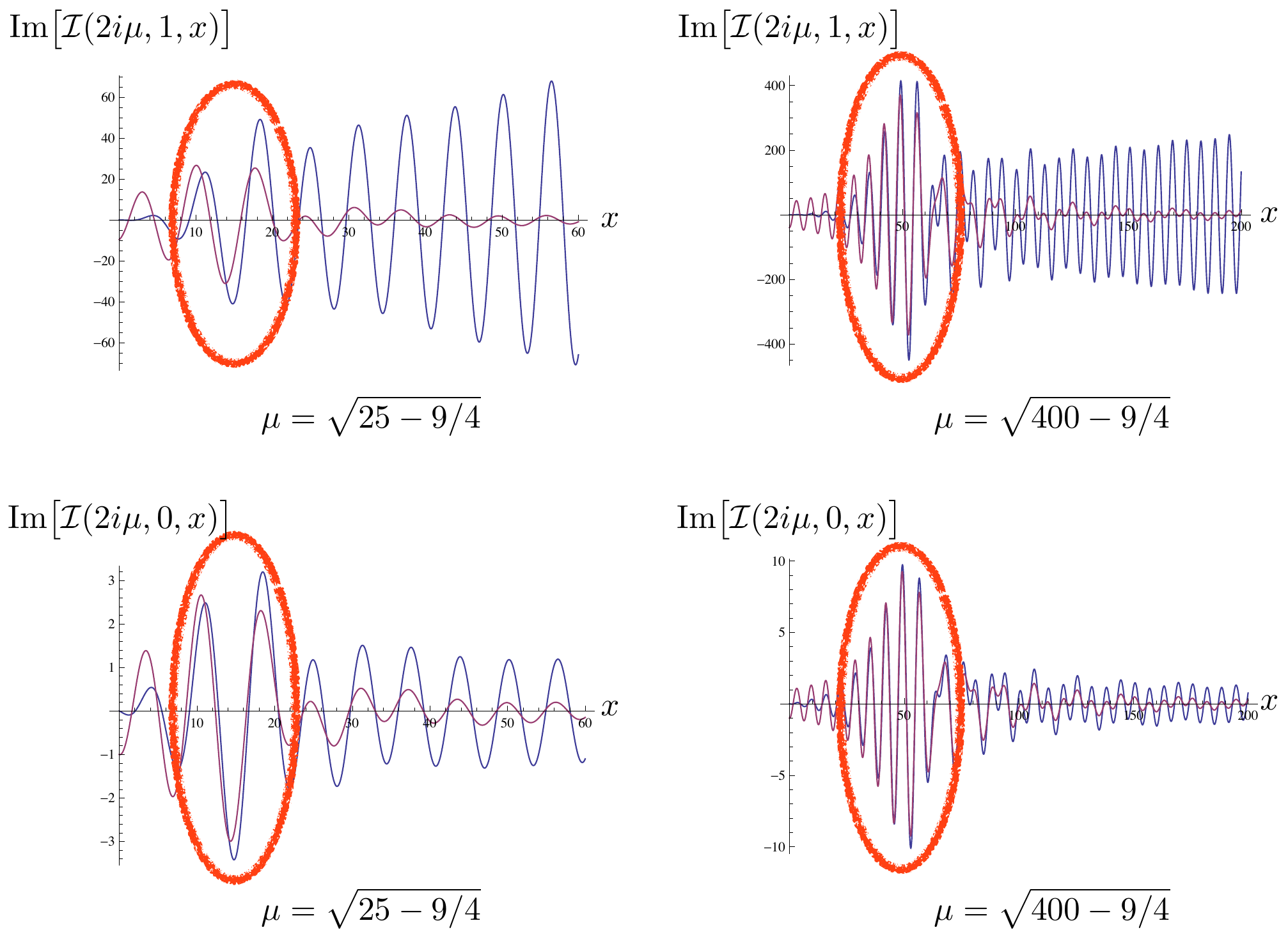}
 \end{center}
\vspace{-5mm}
\caption{
Comparison of $\mathcal{I}_{\,\rm reg}$ (red)
and $\mathcal{I}$ (blue).
The regularized function $\mathcal{I}_{\,\rm reg}$
well reproduces the resonance features
circled with thick red curves
of the original function $\mathcal{I}$
while
singular behaviors
deep inside the horizon
are removed appropriately.}
\label{fig:CalIreg}
\end{figure}
Note that
the resonance features in $\mathcal{I}$
are well reproduced by $\mathcal{I}_{\,\rm reg}$
as in Fig.~\ref{fig:CalIreg}.
In the bispectrum calculation of this paper,
we use the original function $\mathcal{I}$
for $x\lesssim2\mu$
and
the regularized function $\mathcal{I}_{\,\rm reg}$
for $x\gtrsim 2\mu$.
Physically speaking,
this interpolation procedure corresponds to assume
that the Bunch-Davies vacuum is realized
before the resonance era,
where the mode is deep inside the horizon.
In the following subsections,
we discuss the relation
between our procedure and the stationary phase approximation,
which is often used to evaluate integrals of oscillating functions.

\subsection{Stationary phase approximation for $\mathcal{I}(\delta,n,x)$}
\begin{figure}[t]
\begin{center}
\includegraphics[width=160mm, bb=0 0 580 422]{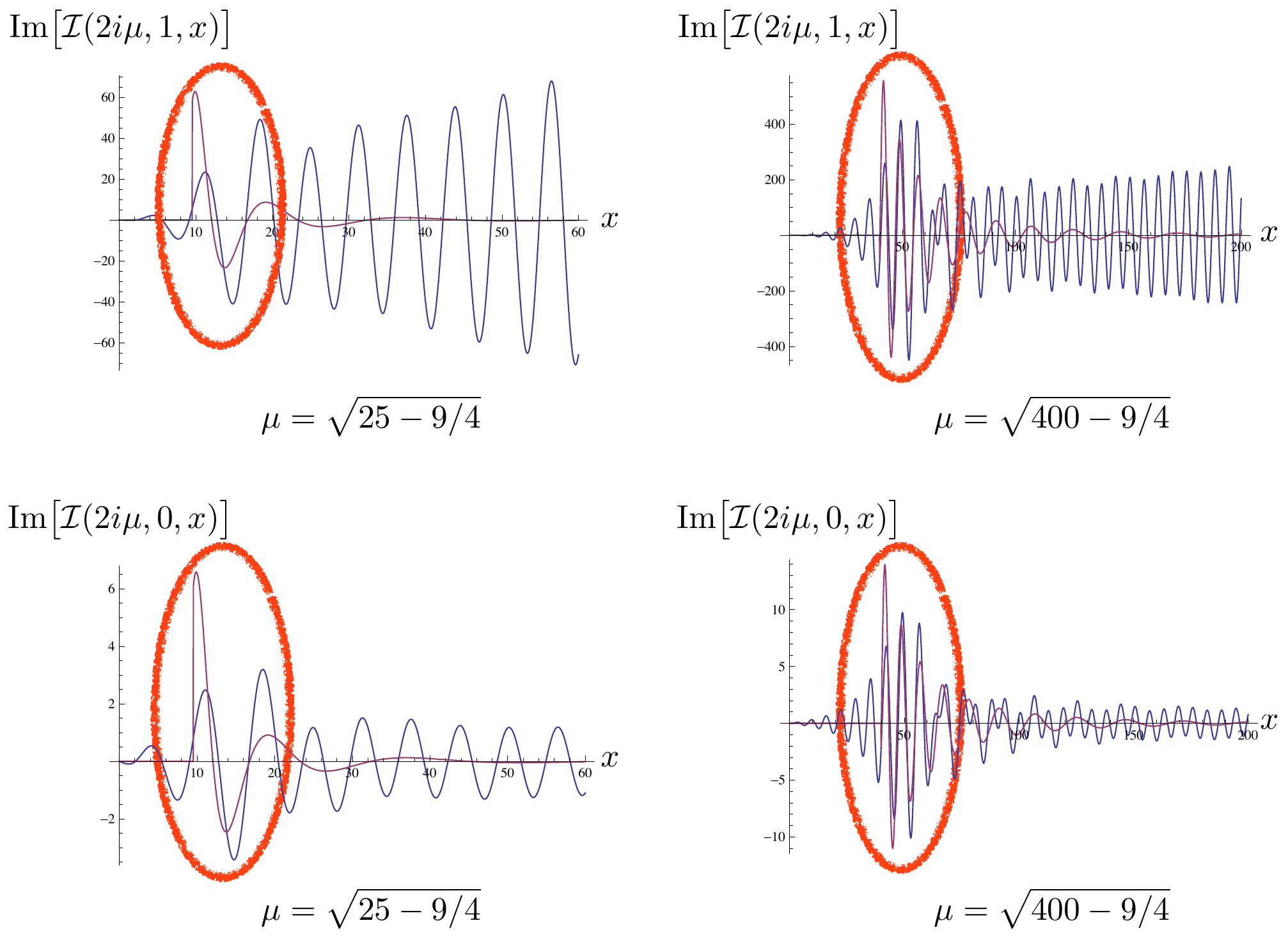}
 \end{center}
\vspace{-5mm}
\caption{
Stationary phase approximation for $\mathcal{I}$.
The red curves are results of the stationary phase approximation and the blue ones are
the original function $\mathcal{I}$.
Resonance features of the original function
are well reproduced except for $x\sim 2\mu$,
where the approximation breaks down.
Also notice that
singular behaviors deep inside the horizon
are removed in this approximation.}
\label{fig:CalISPA}
\end{figure}
Let us then
apply the stationary phase approximation
to the function $\mathcal{I}(\delta,n,x)$:
\begin{align}
\nonumber
\mathcal{I}(\delta,n,x)
&=\int_0^x d\tilde{x}\,(\tilde{x}/x)^{3+\delta}\tilde
{x}^ne^{-i\tilde{x}}\\
&=x^{1+n}\int_0^1dy\,y^{3+n}e^{i\,\Omega(y)}
\quad
{\rm with}
\quad
\Omega(y)=-i\delta \ln y-xy\,.
\end{align}
It is obvious
from this expression
that
the phase factor $e^{i\Omega}$
highly oscillates
for large $x$
and the integral is suppressed by its frequency
unless the phase function $\Omega(y)$
has a stationary point $\Omega^\prime=0$.
While it has no stationary points
for $\delta=0,-2i\mu$,
there is a stationary point for the case $\delta=2i\mu$:
\begin{align}
\Omega^\prime(y_\ast)=0
\quad
\leftrightarrow
\quad
y_\ast=\frac{2\mu}{x}\,.
\end{align}
The integral has a non-negligible contribution
around this stationary point,
which can be observed as the resonance.
Let us then evaluate $\mathcal{I}(2i\mu,n,x)$
using the following Gaussian approximation:
\begin{align}
\nonumber
\mathcal{I}(2i\mu,n,x)
&\simeq
x^{1+n}(y_\ast)^{3+n}e^{i\,\Omega(y_\ast)}
\int_0^1dy\,\exp\left[\frac{i}{2}\,\Omega^{\prime\prime}(y_\ast)(y-y_\ast)^2\right]
\\
\label{Gaussian_app1}
&=x^{1+n}\left(\frac{2\mu}{x}\right)^{3+n+2i\mu}e^{-2i\mu}
\sqrt{\frac{4\mu}{x^2}}\int_{-\sqrt{\mu}}^{\frac{x}{\sqrt{4\mu}}-\sqrt{\mu}}d\tilde{y}\,\exp\left[-i\tilde{y}^2\right]\,.
\end{align}
For $x\gtrsim2\mu\gg1$,
we further approximate it as
\begin{align}
\label{Gaussian_app2}
\int_{-\sqrt{\mu}}^{\frac{x}{\sqrt{4\mu}}-\sqrt{\mu}}d\tilde{y}\,\exp\left[-i\tilde{y}^2\right]
\simeq
\int_{-\infty}^\infty d\tilde{y}\,\exp\left[-i\tilde{y}^2\right]=\sqrt{\frac{\pi}{i}}\,.
\end{align}
Note that
the approximation
becomes worse for $x\sim 2\mu$
because the upper limit of the integral becomes zero
for $x=2\mu$:
$\displaystyle\frac{x}{\sqrt{4\mu}}-\sqrt{\mu}=0$
(see Fig.~\ref{fig:CalISPA}).
Also note that
the integral~(\ref{Gaussian_app1})
vanishes in the limit $\mu\to\infty$
for $x<2\mu$.
By contour deformations of the integral~(\ref{Gaussian_app2}),
we obtain the following expression
of $\mathcal{I}(2i\mu,n,x)$ in the stationary phase approximation:
\begin{align}
\mathcal{I}(2i\mu,n,x)
&\simeq
\left(\frac{2\mu}{x}\right)^{3+2i\mu}
(2\mu)^{n+\frac{1}{2}}\sqrt{2\pi}e^{-2i\mu-\frac{i}{4}\pi}\,.
\end{align}
As depicted in Fig.~\ref{fig:CalISPA},
the resonance effects in $\mathcal{I}(2i\mu,n,x)$
are well reproduced
by the stationary phase
approximation
except for $x\sim2\mu$,
where the approximation breaks down
as discussed above.
Also notice that
singular behaviors deep inside the horizon
are regularized and removed in this approximation.

\subsection{Interpolation procedure vs stationary phase approximation}
\begin{figure}[t]
\begin{center}
\includegraphics[width=160mm, bb=0 0 580 396]{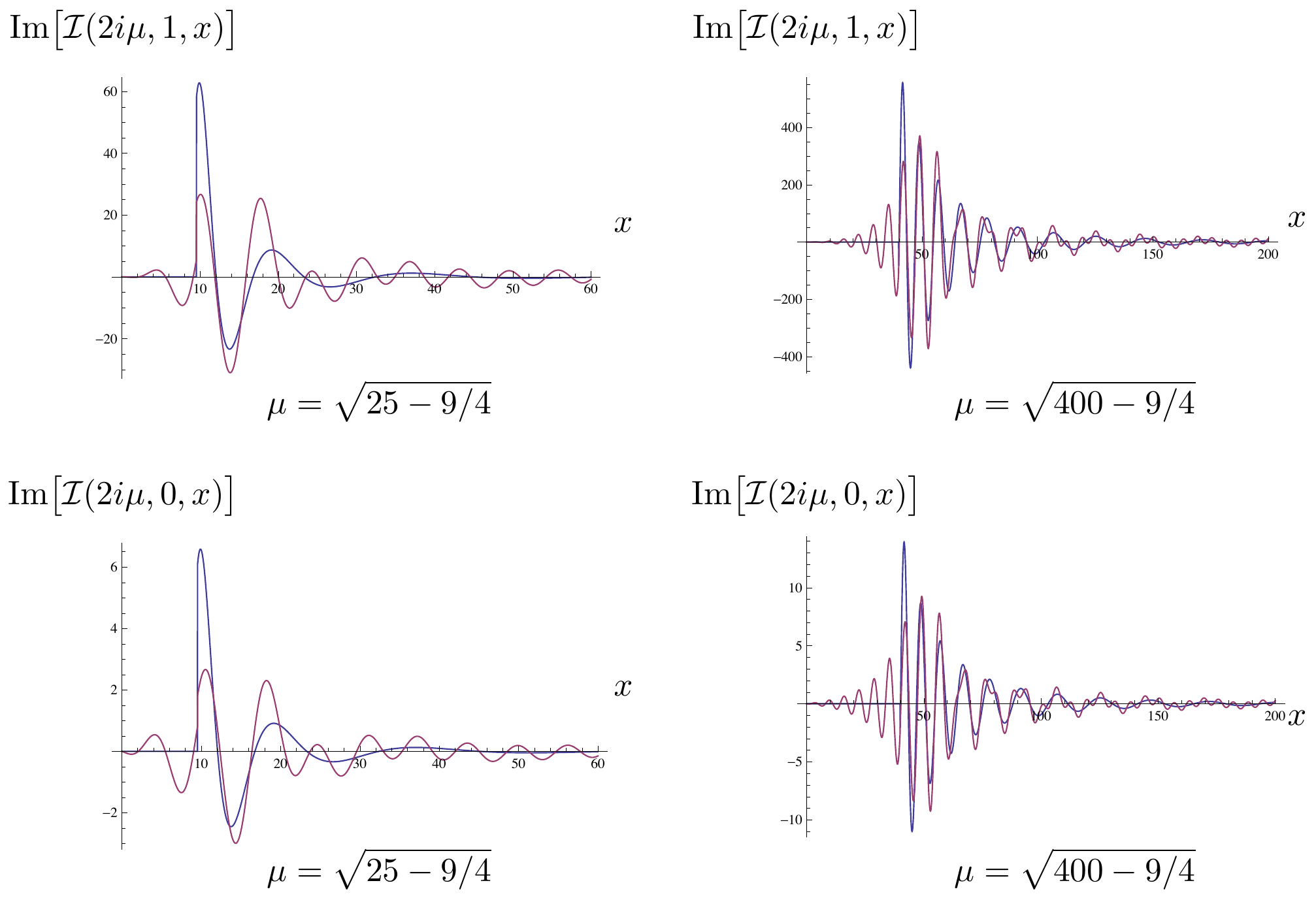}
 \end{center}
\vspace{-5mm}
\caption{
Comparison of our interpolation procedure (red)
and the stationary phase approximation (blue).
Both results well agree except for $x\sim2\mu$,
where the latter does not provide a good approximation.
The agreement for $\mu=\sqrt{400-9/4}$
is better than that for $\mu=\sqrt{25-9/4}$
because the use of these procedures
is justified in the large $\mu$ region.}
\label{fig:reg_vs_SPA}
\end{figure}
Finally,
we compare our interpolation procedure
and the stationary phase approximation.
As we have discussed,
both the procedures
extract the resonance features in the integral
by removing singular behaviors deep inside the horizon.
Our procedure is therefore considered
to be essentially the same
as the stationary phase approximation.
In fact, as depicted in Fig.~\ref{fig:reg_vs_SPA},
their results well agree except for $x\sim2\mu$,
where the stationary phase approximation
becomes worse.
It should be noticed, however,
the use of these procedures
is not necessarily justified in the small $\mu$ region:
First,
the stationary phase approximation
does not provides a good approximation
in such a region
because higher order terms in $1/\mu$-expansions
are dropped.
Second,
our procedure
is based on the assumption
that
the Bunch-Davies vacuum is realized
deep inside the horizon.
For small $\mu$,
however,
the mode is not so deep inside the horizon
at the resonance era.
Therefore,
in order to justify the use of our procedure,
it should be
specified how deep inside the horizon
the Bunch-Davies vacuum is realized.
In this sense,
our procedure depends on the details
of the assumption in the small mass region.
Reflecting these subtleties,
the agreement between both the procedures
in the small $\mu$ region
is not so well compared to the large $\mu$ region.

\medskip
To summarize,
our interpolation procedure
is essentially the same as
the stationary phase approximation
and its use would be justified
at least in the heavy mass region $\mu\gtrsim10$.
For small $\mu$,
however,
it may be not suitable for quantitative discussions
because of the subtleties in removing singular behaviors
deep inside the horizon
although it should be useful for qualitative discussions
even in the small $\mu$ region.

\end{document}